\documentclass[structabstract]{aa}
\usepackage[utf8]{inputenc}
\usepackage{txfonts} 
\usepackage{graphicx} 
\usepackage{xcolor}
\usepackage{natbib} 
\usepackage{hyperref} 

\bibpunct{(}{)}{;}{a}{}{,} 

\newcommand{\ie}{{i.e.}} 
\newcommand{\eg}{{e.g.}}

\newcommand{\emm}[1]{\ensuremath{#1}}
\newcommand{\emr}[1]{\emm{\mathrm{#1}}}
\newcommand{\chem}[1]{\emr{#1}} 
\newcommand{\unit}[1]{\emr{\,#1}}

\newcommand{\radec}[6]{\emr{#1^{h}#2^{m}#3^{s},#4^{\circ}#5^{'}#6^{''}}}

\newcommand{\pc}{\unit{pc}} 
\newcommand{\kpc}{\unit{kpc}} 
\newcommand{\mum}{\unit{\mu m}}

\newcommand{\pccm}{\unit{cm^{-3}}} 
\newcommand{\pscm}{\unit{cm^{-2}}} 

\newcommand{\kms}{\unit{km\,s^{-1}}}

\newcommand{\K}{\unit{K}}
\newcommand{\GHz}{\unit{GHz}} 
\newcommand{\Kkms}{\unit{K\,km\,s^{-1}}}
\newcommand{\asinh}{\unit{asinh}} 
\newcommand{\thCO}{\chem{^{13}CO}}
\newcommand{\twCO}{\chem{^{12}CO}} 
\newcommand{\CeiO}{\chem{C^{18}O}}
\newcommand{\HCOp}{\chem{HCO^{+}}}
\newcommand{\HNC}{\chem{HNC}}
\newcommand{\CCH}{\chem{CCH}}
\newcommand{\twCS}{\chem{^{12}CS}}
\newcommand{\NNHp}{\chem{N_{2}H^{+}}}
\newcommand{\methanol}{\chem{CH_{3}OH}}
\newcommand{\Jone}{(1--0)}
\newcommand{\Jtwo}{(2--1)}

\newcommand{\Ht}{\emr{H_2}} 

\newcommand{\Av}{\emr{A_v}} 
\newcommand{\NHt}{\emm{(N_{\Ht} / \pscm)}}
\newcommand{\Ndust}{\emm{N^\emr{d}_\Ht}}

\newcommand{\Xco}{\emm{X_\emr{CO}}}

\newcommand{\Hii}{\ion{H}{ii}} 
\newcommand{\dix}[2]{\emm{#1 \times 10^{#2}}}

\newcommand{\mycaption}[1]{%
  \rotatebox{270}{\parbox{\PanelHeight}{%
      \centering{} #1}}}

\newcommand{\mycaptionbis}[2]{%
  \rotatebox{90}{\parbox{#1}{%
      \centering{} #2}}}

\definecolor{vs}{rgb}{0.1,0.4,0.1}                 

\newcommand{\cbrace}[1] {\emm{\left\{ #1 \right\}}}

\newcommand{\referee}[1]{{#1}}

\newcommand{\FigDataArea}{%
  \begin{figure*}
    \centering %
    \includegraphics[width=\linewidth]{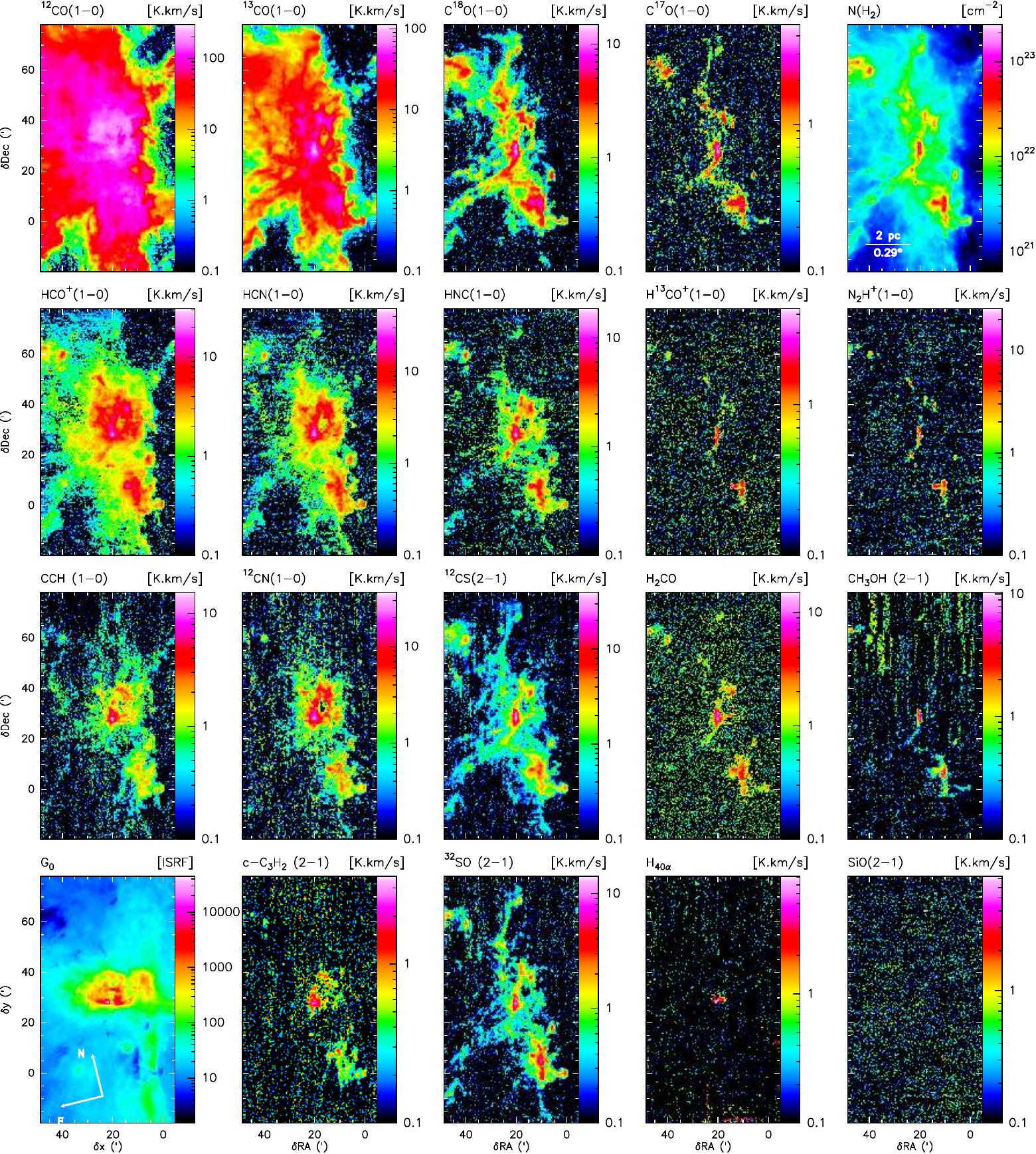}
    \caption{Spatial distribution of the line integrated intensity for some
      of the detected lines in the 3\,mm band, plus the dust-traced far-UV
      illumination (bottom left panel) and the dust-traced H$_2$ column
      density (top right corner).  The color-scales are logarithmic to
      reveal the distribution of faint signal and positive noise. The maps
      are rotated counter-clockwise by 14 degrees from the RA/DEC J2000
      reference frame. The spatial offsets are given in arcsecond from the
      projection center located at \radec{05}{40}{54.270}{-02}{28}{00.00}.}
    \label{fig:data:area}
  \end{figure*}}

\newcommand{\FigDataPeak}{%
  \begin{figure*}
    \centering %
    \includegraphics[width=\linewidth]{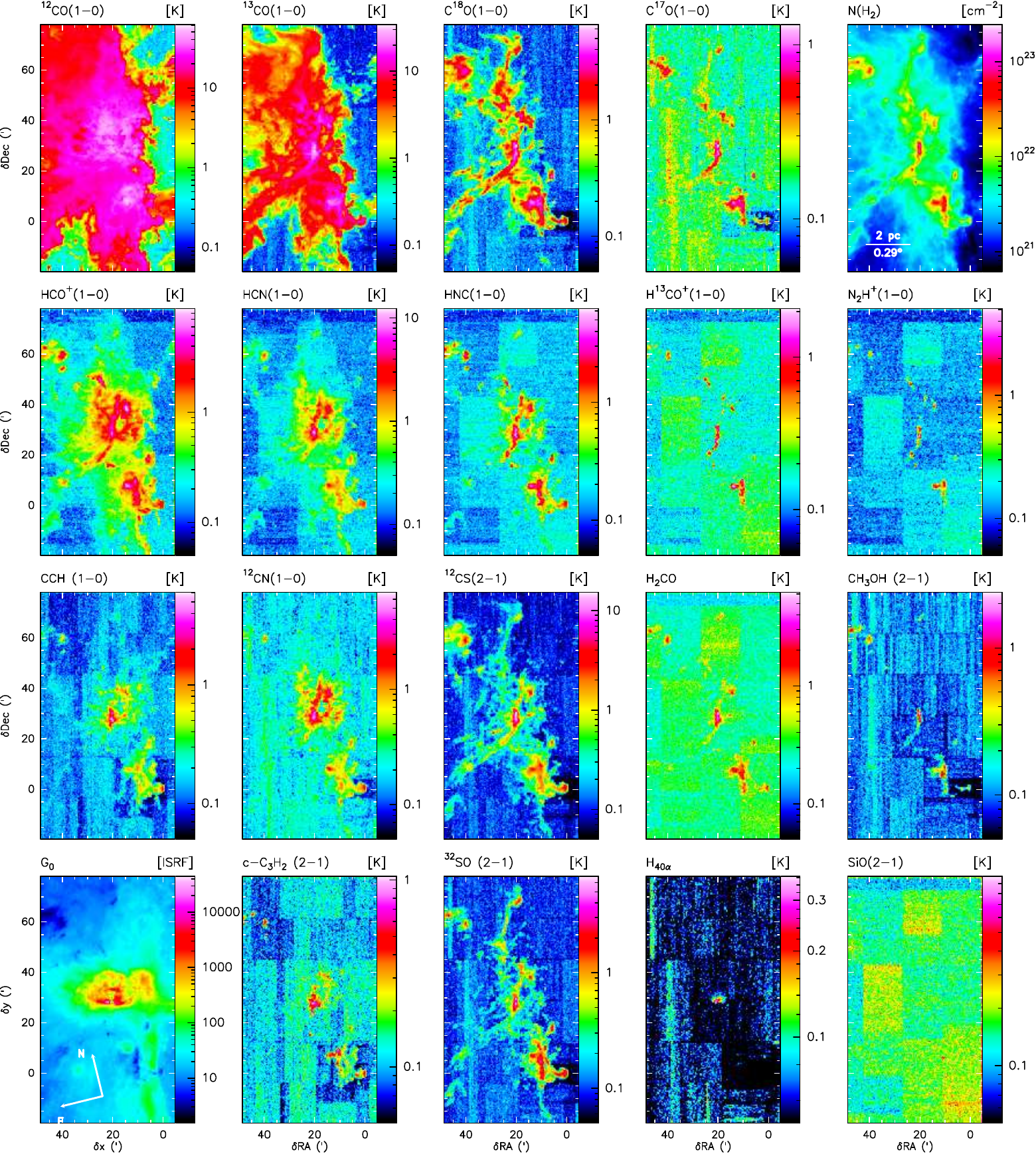}
    \caption{Spatial distribution of the line peak temperature for some of
      the detected lines in the 3\,mm band, plus the dust-traced far-UV
      illumination (bottom left panel) and the dust-traced H$_2$ column
      density (top right corner).  The color-scales are logarithmic to
      reveal the distribution of faint signal and positive noise. The maps
      are rotated counter-clockwise by 14 degrees from the RA/DEC J2000
      reference frame. The spatial offsets are given in arcsecond from the
      projection center located at \radec{05}{40}{54.270}{-02}{28}{00.00}.}
    \label{fig:data:peak}
  \end{figure*}}

\newcommand{\FigHorsehead}{%
  \begin{figure*}
    \centering %
    \includegraphics[width=0.9\linewidth]{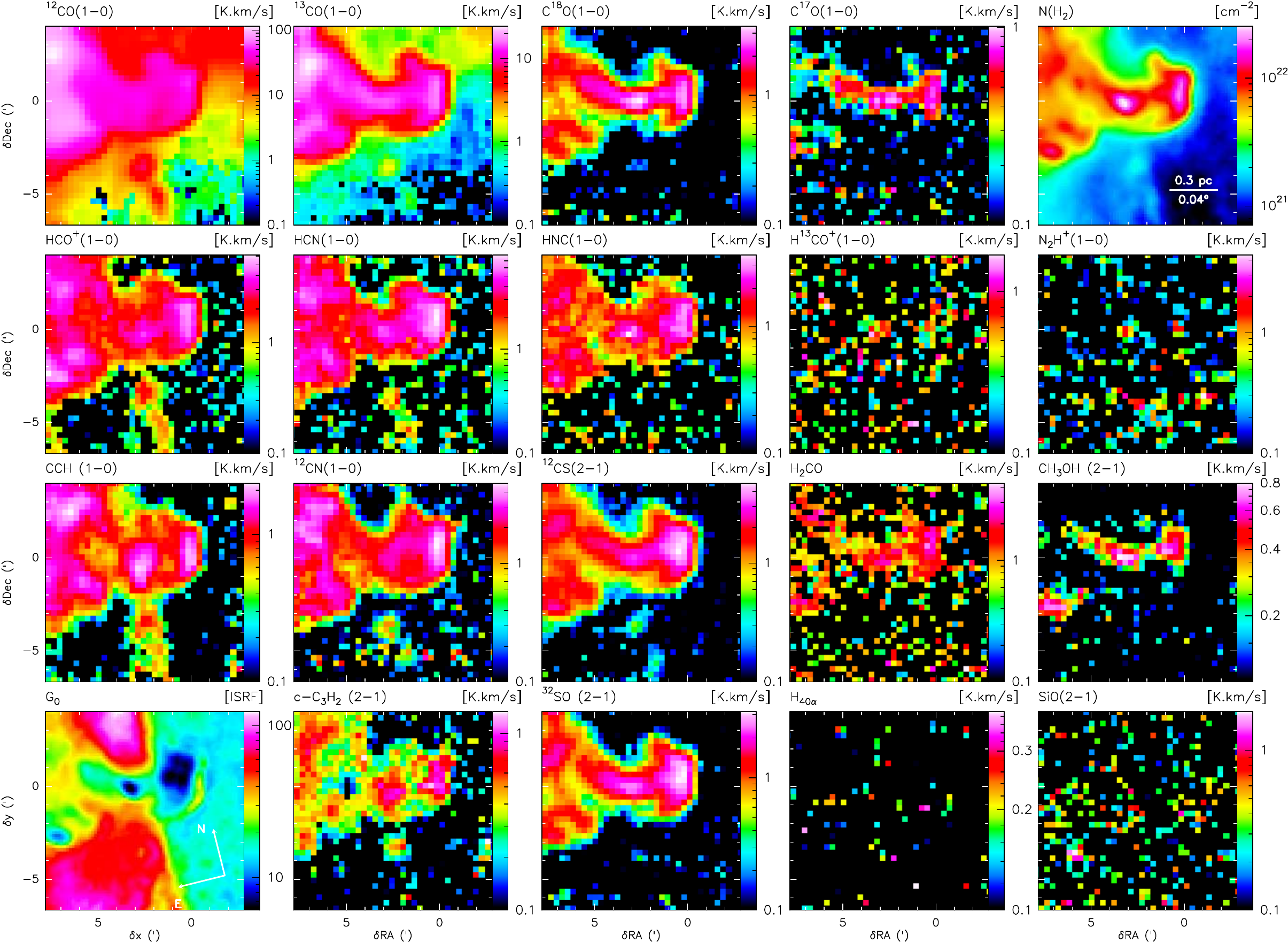}
    \mbox{} %
    \vspace*{\medskipamount} %
    \mbox{} %
    \includegraphics[width=0.9\linewidth]{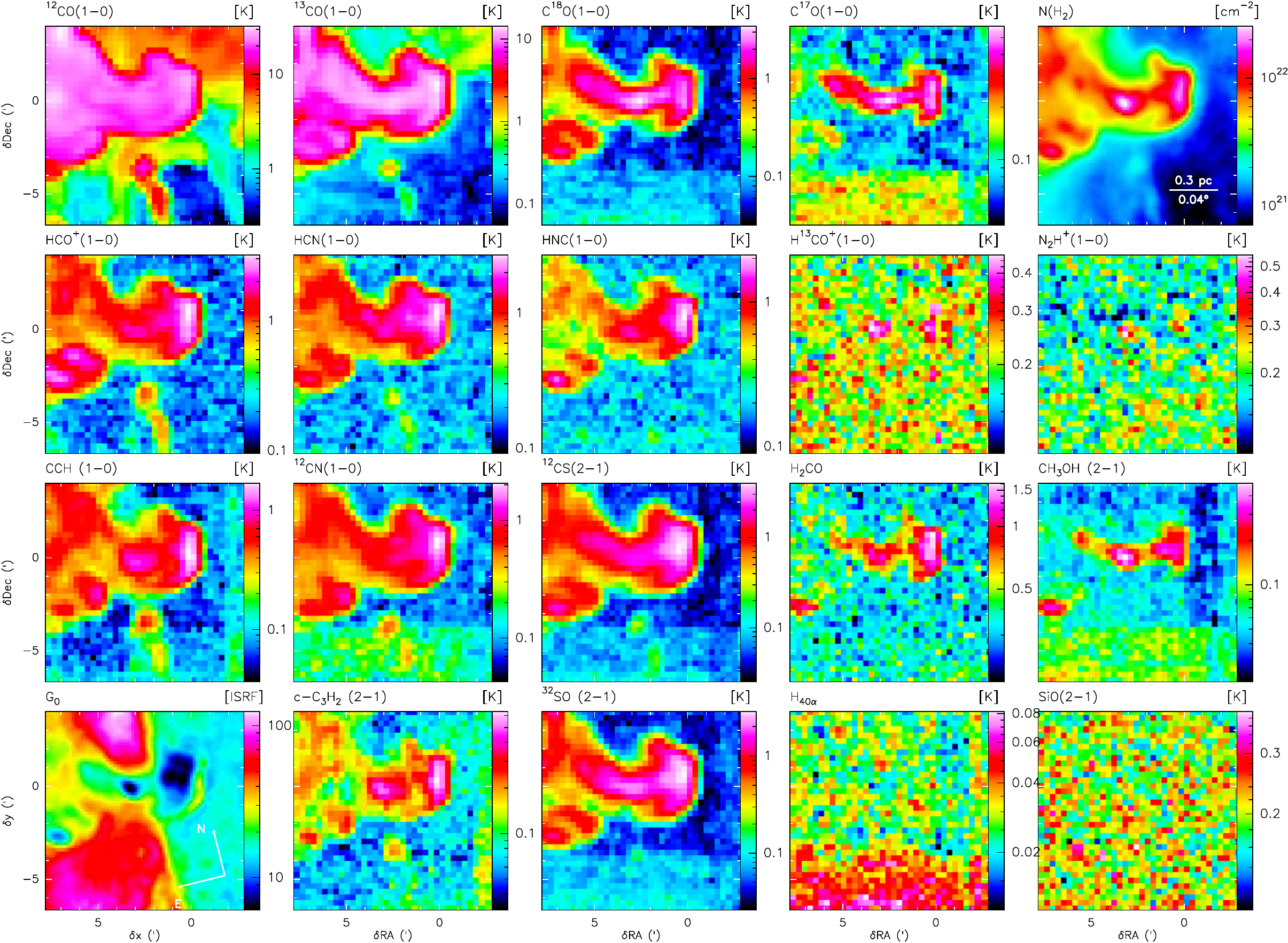}
    \caption{Zoom of the spatial distribution of the integrated intensity
      \textbf{(top)} and peak temperature \textbf{(bottom)} towards the
      Horsehead nebula.}
    \label{fig:Horsehead}
  \end{figure*}}

\newcommand{\FigPredicted}{%
  \begin{figure}
    \centering %
    \includegraphics[width=\linewidth]{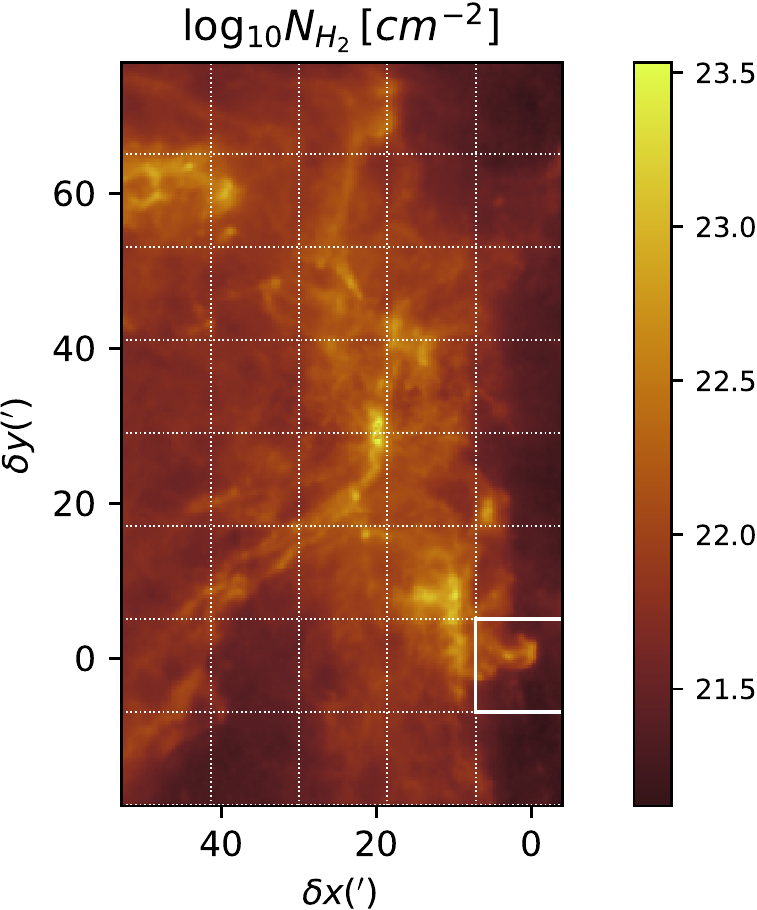}
    \caption{Spatial distribution of the dust-traced \Ht\ column density
      derived from Herschel data~\citep{Andre.2010,Lombardi.2014}. The
      dotted grid is used to define the training and test sets. The white
      square in the bottom right around the Horsehead region corresponds to
      the test set where the random forest predictions will be compared to
      the observations. This subset is never used during the training
      phase. Only the remainder of the map is used as the training set.}
    \label{fig:data:dust}
  \end{figure}}

\newcommand{\FigOptimization}{%
  \begin{figure*}
    \centering
    \begin{tabular}{lll}
      \includegraphics[height=4.5cm,trim={0.2cm 0 3.0cm 0},clip]{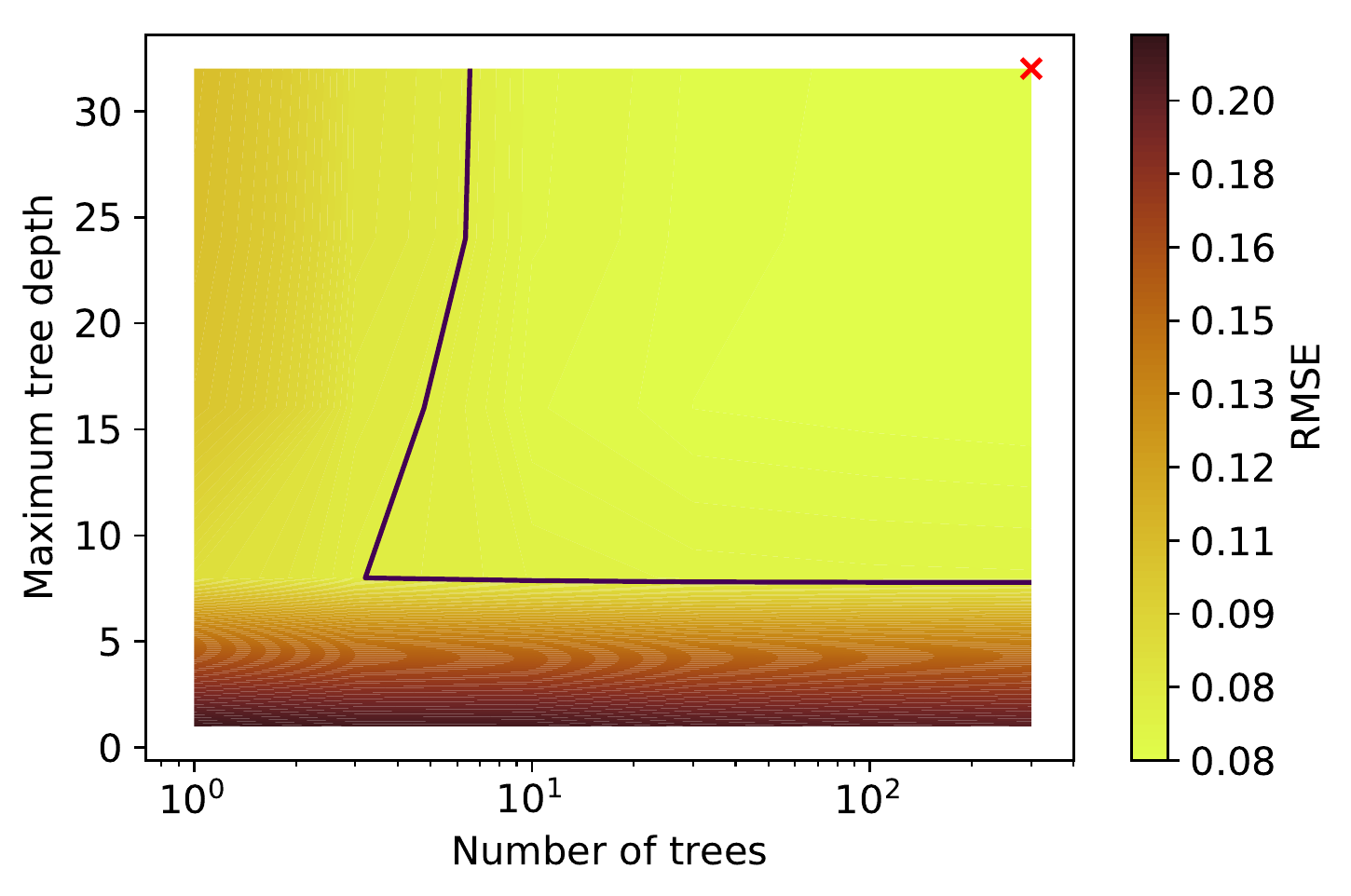}
      &
      \includegraphics[height=4.5cm,trim={0.2cm 0 3.0cm 0},clip]{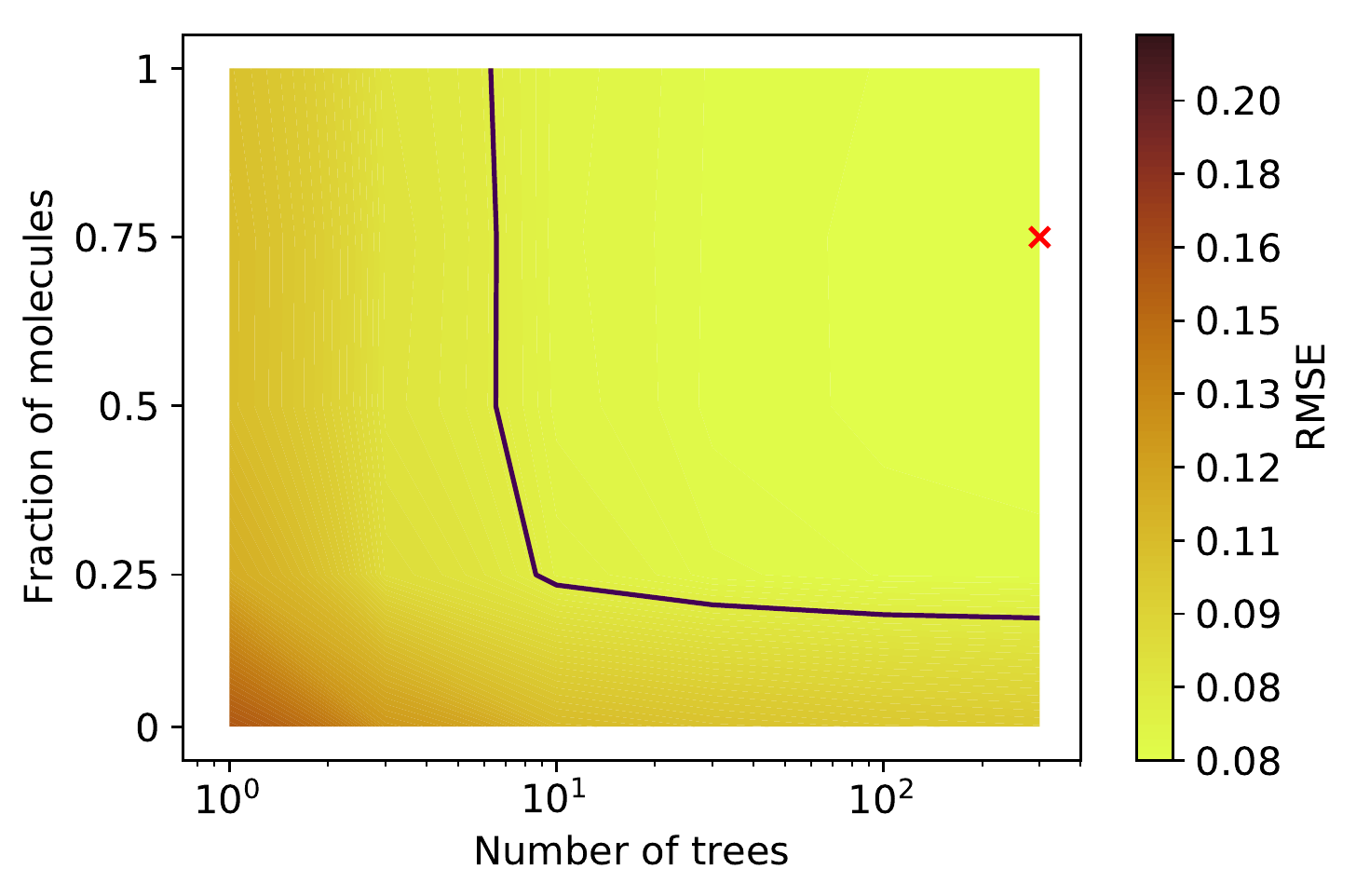}        
      &
      \includegraphics[height=4.5cm,trim={0.2cm 0 0.0cm 0},clip]{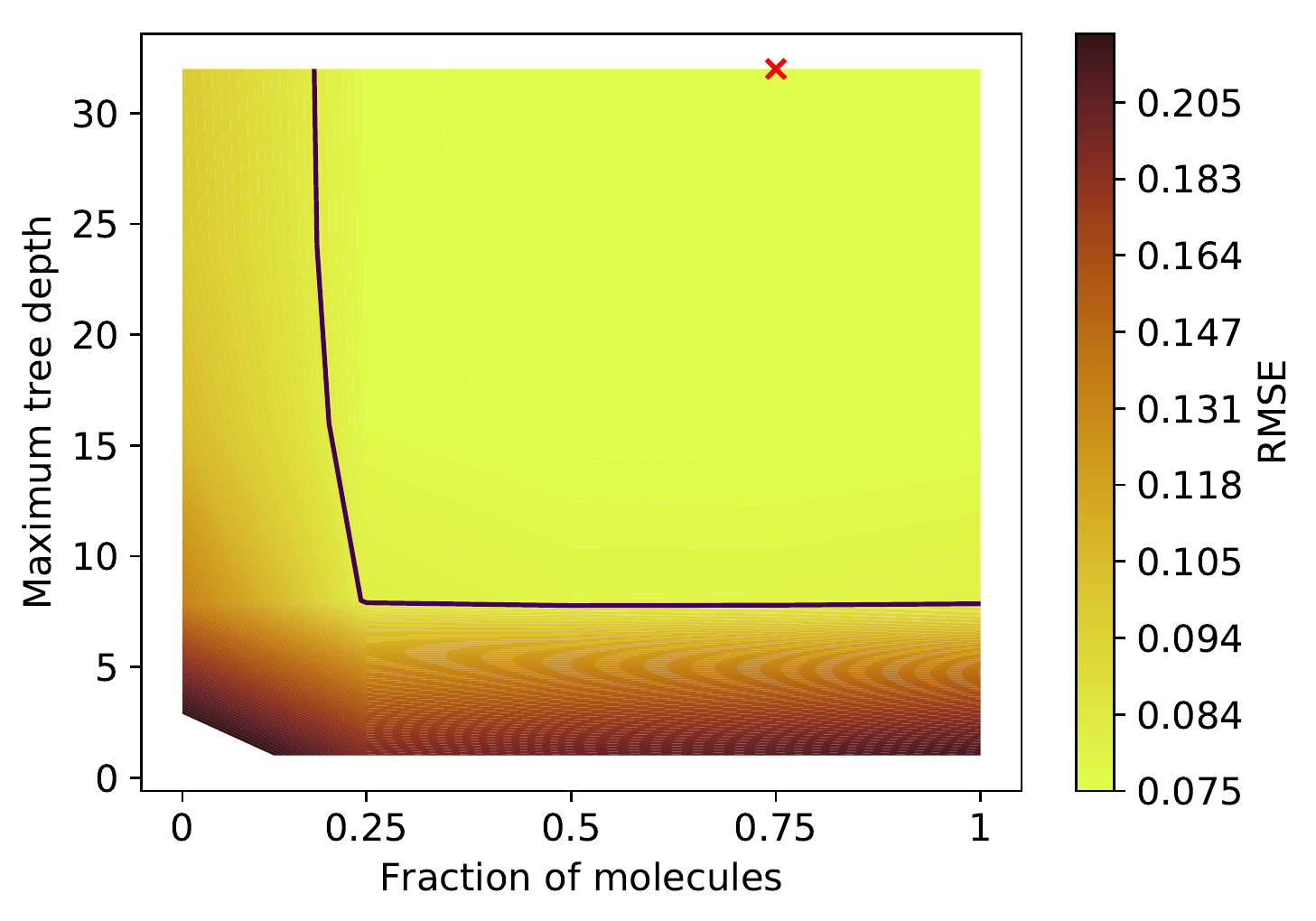}
      \\
      \includegraphics[height=3.5cm,trim={0.0cm 0 0 0},clip]{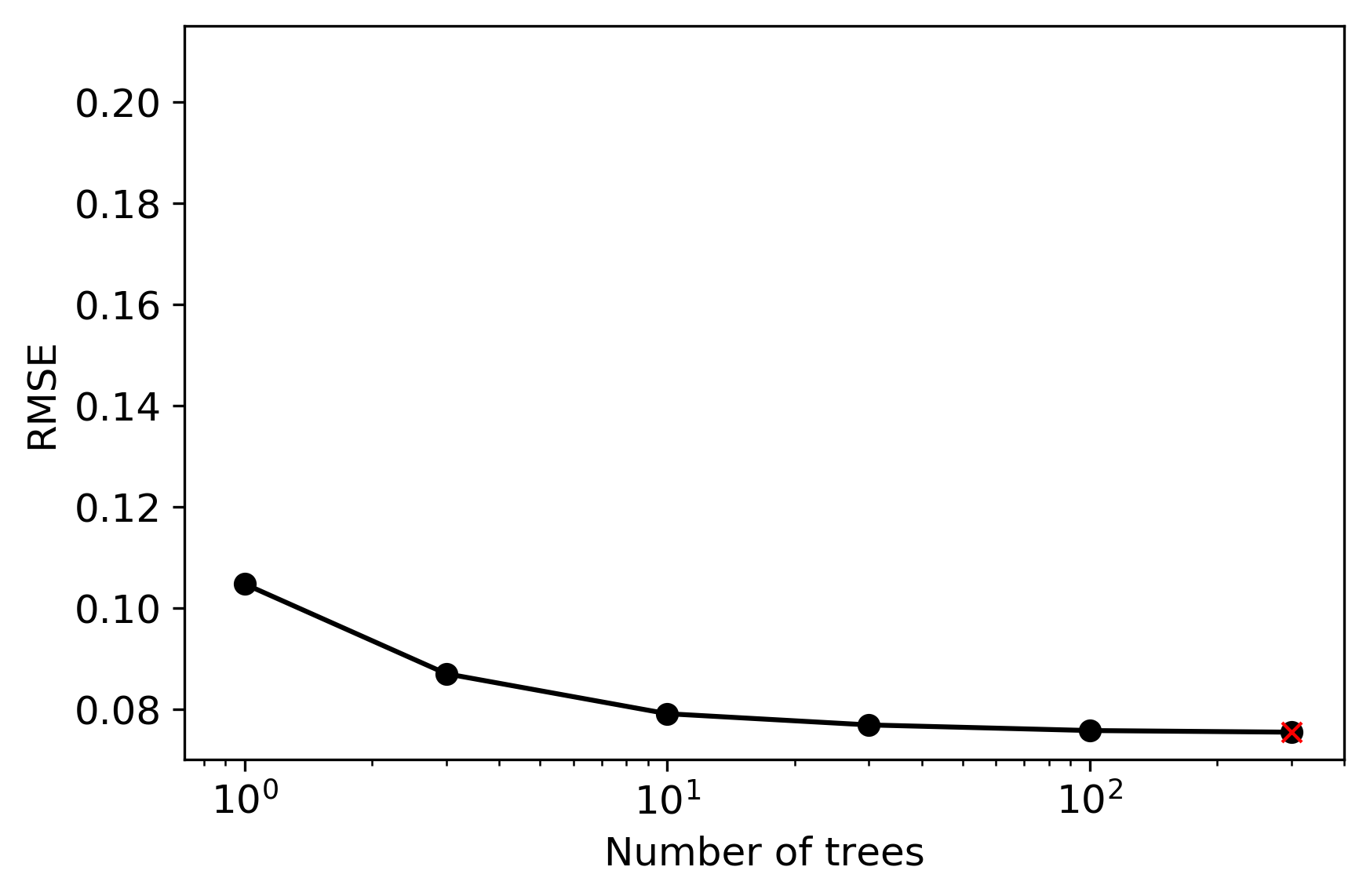}
      &
      \includegraphics[height=3.5cm,trim={0.0cm 0 0 0},clip]{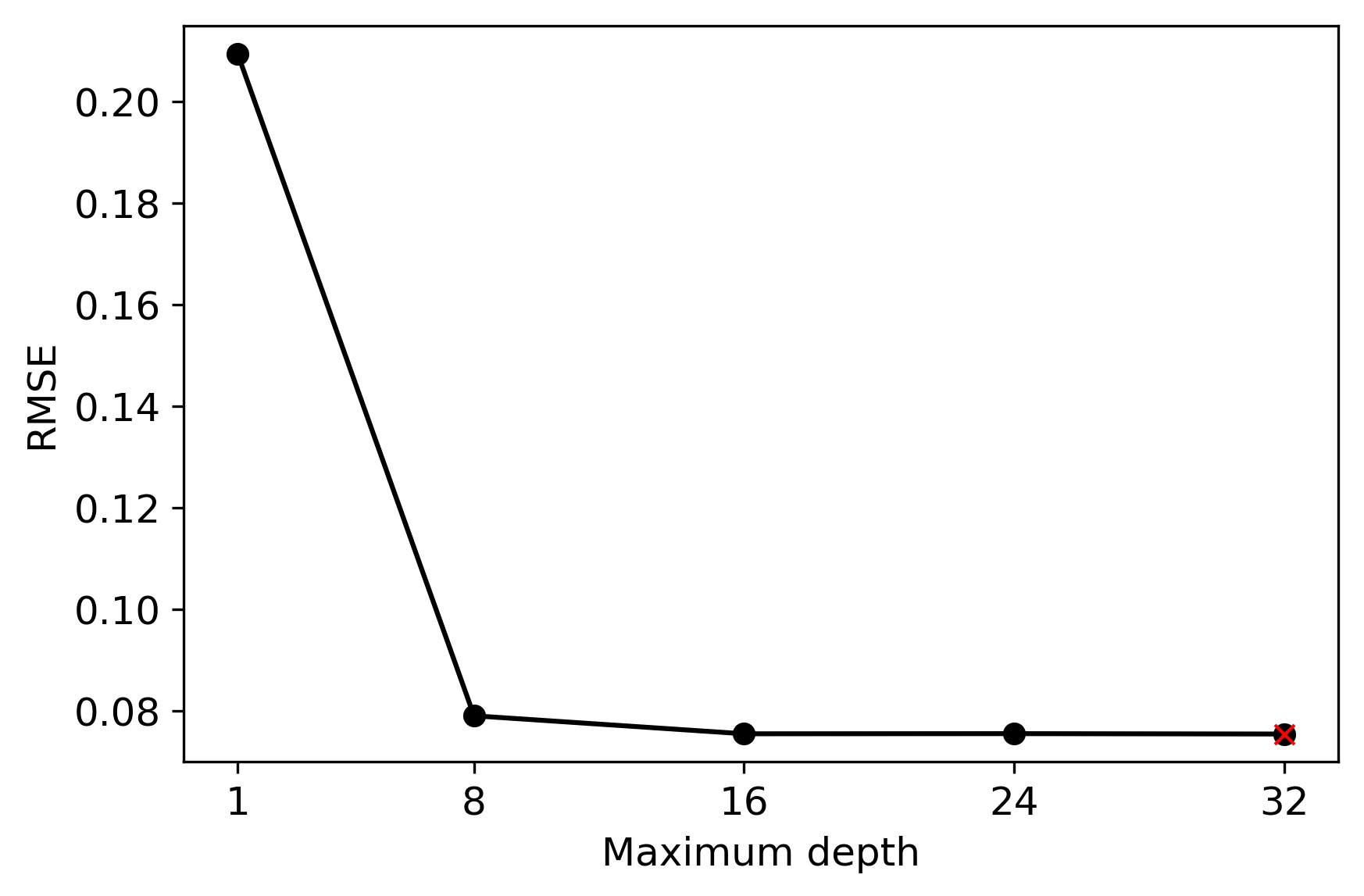}
      &
      \includegraphics[height=3.5cm,trim={0.0cm 0 0 0},clip]{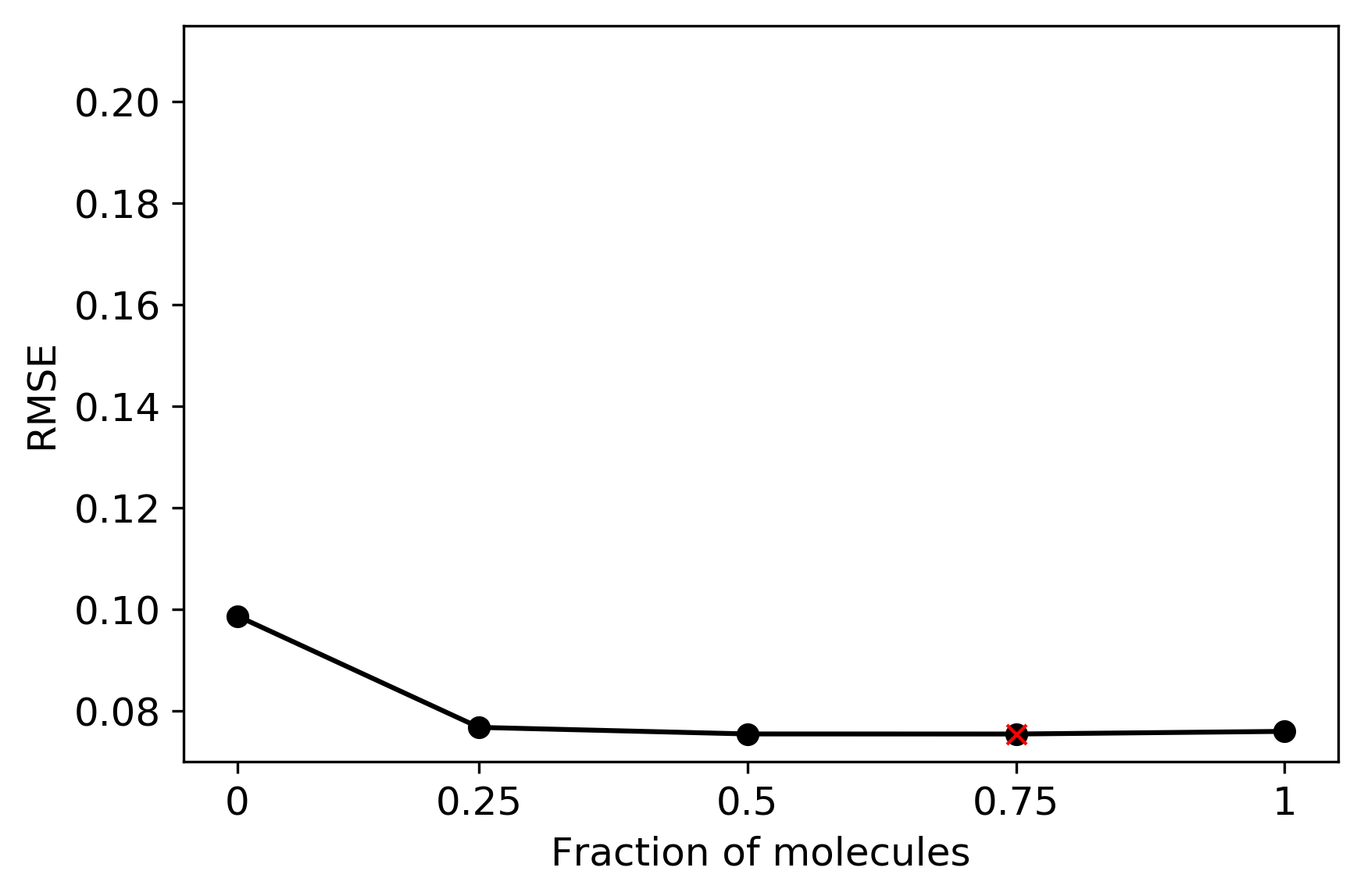}
    \end{tabular}
    \caption{Variations of the root mean square error (RMSE) between the
      predicted and the observed \Ndust{} computed on the validation set
      when optimizing the random forest hyper-parameters: 1) The number of
      trees in the forest, 2) the maximum depth of one tree, and 3) the
      number of features (line peak temperatures and integrated
      intensities) randomly chosen to train each individual regression
      tree of the forest. Bi-dimensional (\textbf{top}) and
      mono-dimensional (\textbf{bottom}) cuts going through the minimum
      RMSE over the full cube. The space of acceptable parameters is shown
      inside the black contours (minimum plus 10\%).  The minimum
      values are shown as the red crosses in all cases.}
    \label{fig:hyperparams}
  \end{figure*}}

\newcommand{\FigUncertainty}{%
  \begin{figure}
    \centering %
    \includegraphics[width=\linewidth]{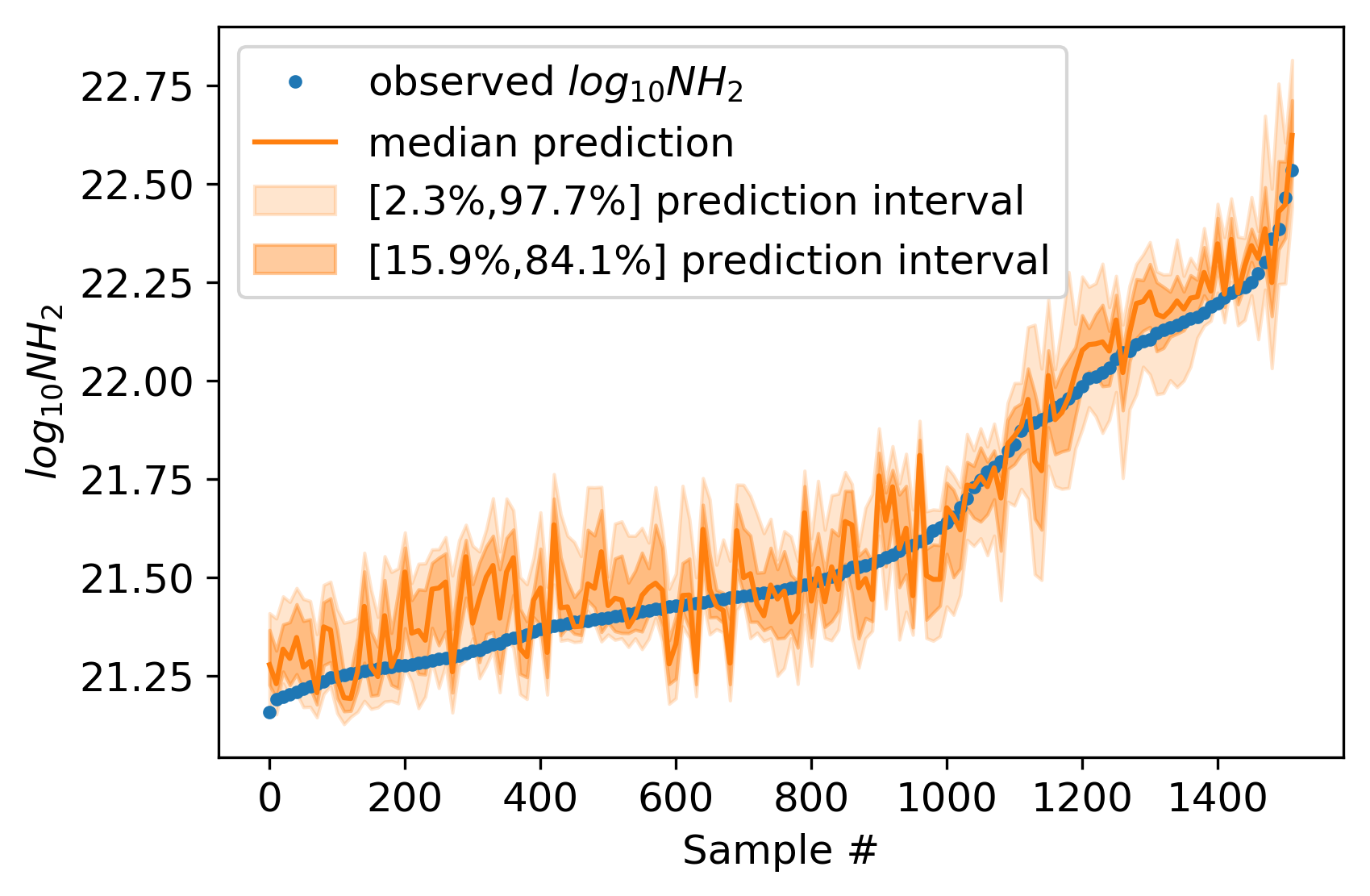}
    \caption{Median column density and quantile intervals for each pixel of
      the test set ordered by increasing observed column density. The blue
      dots display the associated observed column densities.}
    \label{fig:rf:uncertainty}
  \end{figure}}

\newcommand{\FigGaussianMixture}{%
  \begin{figure}
    \centering %
    \includegraphics[width=\linewidth,trim={0 0 0 0.7cm},clip]{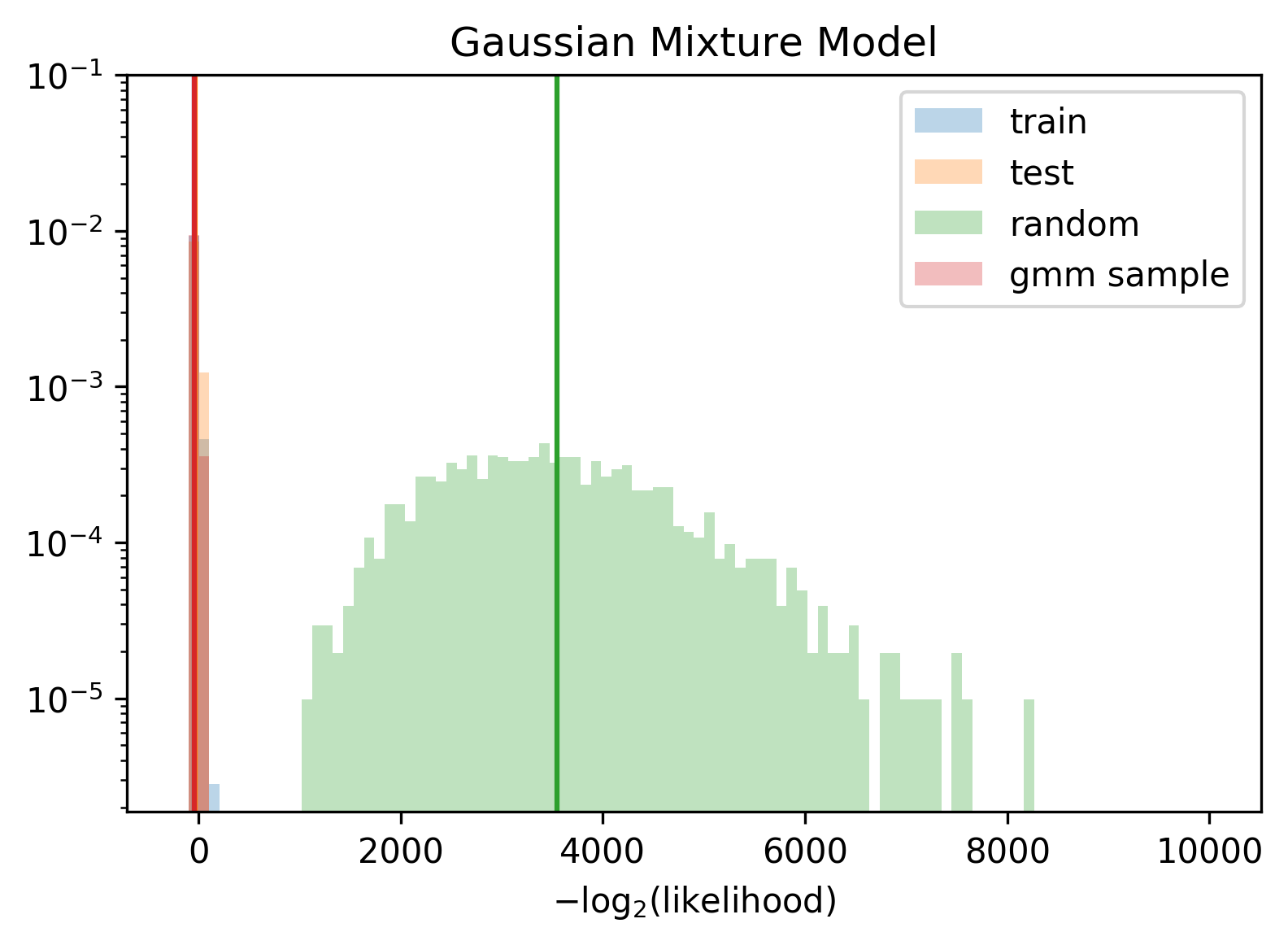}
    \includegraphics[width=\linewidth,trim={0 0 0 0.7cm},clip]{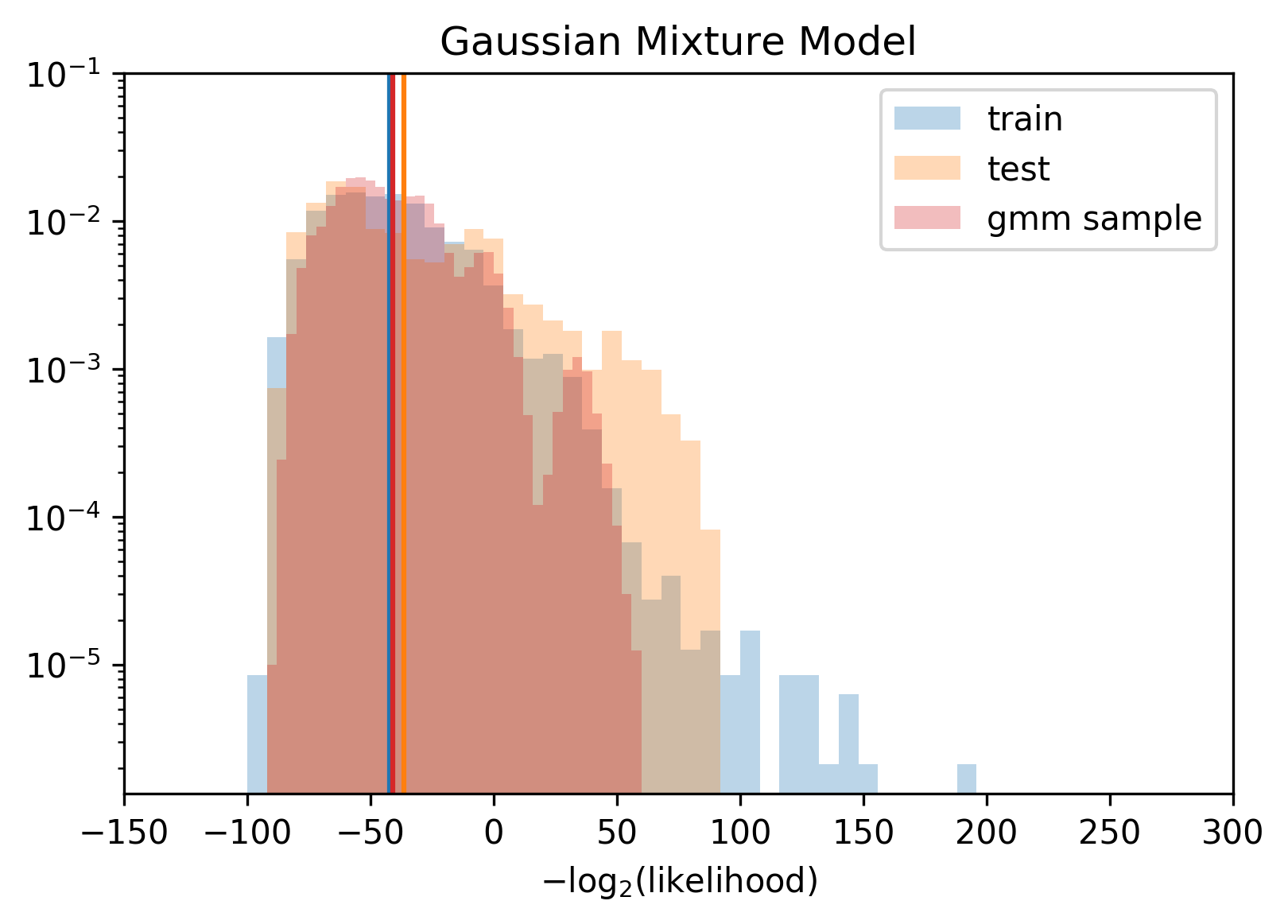}
    \caption{Probability distribution function of the value of the Gaussian
      mixture fitted on the training set, for each data point. The
      log$_2$-likelihood shown on the horizontal axis can be interpreted as
      the number of bits (to within a constant value) required to encode
      each point of the sample. The blue, orange, green, and pink colors
      show the distributions for the training set, the test set, a set of
      points uniformly drawn in the same multi-dimensional space, and a set
      of points randomly drawn in the Gaussian Mixture, respectively.  The
      vertical lines show the associated means of the negative
      log$_2$-likelihood.  The bottom panel is a zoom of the top one. }
    \label{fig:gaussian-mixture}
  \end{figure}}

\newcommand{\FigResultsMaps}{%
  \setlength{\tabcolsep}{2pt}
  \begin{figure*}
    \begin{minipage}{0.33\linewidth}
      \centering{} %
      \Large{} %
      Observed
    \end{minipage}
    \begin{minipage}{0.33\linewidth}
      \centering{} %
      \Large{} %
      Predicted
    \end{minipage}
    \begin{minipage}{0.33\linewidth}
      \centering{} %
      \Large{} %
      Predicted/Observed
    \end{minipage}
    \\[\smallskipamount]
    \begin{tabular}{rc}
      \mycaptionbis{5.4cm}{\Large Linear Method}                      & \includegraphics[width=0.975\linewidth,trim={0 1.6cm 0 0.75cm},clip]{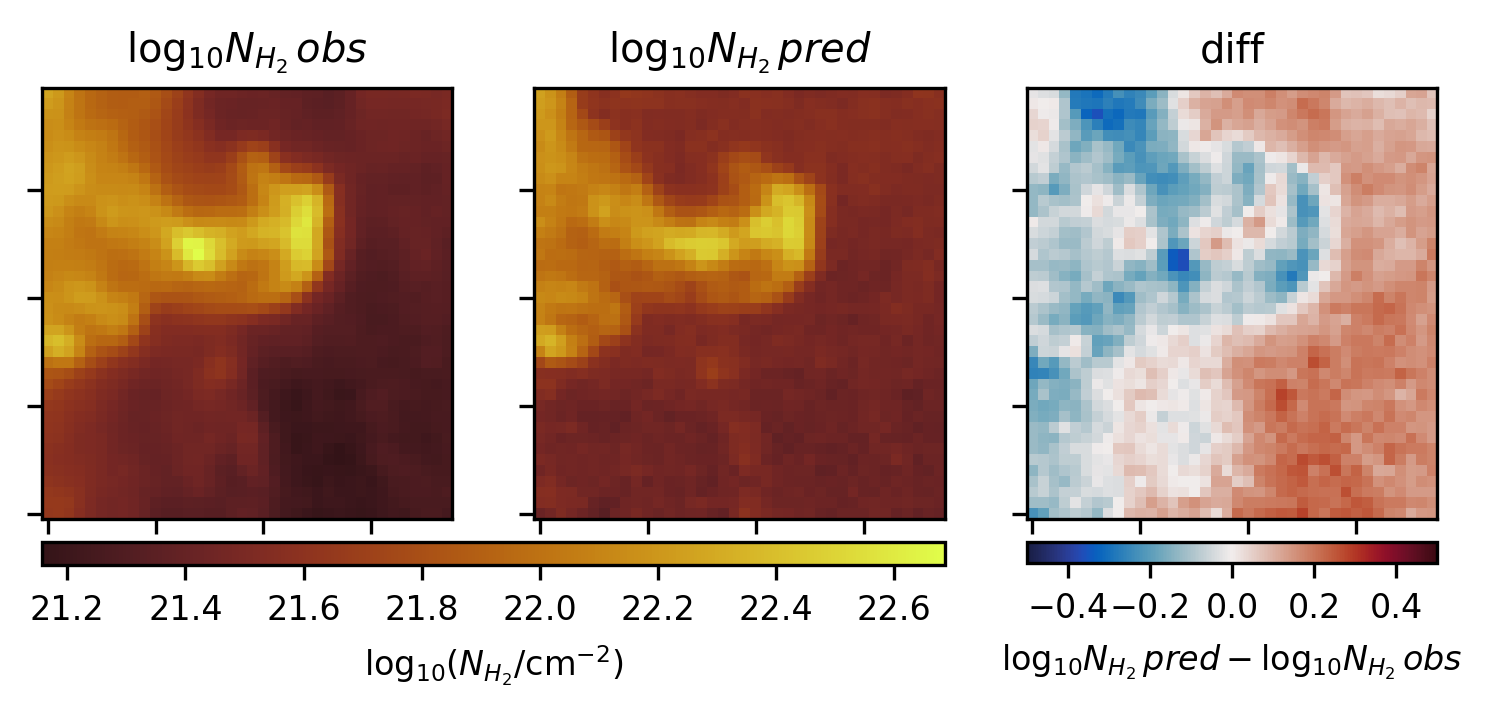}\\
      \mycaptionbis{5.4cm}{\Large asinh Pre-Processing}               & \includegraphics[width=0.975\linewidth,trim={0 1.6cm 0 0.75cm},clip]{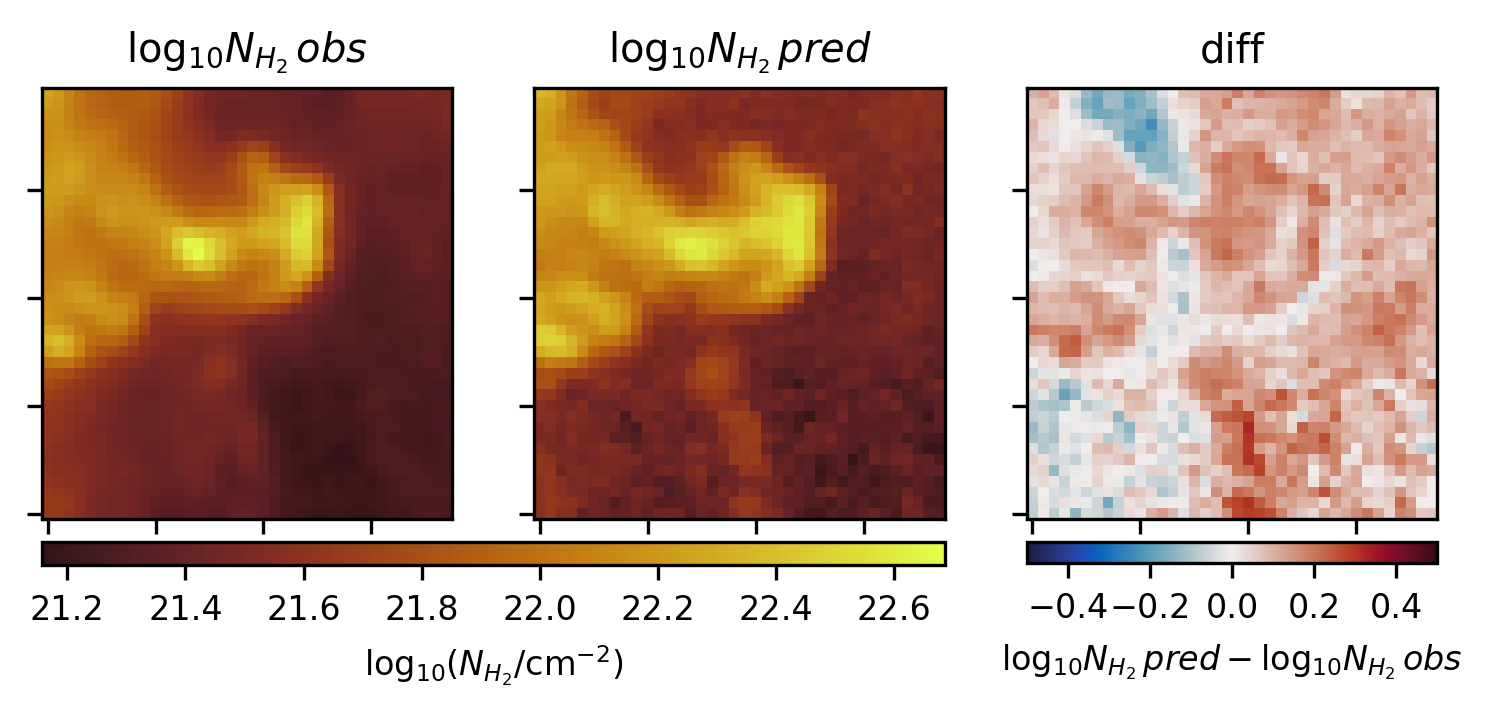}\\
      \mycaptionbis{7.0cm}{\Large \mbox{} \hspace{2cm} Random Forest} & \includegraphics[width=0.975\linewidth,trim={0 0     0 0.75cm},clip]{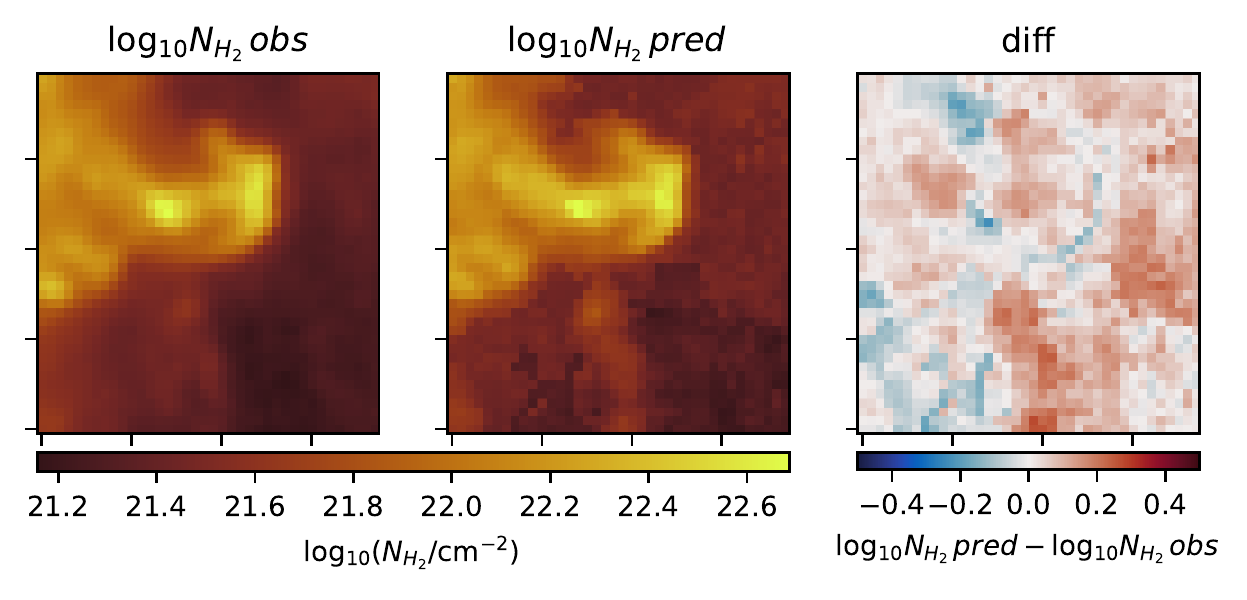}
    \end{tabular}
    \caption{Comparison of the generalization performances of three
      predictors.  \textbf{Top row:} Linear method. \textbf{Middle row:}
      Linear method with a non-linear pre-processing. \textbf{Bottom row:}
      Random forest. All results are computed on the Horsehead pillar,
      i.e., the test set. \textbf{Left and middle columns:} Spatial
      distribution of the observed and predicted column density. Both
      images share the same color scale. \textbf{Right column:} Ratio of
      the predicted column density over the observed one. The limits of the
      color scale correspond to a ratio interval from $1/3$ to $3$.}
    \label{fig:results:maps}
  \end{figure*}}

\newcommand{\FigResultsHistos}{%
  \begin{figure*}
    \centering %
    \begin{tabular}{rc}
      \mycaptionbis{7.0cm}{\Large Linear Method}                      & \includegraphics[width=0.9\linewidth,trim={0 1.0cm 0 1.75cm},clip]{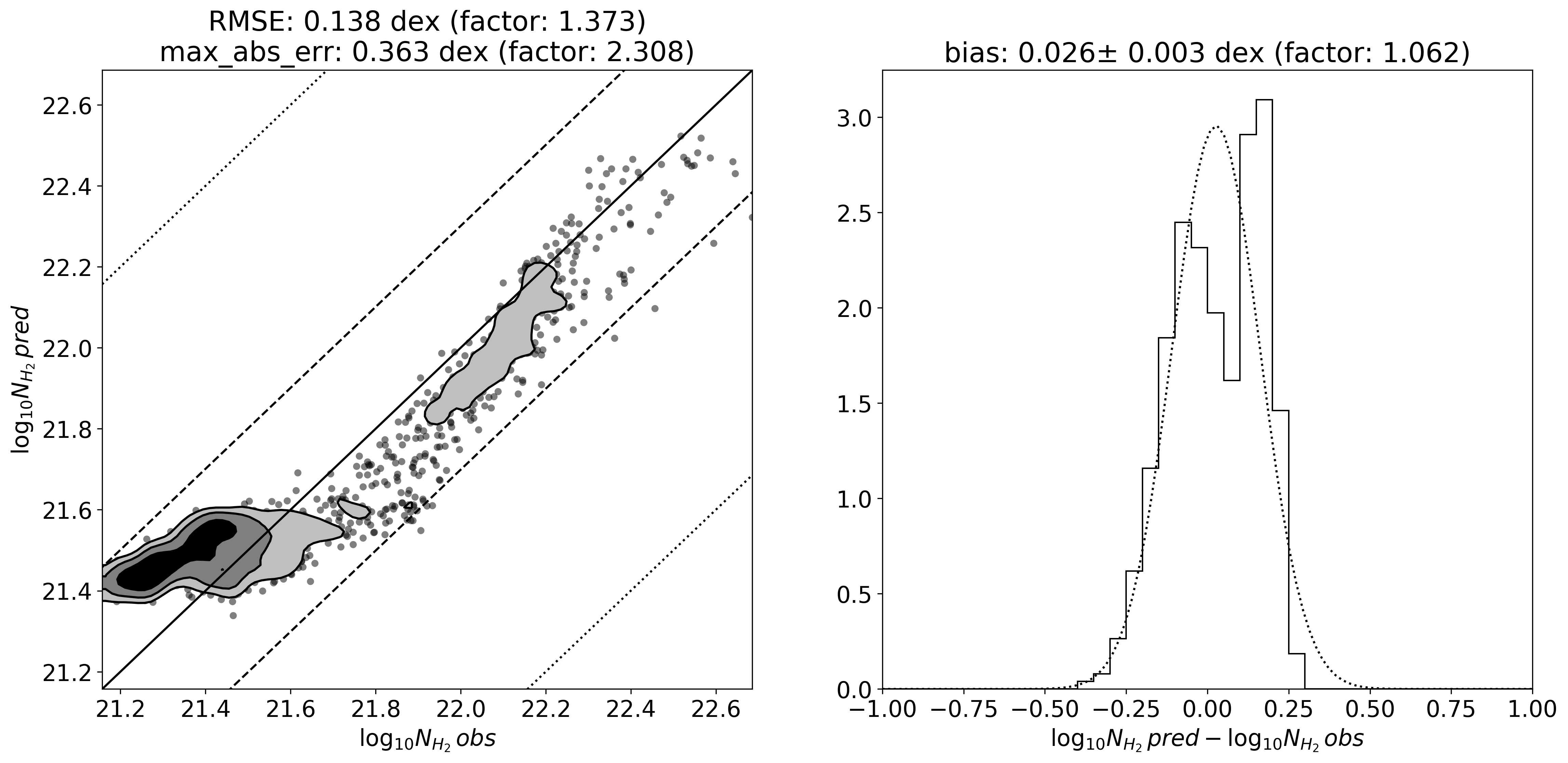}   \\
      \mycaptionbis{7.0cm}{\Large asinh Pre-Processing}               & \includegraphics[width=0.9\linewidth,trim={0 1.0cm 0 1.75cm},clip]{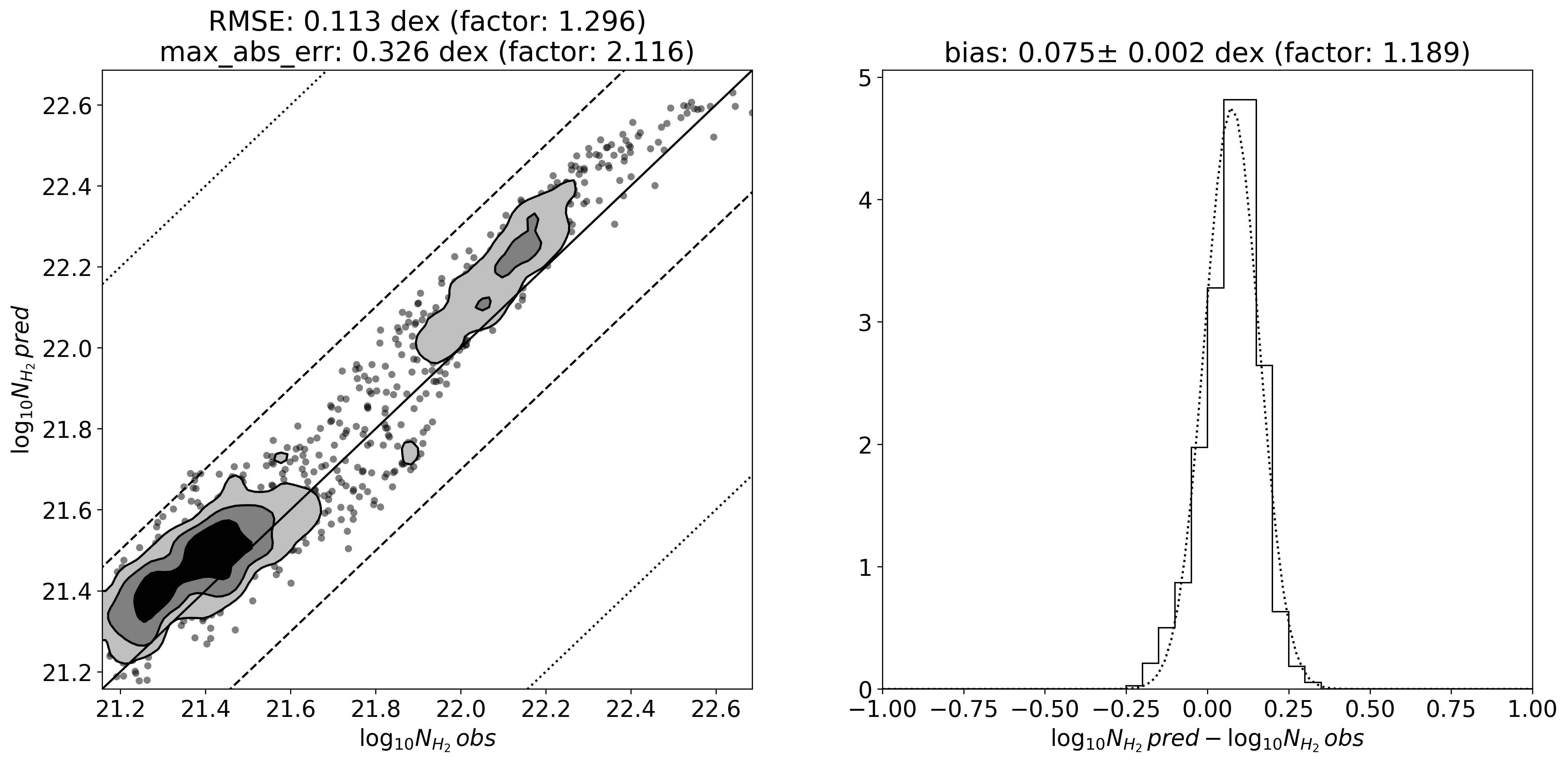} \\
      \mycaptionbis{7.0cm}{\Large \mbox{} \hspace{1cm} Random Forest} & \includegraphics[width=0.9\linewidth,trim={0 0.0cm 0 1.70cm},clip]{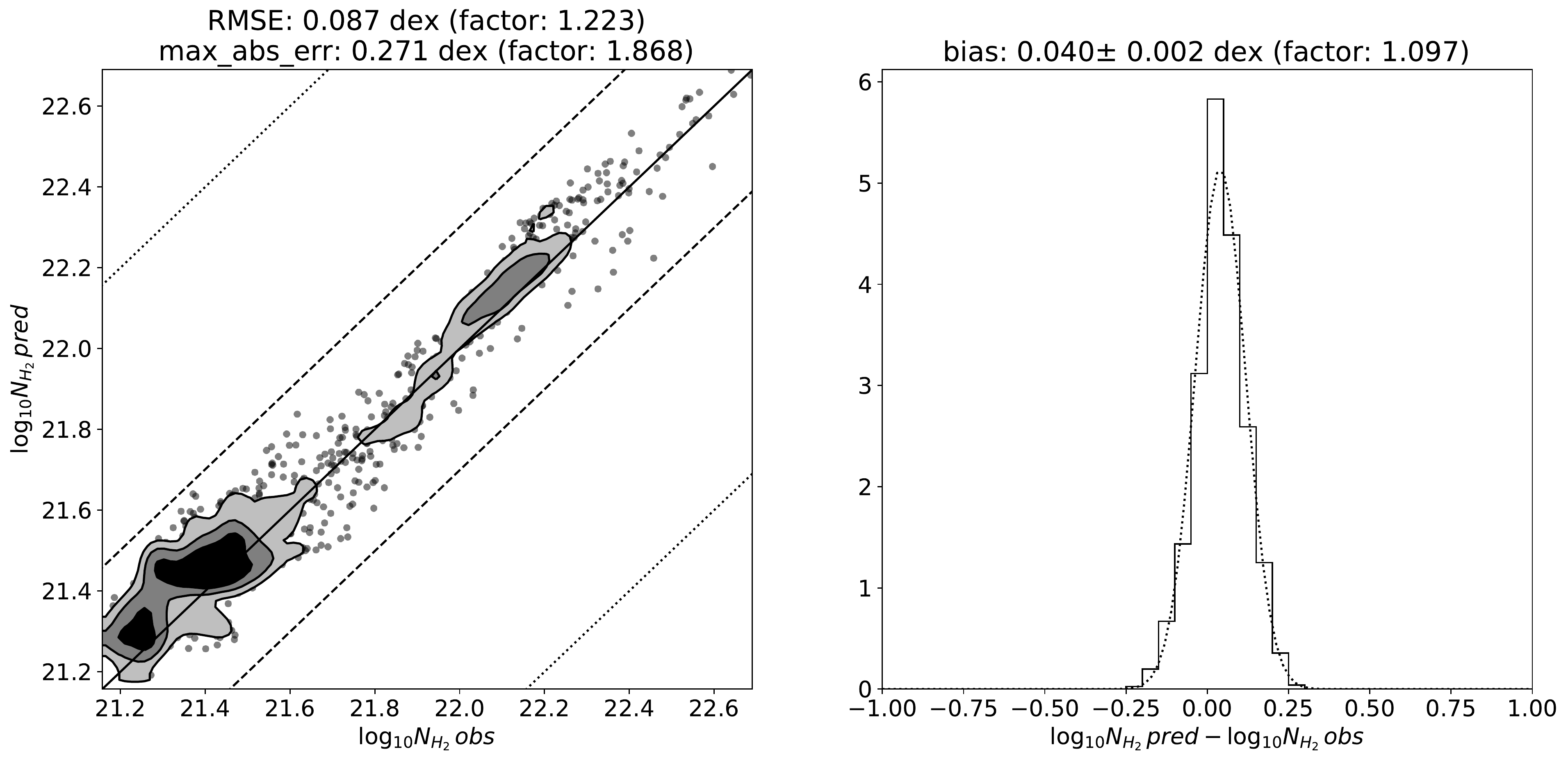}    \\
    \end{tabular}
    \caption{Comparison of the generalization performances of three
      predictors.  \textbf{Top row:} Linear method. \textbf{Middle row:}
      Linear method with a non-linear pre-processing. \textbf{Bottom row:}
      Random forest. All results are computed on the Horsehead pillar,
      i.e., the test set.  \textbf{Left column:} Joint Probability
      Distribution Function (PDF) of the predicted column density and of
      the observed one. The contours are the PDF isocontours enclosing 25,
      50, and 75\% of the datapoints. Points whose density falls below
      these values are shown as black dots. The oblique lines have a slope
      of 1. They indicates ratio values of 1.0 (plain), 0.5 and 2.0
      (dashed), 0.1 and 10.0 (dotted).  \textbf{Right column:} Histogram of
      the ratio of the predicted column density over the observed one on a
      logarithmic scale. The dotted lines show the Gaussian of same mean
      and width.}
    \label{fig:results:histos}
  \end{figure*}}

\newcommand{\FigLineImportanceLatex}{%
  \begin{figure}
    \includegraphics[width=\linewidth,trim={0 0 0.25cm 0.7cm},clip]{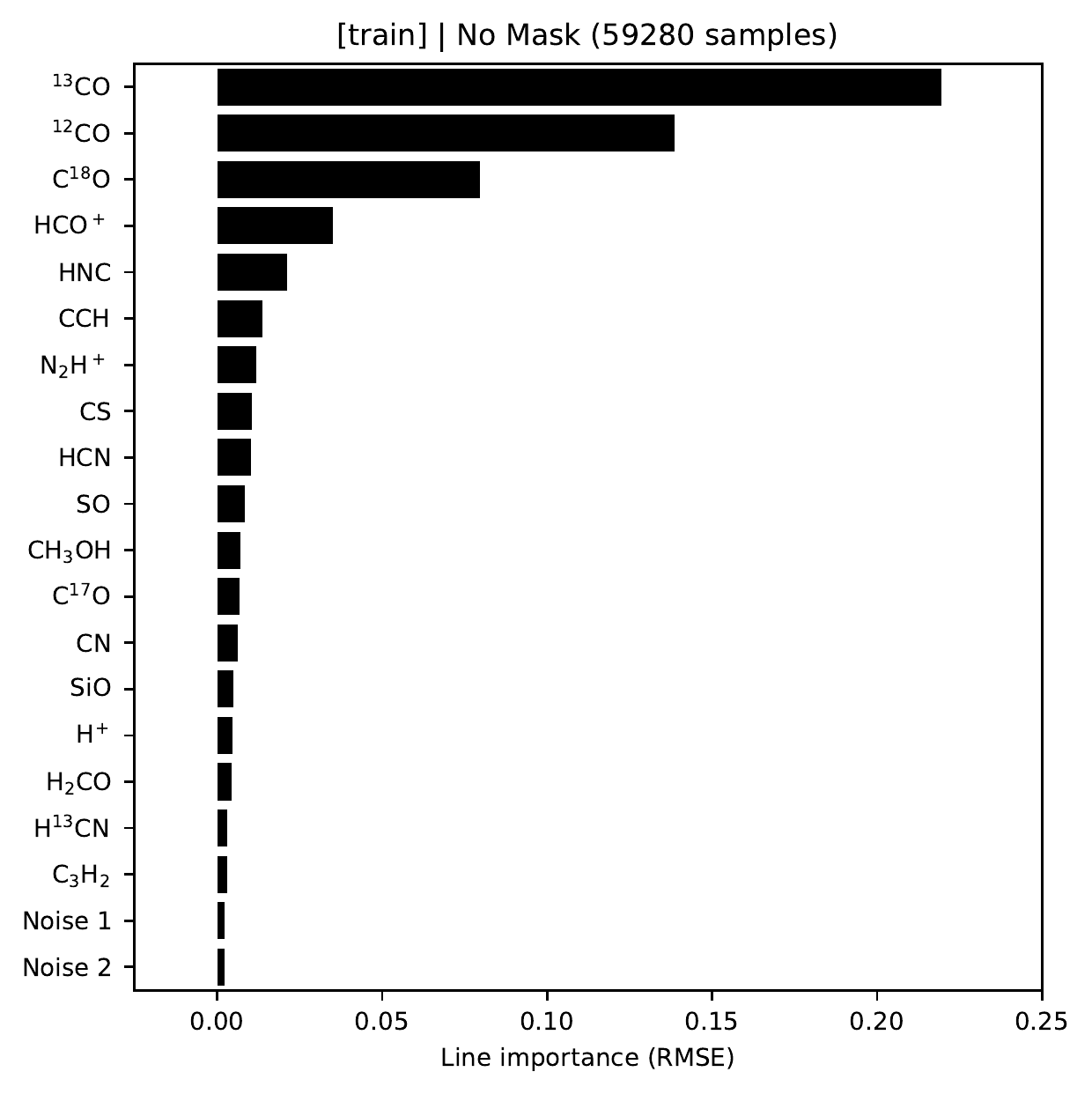}
    \hfill{}
    \includegraphics[width=\linewidth]{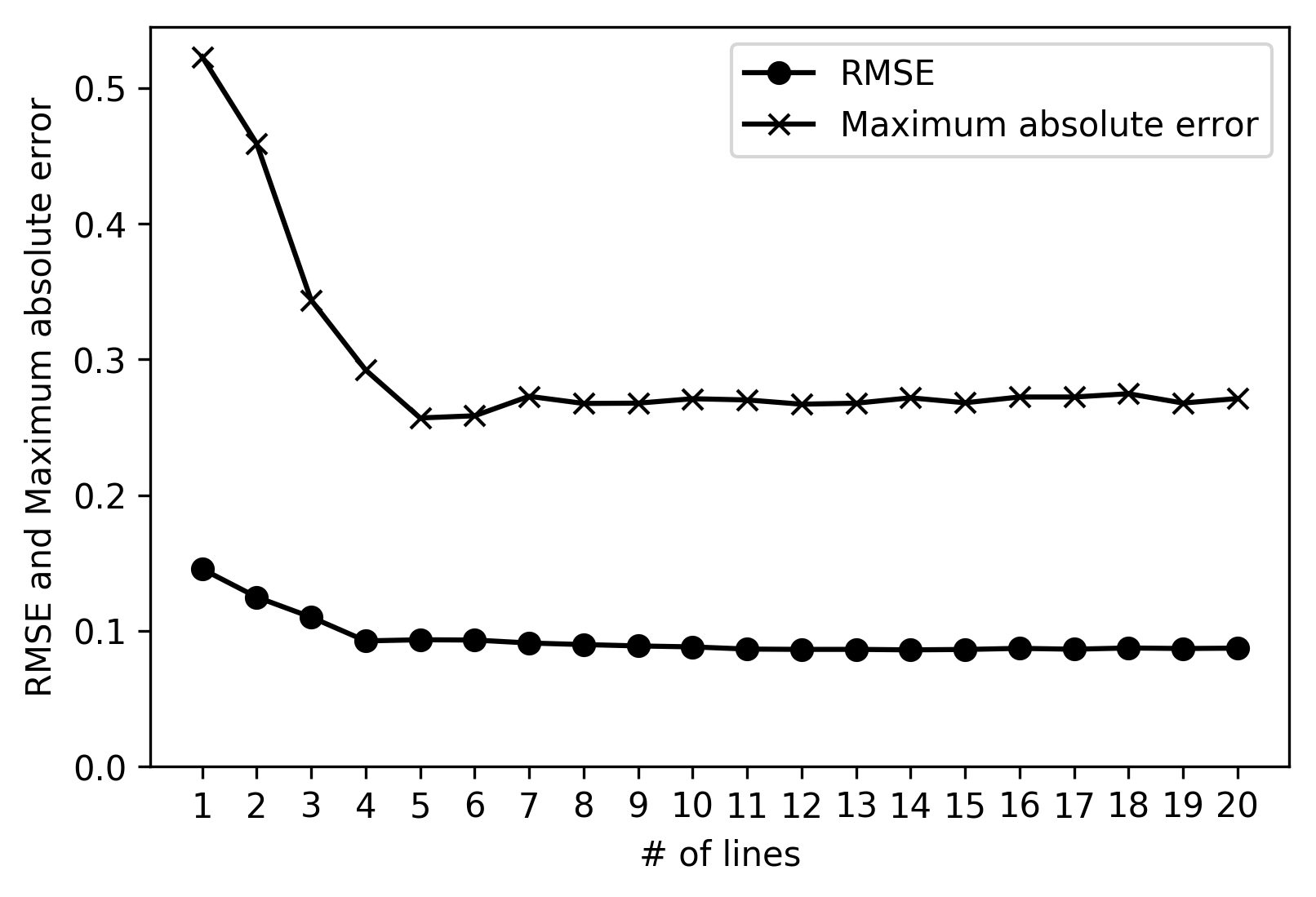} 
    \caption{Contribution of the different lines (both integrated intensity
      and temperature peak) to the quality of the random forest fit of the
      training data set. Noise \#1 and 2 are two additional random sets of
      input data.  \textbf{Top:} Quantitative improvement of the quality of
      the fit (RMSE feature importance) for each available line.
      \textbf{Bottom:} Evolution of the RMSE (filled circles) and maximum
      absolute error when each line is progressively added into the
      training phase in the order defined in the top panel. These results
      are computed on the training set. The RMSE values on each diagram
      are commensurate to log\NHt.}
    \label{fig:line-importance}
  \end{figure}}

\newcommand{\FigResultsMapsOneByOne}{%
  \def\PanelHeight{2.65cm}
  \begin{figure}
    \begin{minipage}{0.94\linewidth}
      \begin{minipage}{0.29\linewidth}
        \centering{} %
        Observed
      \end{minipage}
      \begin{minipage}{0.31\linewidth}
        \centering{} %
        Predicted
      \end{minipage}
      \begin{minipage}{0.33\linewidth}
        \centering{} %
        Predicted/Observed
      \end{minipage}
      \centering{}%
      \\
      \includegraphics[width=\linewidth,trim={1.5cm 2.5cm 1.3cm 3.8cm},clip]{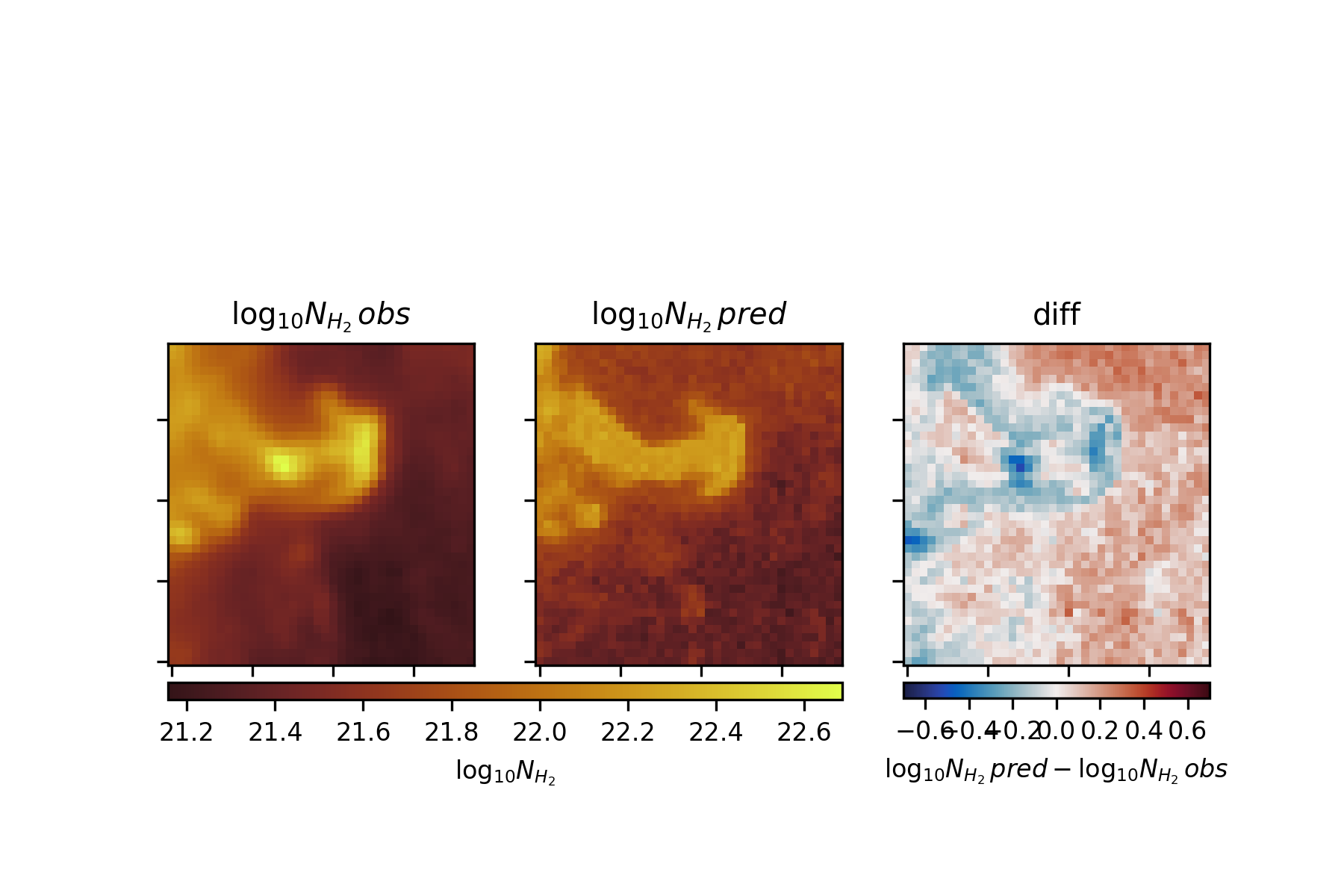}
      \includegraphics[width=\linewidth,trim={1.5cm 2.5cm 1.3cm 3.8cm},clip]{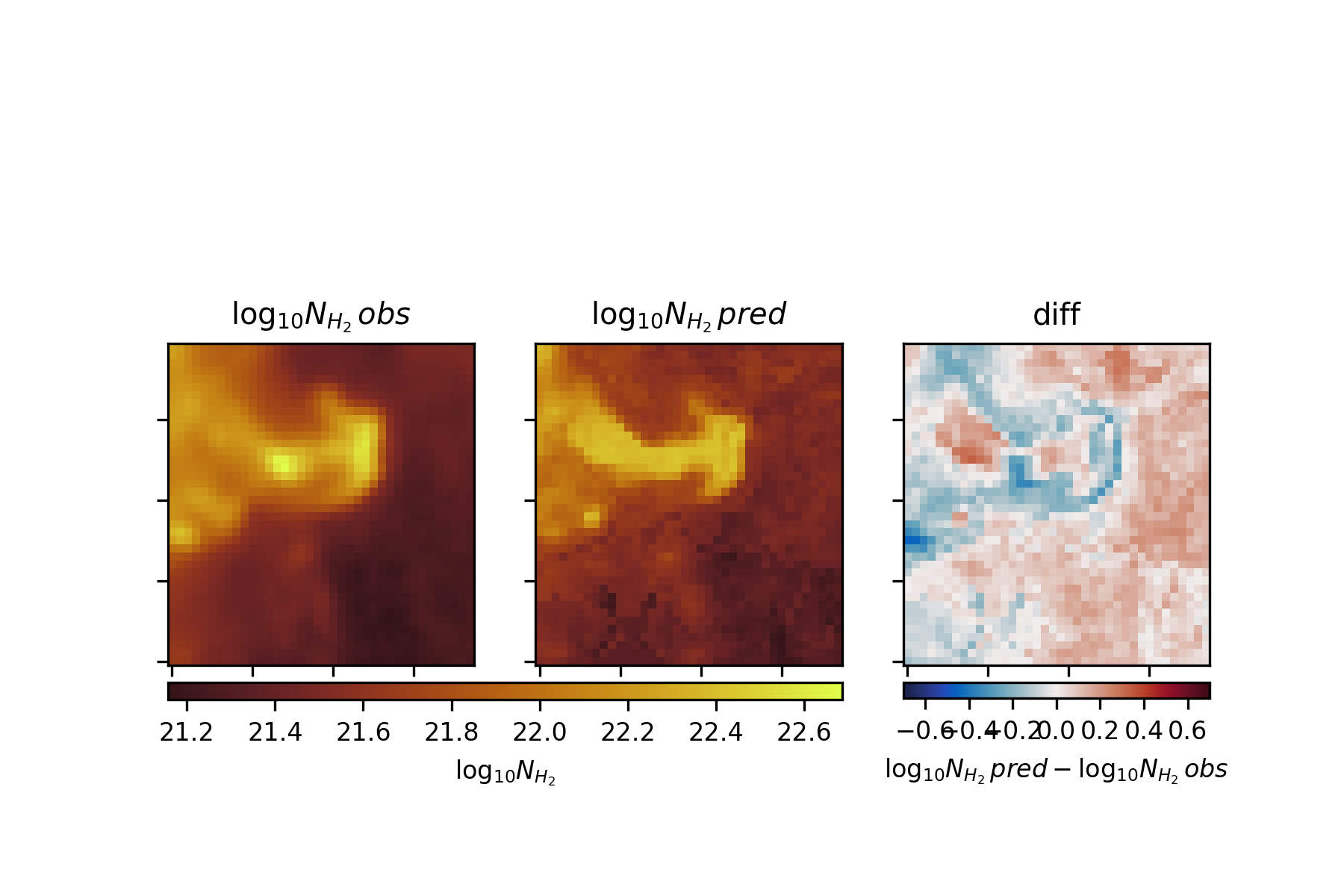}
      \includegraphics[width=\linewidth,trim={1.5cm 2.5cm 1.3cm 3.8cm},clip]{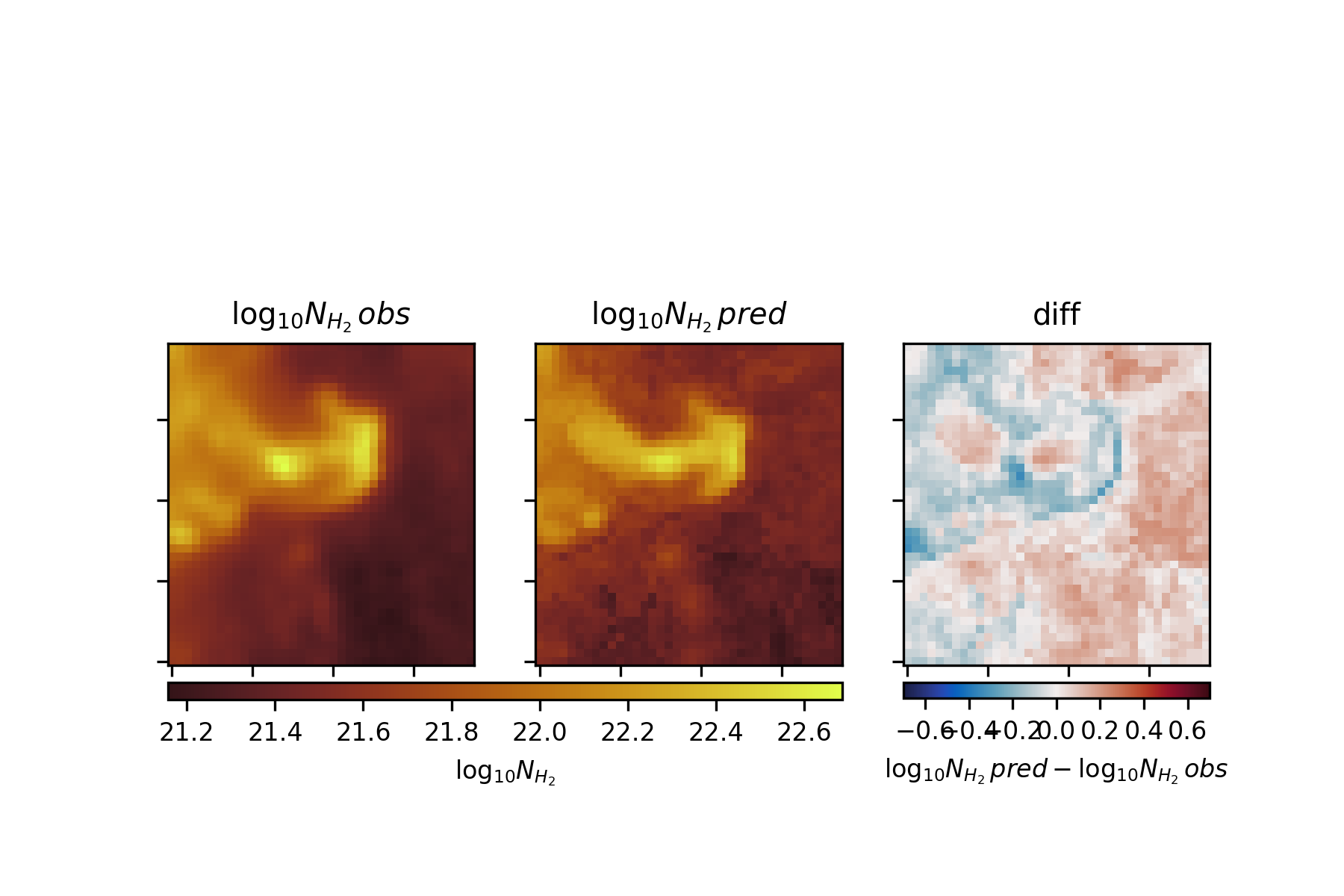}
      \includegraphics[width=\linewidth,trim={1.5cm 2.5cm 1.3cm 3.8cm},clip]{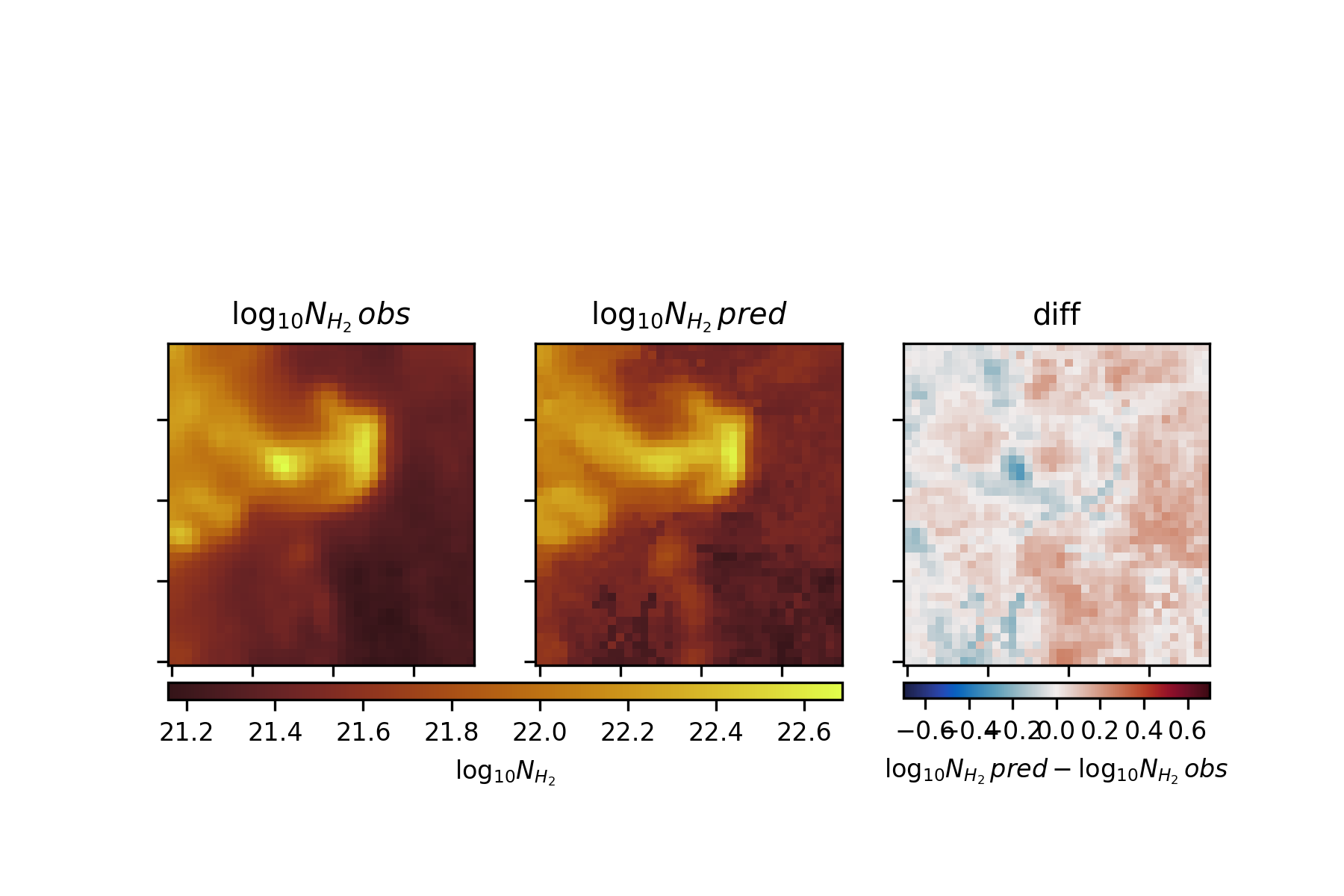}
      \includegraphics[width=\linewidth,trim={1.5cm 2.5cm 1.3cm 3.8cm},clip]{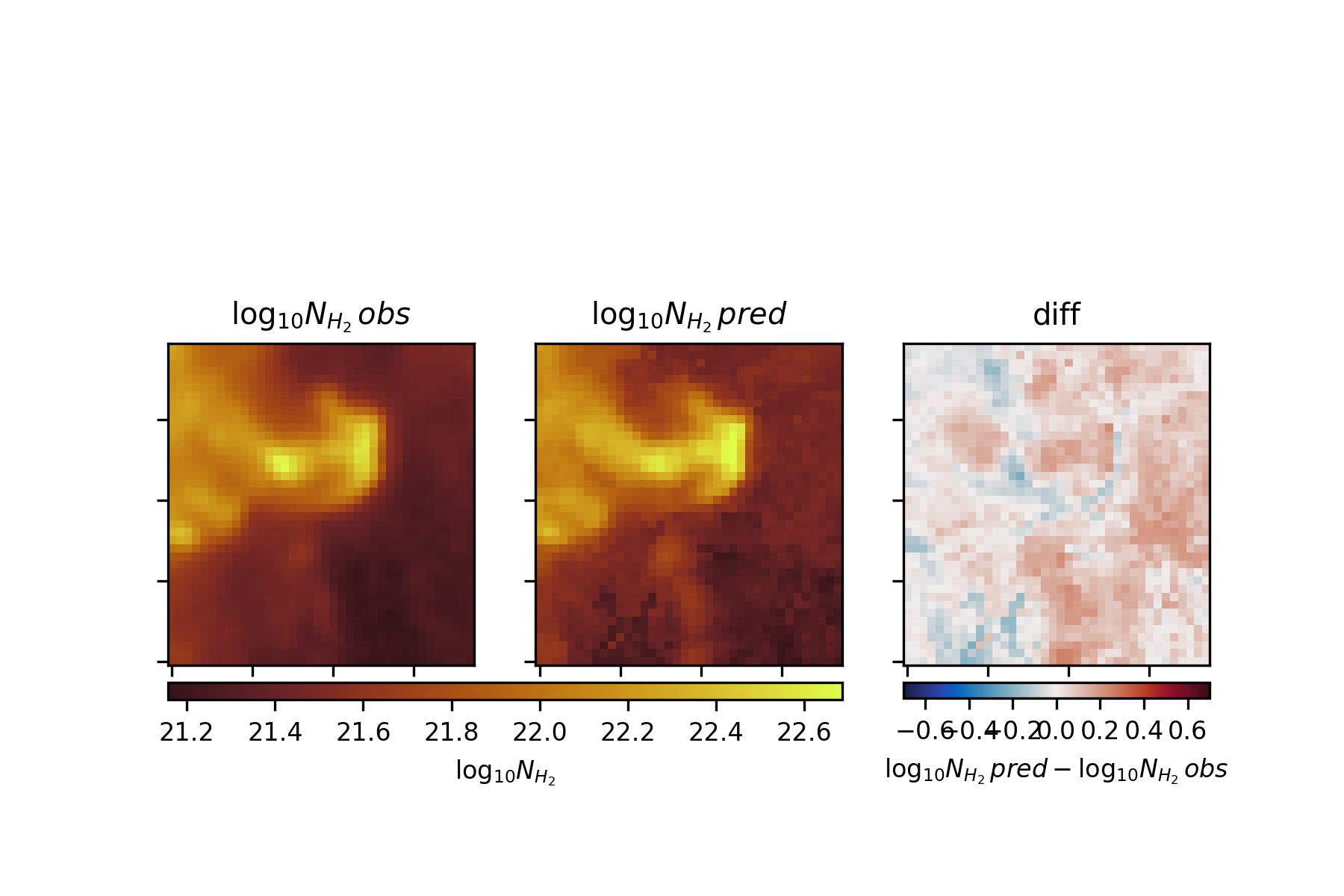}
      \includegraphics[width=\linewidth,trim={1.5cm 0.0cm 1.3cm 3.8cm},clip]{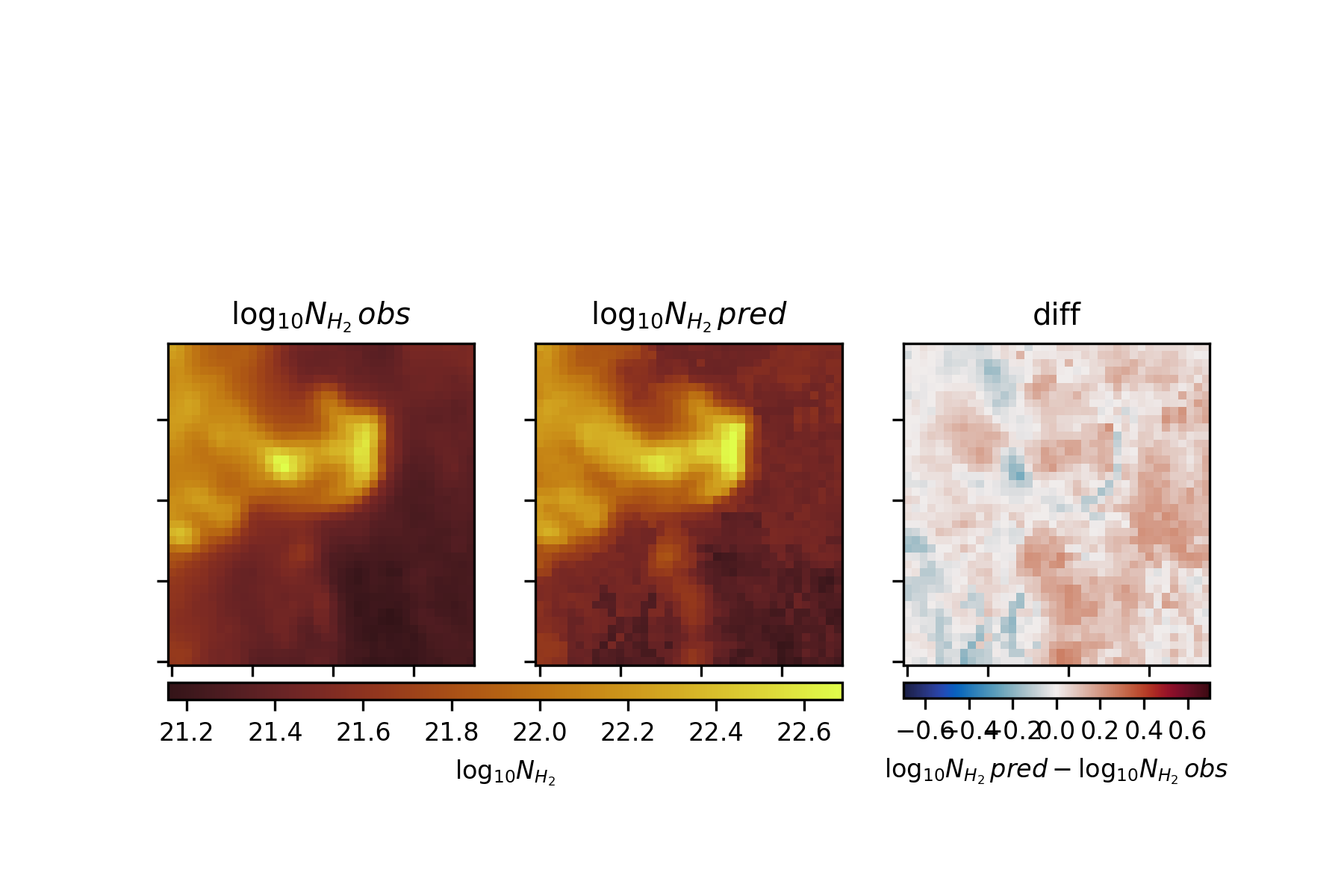}
    \end{minipage}
    \begin{minipage}{0.05\linewidth}
      \mycaption{\thCO{} \Jone{}}
      \mycaption{$+$ \twCO{} \Jone{}}
      \mycaption{$+$ \CeiO{} \Jone{}}
      \mycaption{$+$ \HCOp{} \Jone{}}
      \mycaption{$+$ \HNC{}  \Jone{}}
      \mycaption{$+$ \CCH{} \Jone{}}
      \mbox{}
      \vspace*{1.5cm}
      \mbox{}
    \end{minipage}
    \caption{Evolution of the prediction of the column density when adding
      molecular tracers one by one during the training phase.  \textbf{Left
        and middle columns:} Spatial distribution of the observed and
      predicted column densities. Both images share the same color
      scale. \textbf{Right column:} Ratio of the predicted column density
      over the observed one. The limits of the color scale correspond to a
      ratio interval from $1/5$ to $5$.}
    \label{fig:results:maps:onebyone}
  \end{figure}}

\newcommand{\FigContributionMaps}{%
  \begin{figure*}
    \centering{} %
    {\Large Integrated intensity \hspace{5cm} Peak temperature}
    \vspace*{0.1cm}
    \mbox{}
    \begin{minipage}{0.49\linewidth}
      \centering{} %
      \thCO{} \Jone{} \hspace{2cm} \twCO{} \Jone{} \\
      \includegraphics[height=0.45\linewidth,trim={0 0 2.5cm 0.65cm},clip]{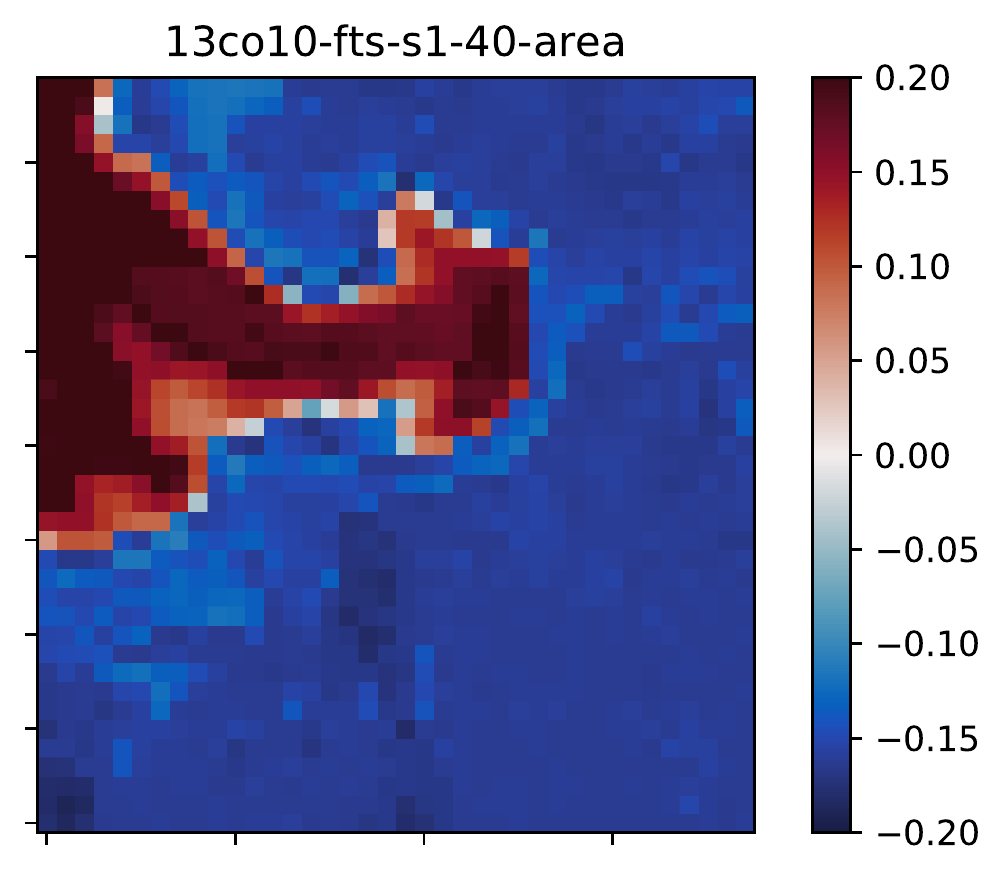}
      \includegraphics[height=0.45\linewidth,trim={0.0 0 2.5cm 0.65cm},clip]{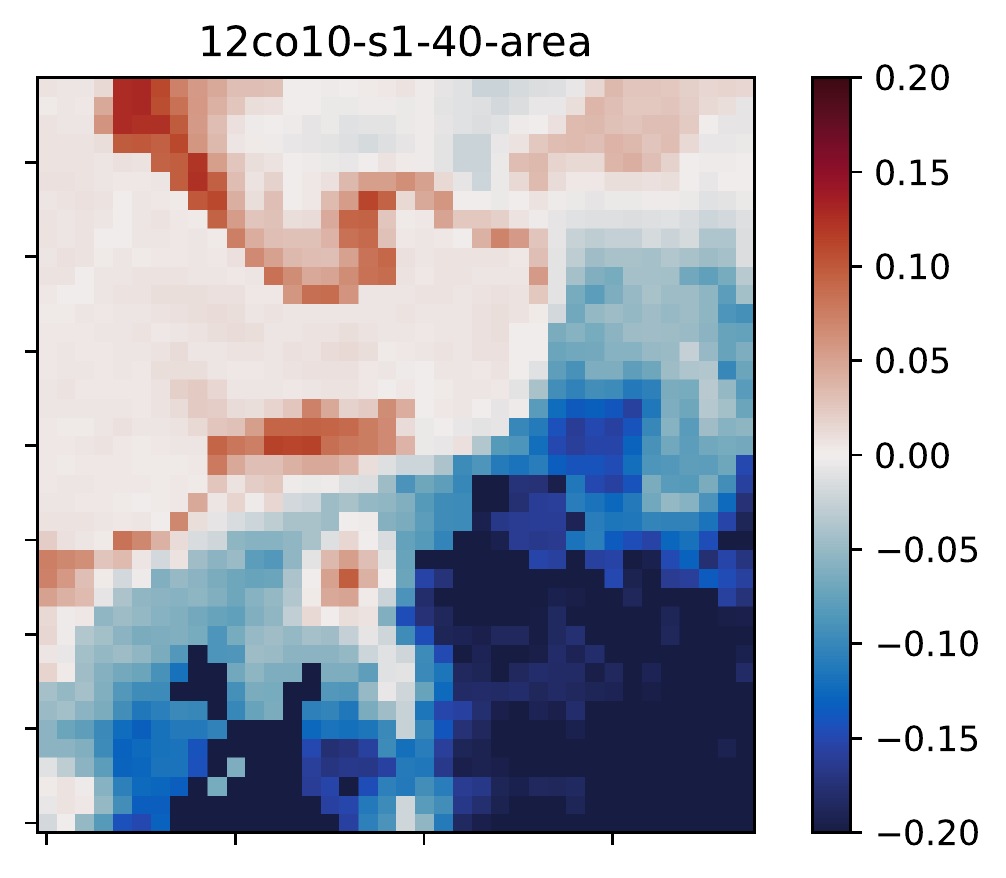}\\
      \CeiO{} \Jone{} \hspace{2cm} \HCOp{} \Jone{} \\
      \includegraphics[height=0.45\linewidth,trim={0 0 2.5cm 0.65cm},clip]{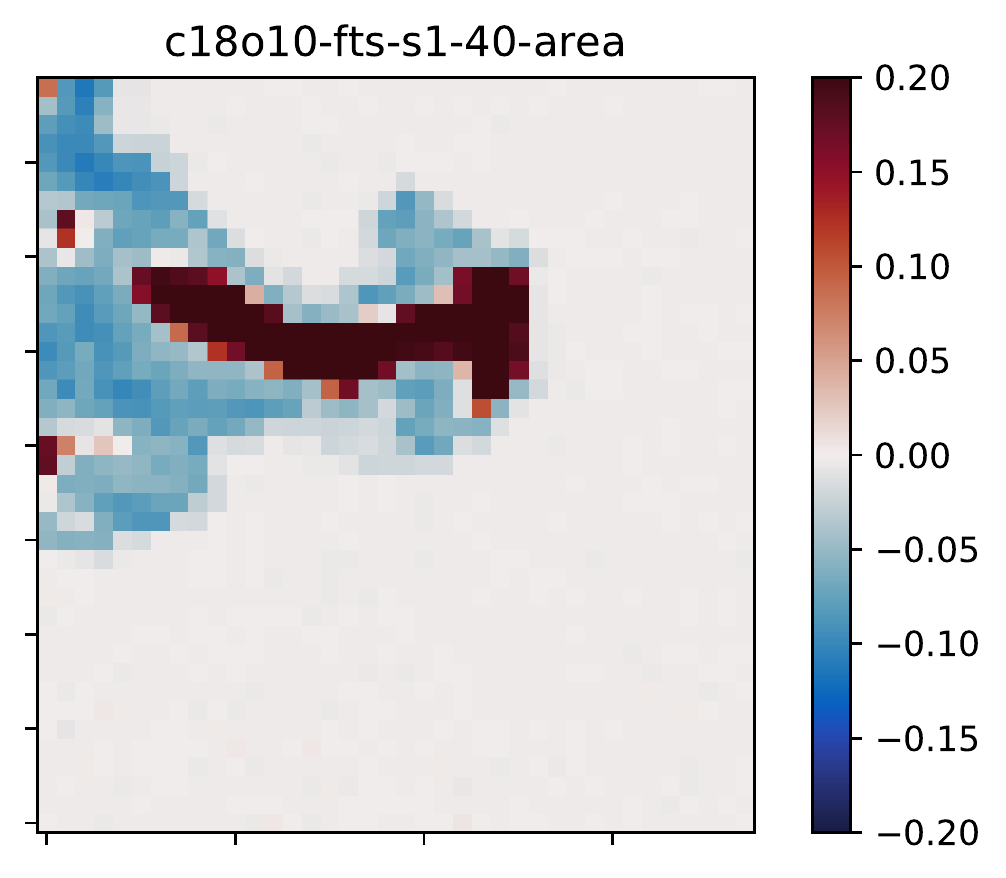}
      \includegraphics[height=0.45\linewidth,trim={0 0 2.5cm 0.65cm},clip]{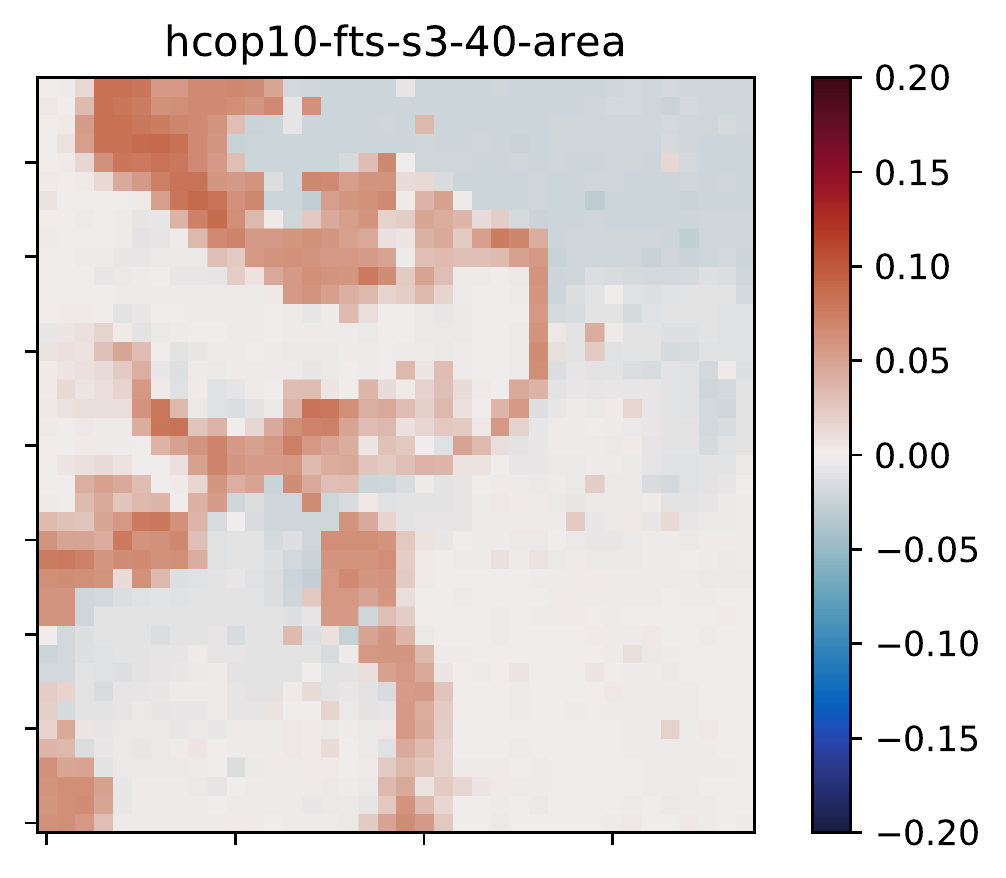}\\
      \HNC{} \Jone{} \hspace{2cm} \NNHp{} \Jone{} \\
      \includegraphics[height=0.45\linewidth,trim={0 0 2.5cm 0.65cm},clip]{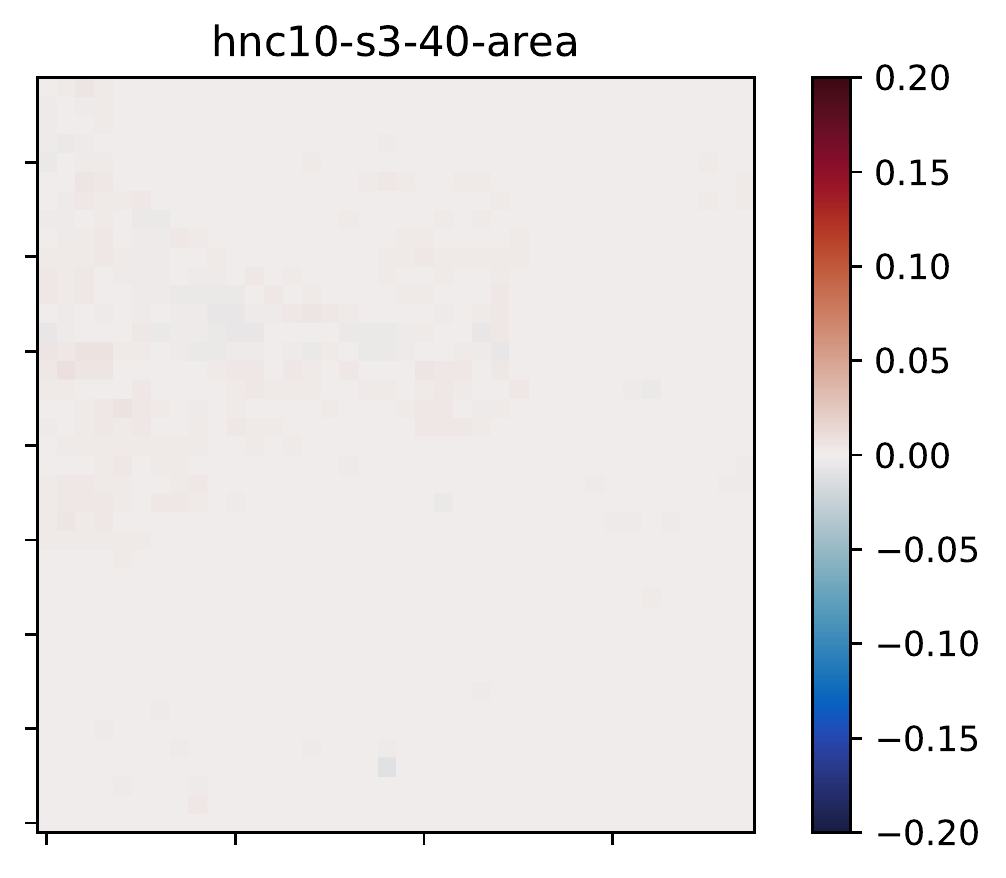}
      \includegraphics[height=0.45\linewidth,trim={0 0 2.5cm 0.65cm},clip]{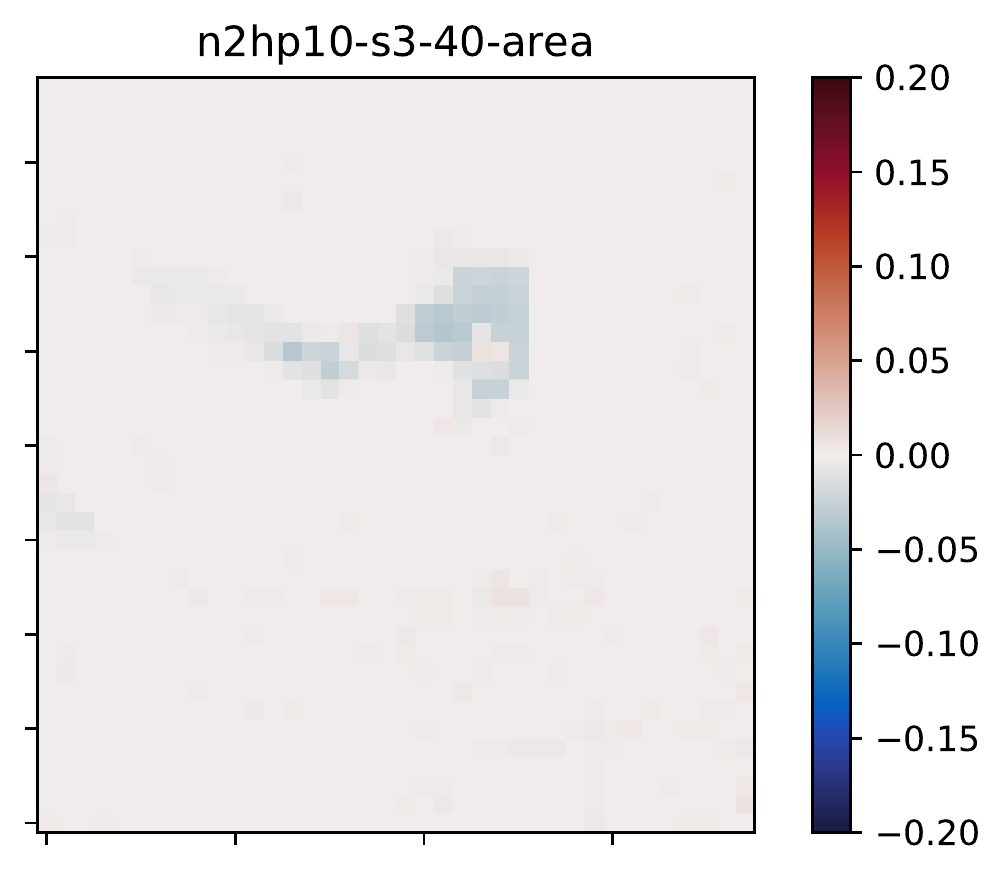}\\
      CCH \Jone{} \hspace{2cm} \twCS{} \Jtwo{} \\
      \includegraphics[height=0.45\linewidth,trim={0 0 2.5cm 0.65cm},clip]{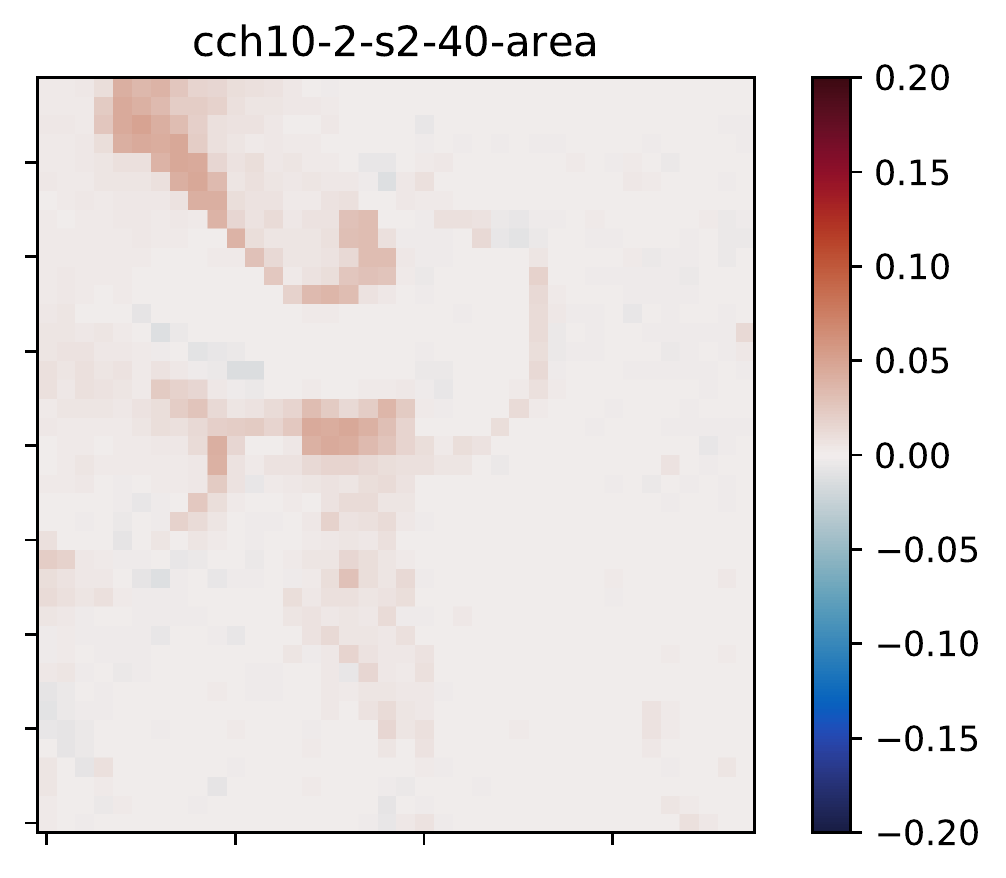}      
      \includegraphics[height=0.45\linewidth,trim={0 0 2.5cm 0.65cm},clip]{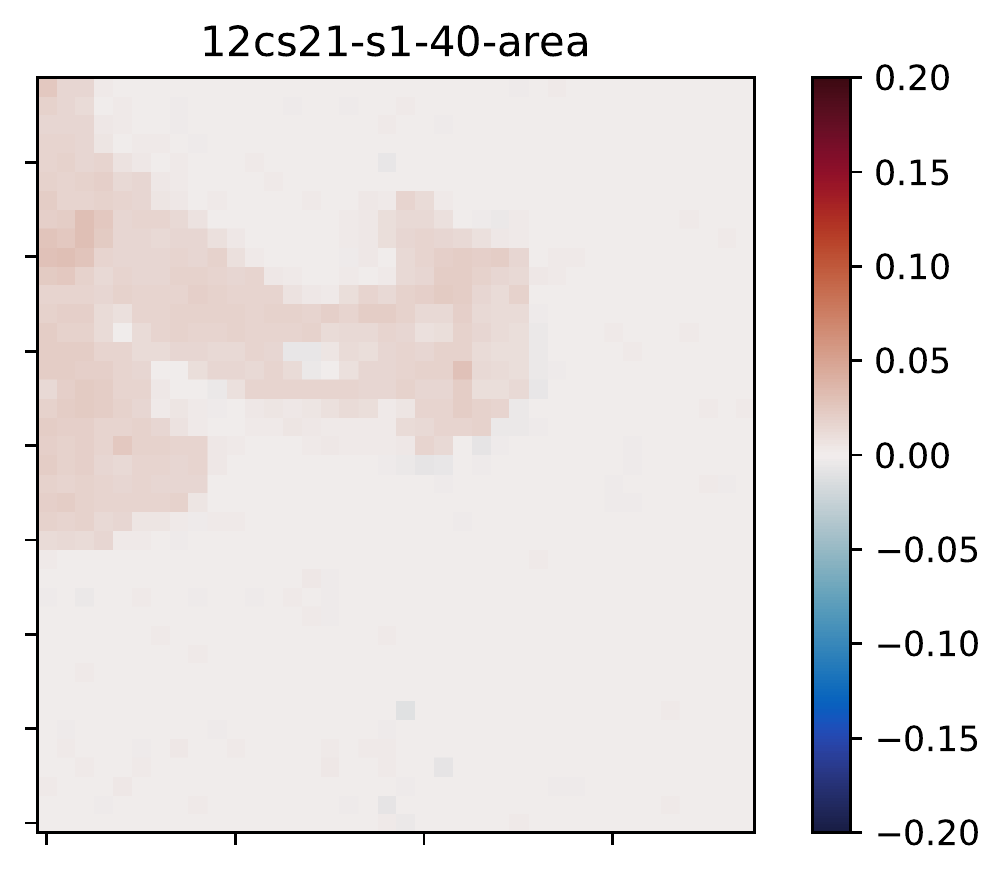}      
    \end{minipage}
    \hfill{} %
    \begin{minipage}{0.49\linewidth}
      \centering{} %
      \thCO{} \Jone{} \hspace{2cm} \twCO{} \Jone{} \hspace{0.7cm} \mbox{} \\
      \includegraphics[height=0.45\linewidth,trim={0 0 2.5cm 0.65cm},clip]{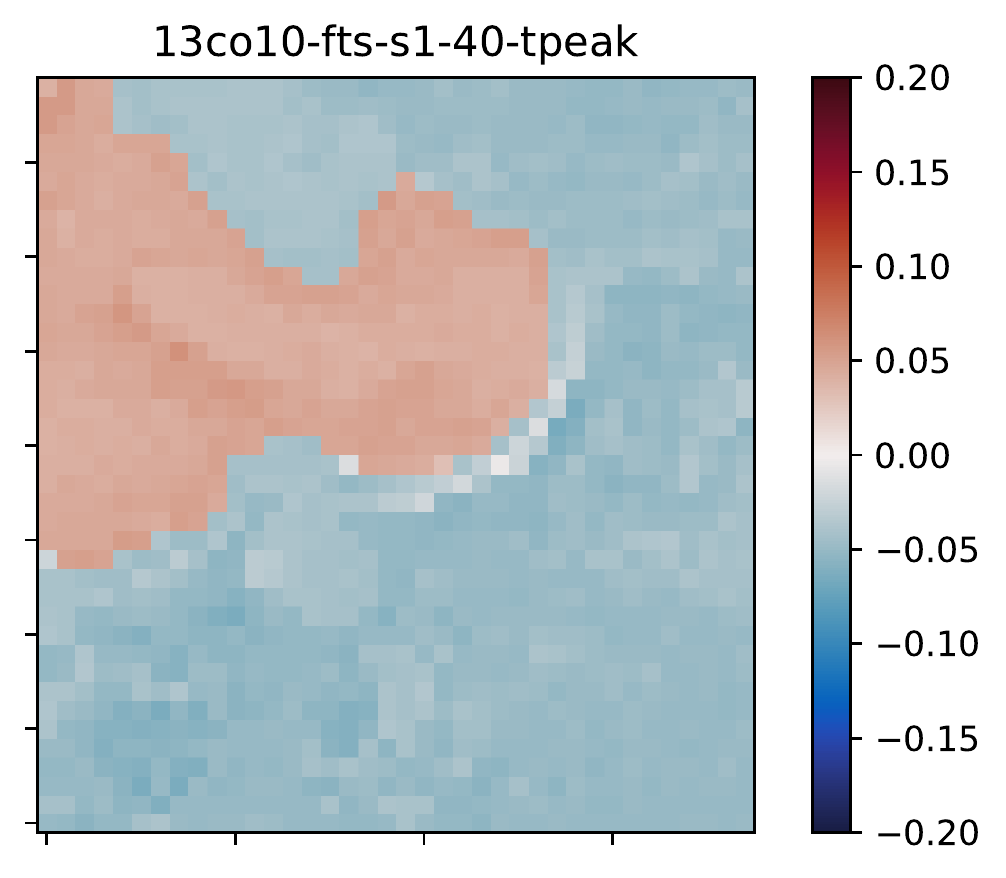}
      \includegraphics[height=0.45\linewidth,trim={0 0 0.3cm 0.65cm},clip]{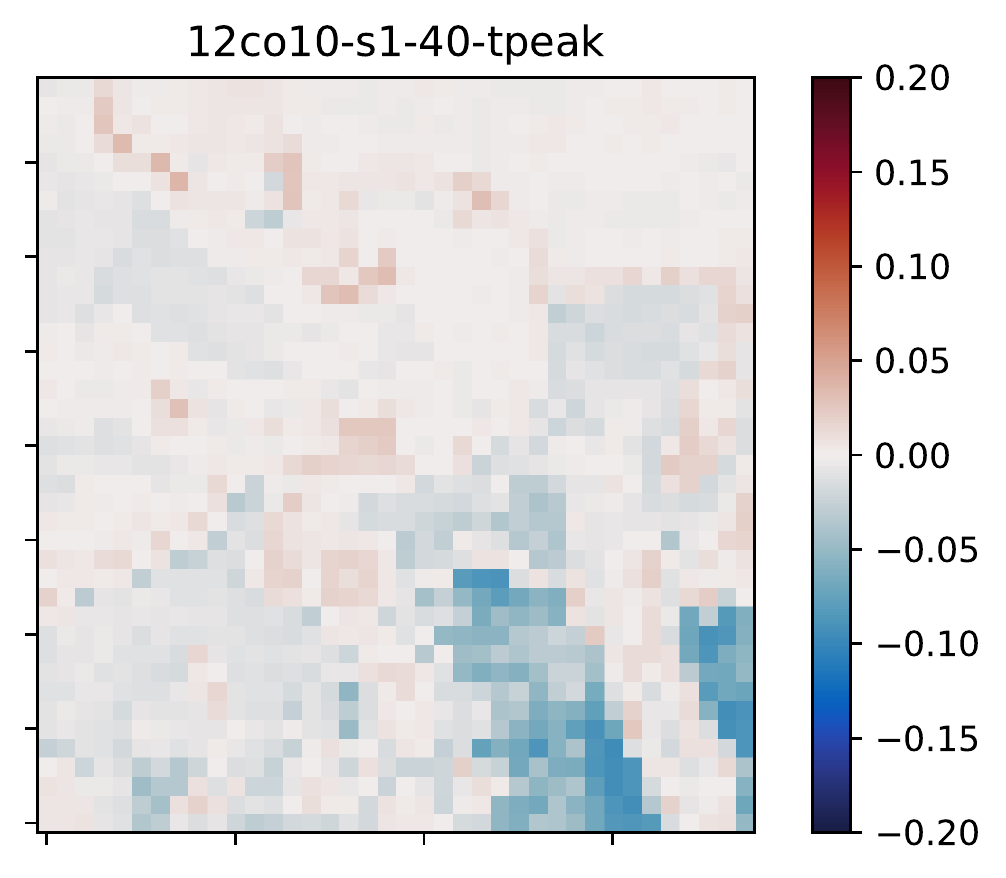}
      \CeiO{} \Jone{} \hspace{2cm} \HCOp{} \Jone{} \hspace{0.7cm} \mbox{} \\
      \includegraphics[height=0.45\linewidth,trim={0 0 2.5cm 0.65cm},clip]{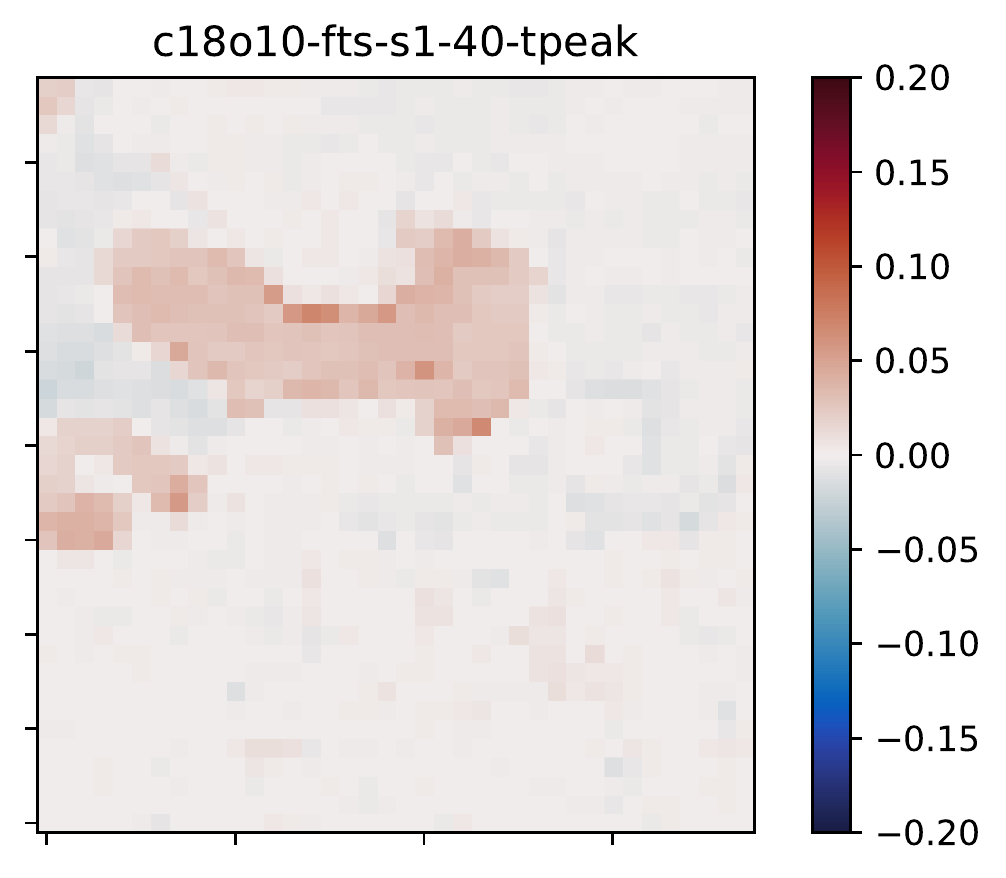}
      \includegraphics[height=0.45\linewidth,trim={0 0 0.3cm 0.65cm},clip]{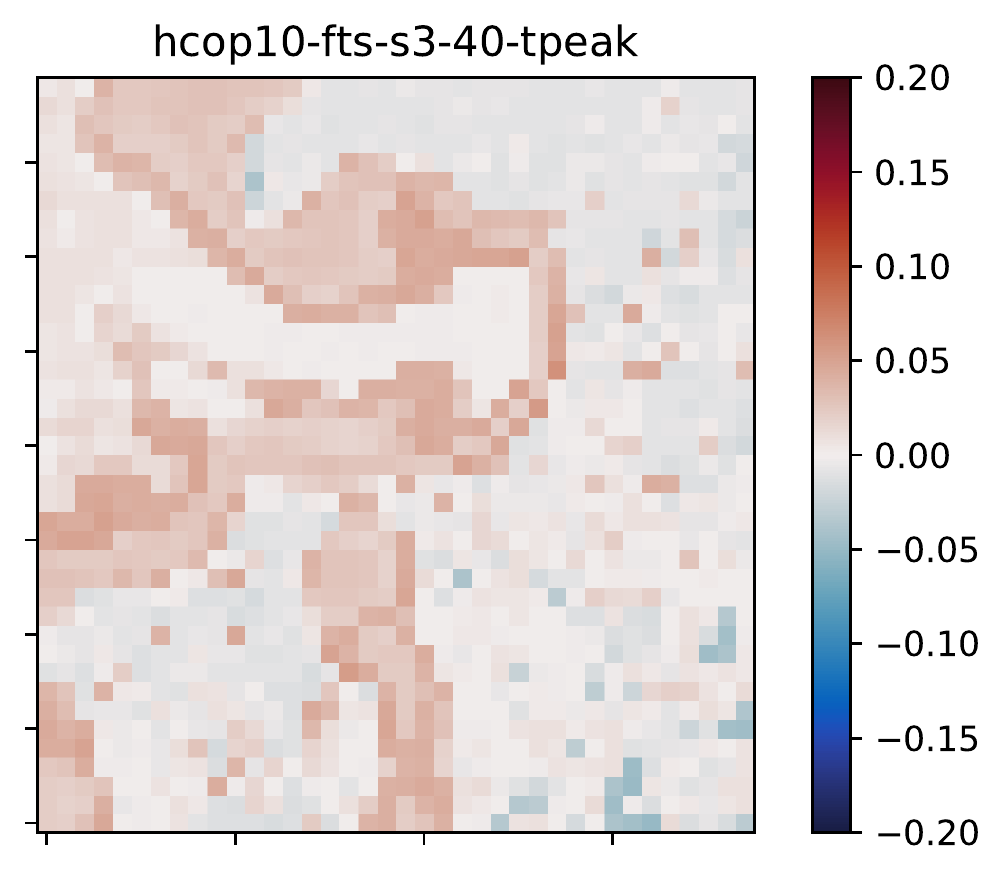}
      \HNC{} \Jone{} \hspace{2cm} \NNHp{} \Jone{} \hspace{0.7cm} \mbox{} \\
      \includegraphics[height=0.45\linewidth,trim={0 0 2.5cm 0.65cm},clip]{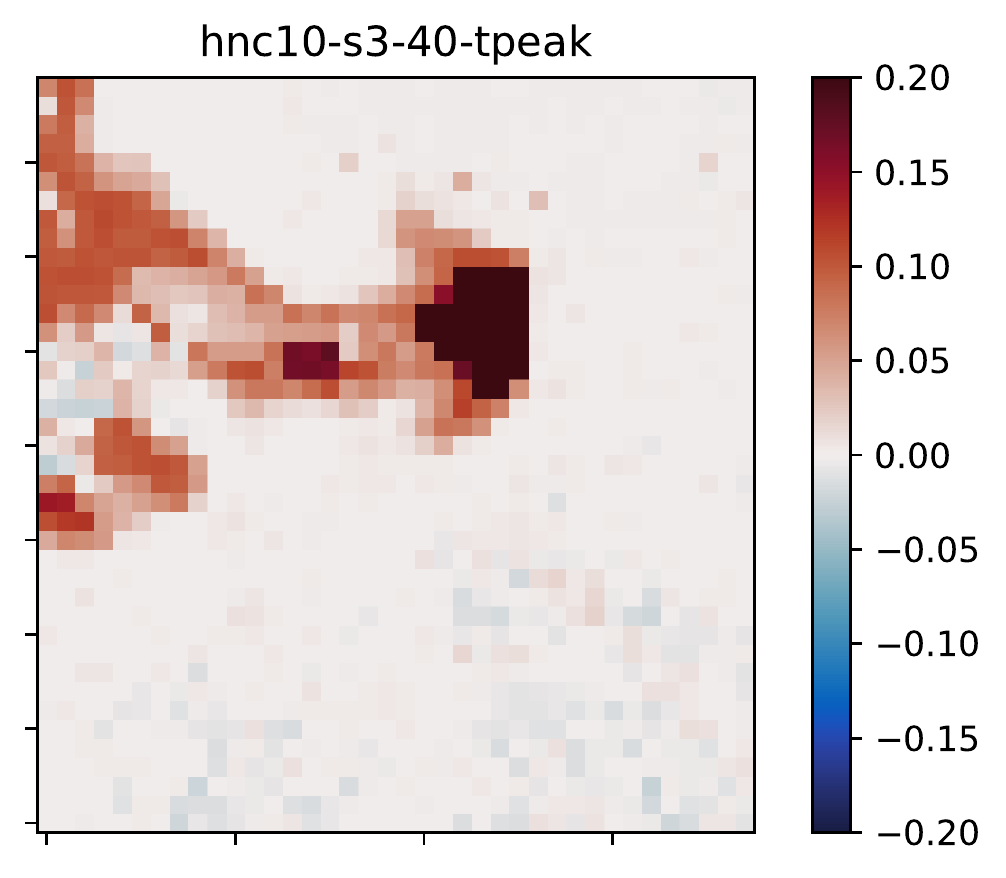}
      \includegraphics[height=0.45\linewidth,trim={0 0 0.3cm 0.65cm},clip]{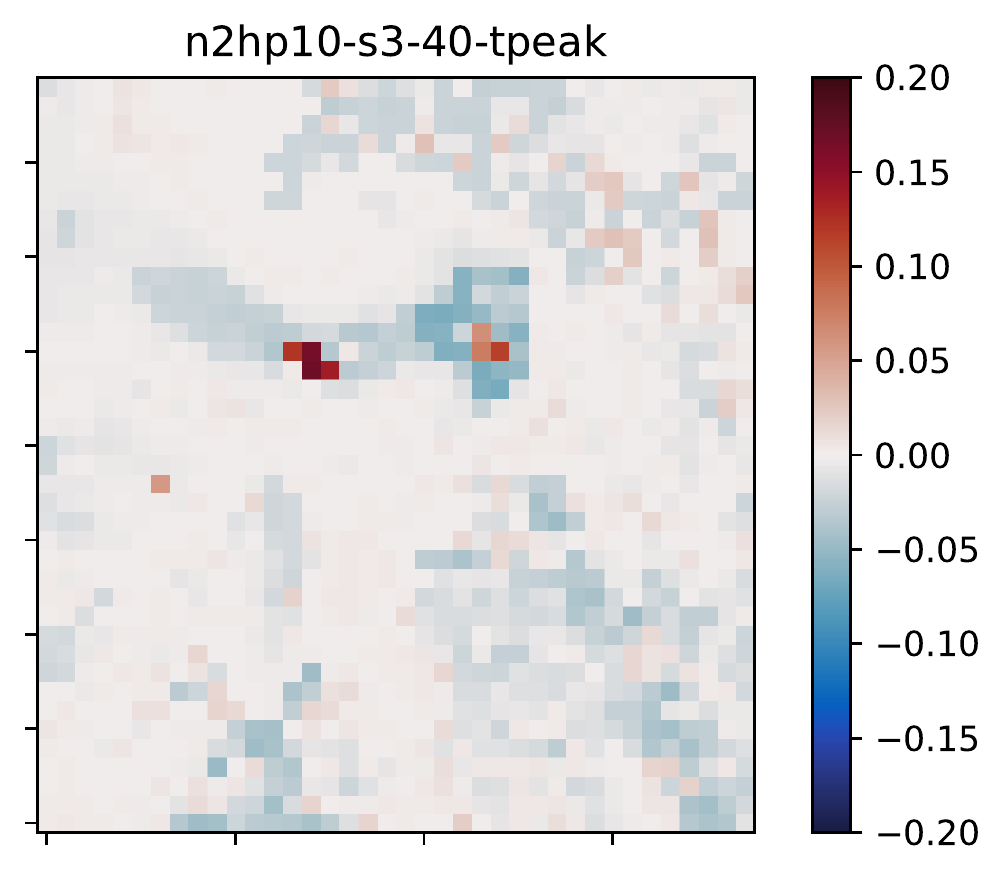}
      CCH \Jone{} \hspace{2cm} \twCS{} \Jtwo{} \hspace{0.7cm} \mbox{} \\
      \includegraphics[height=0.45\linewidth,trim={0 0 2.5cm 0.65cm},clip]{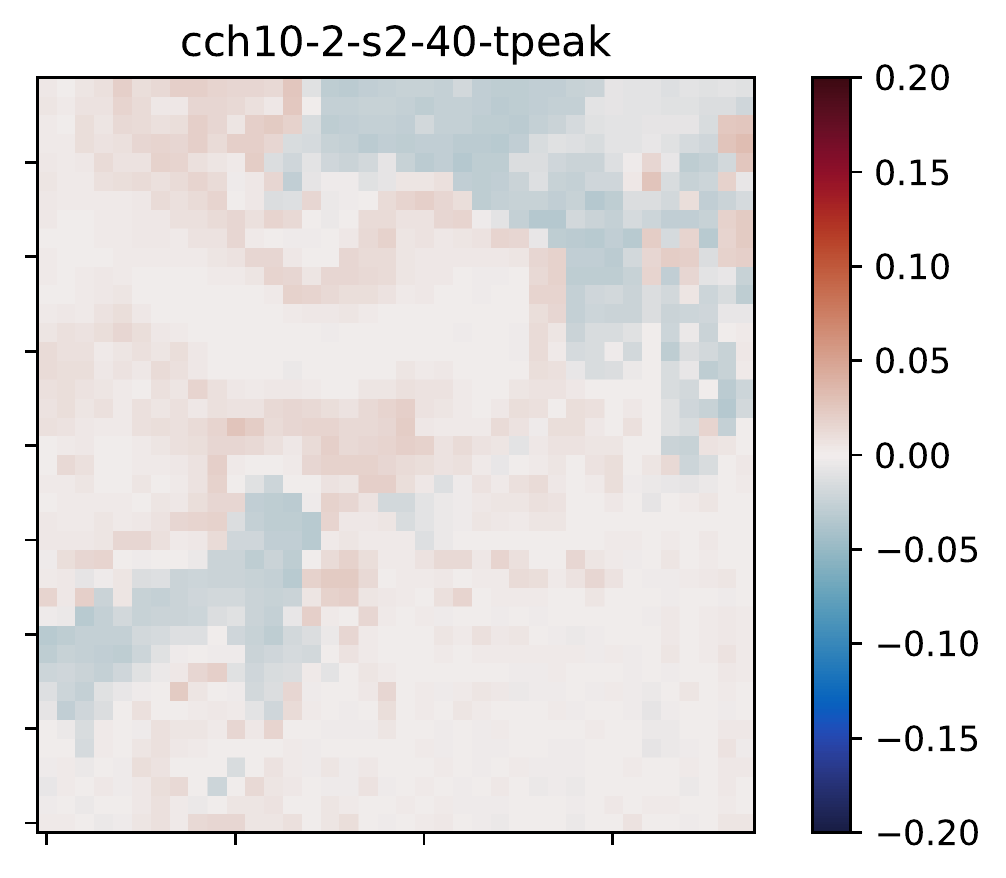}      
      \includegraphics[height=0.45\linewidth,trim={0 0 0.3cm 0.65cm},clip]{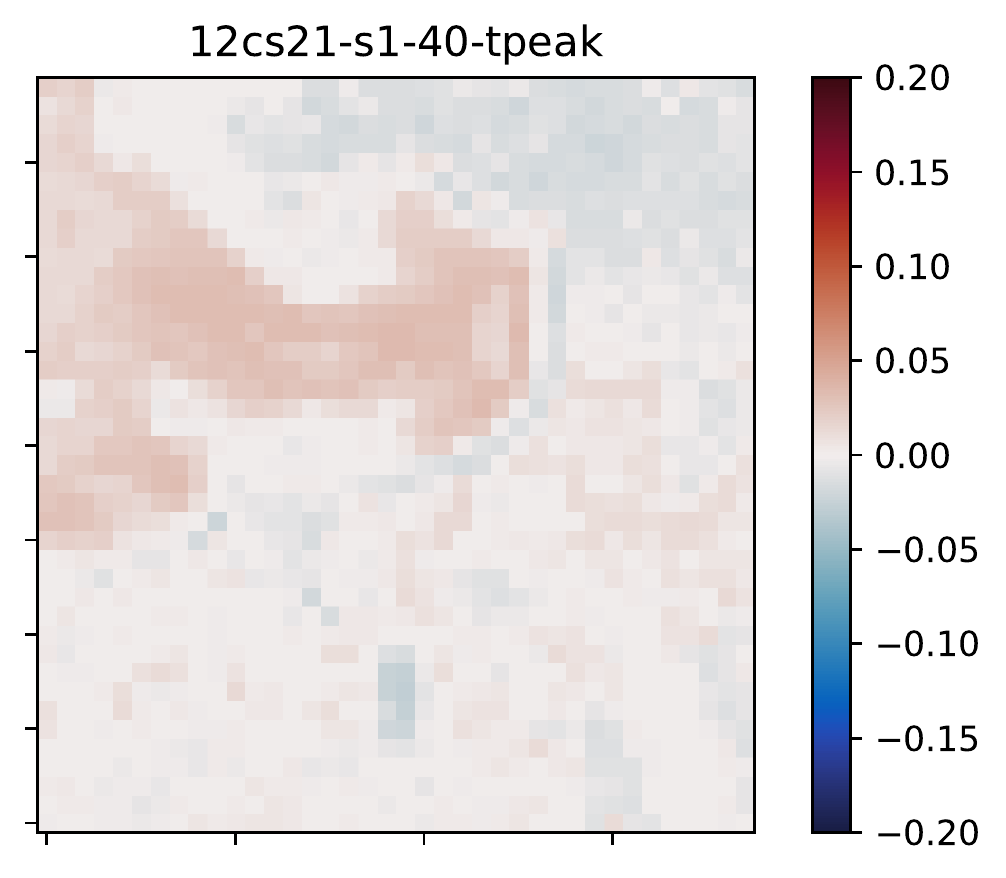}
    \end{minipage}
    \caption{Spatial distribution of the contribution of the integrated
      intensity and the peak temperature of the eight most important lines
      to the estimation of $\log \NHt$. All the maps share the same color
      look-up table to emphasize the relative contributions across pixels
      and lines.}
    \label{fig:contribution-maps}
  \end{figure*}}

\newcommand{\FigAvMasks}{%
  \begin{figure}
    \centering %
    \includegraphics[width=\linewidth,trim={6.7cm 1.1cm 3.9cm 1.0cm},clip]{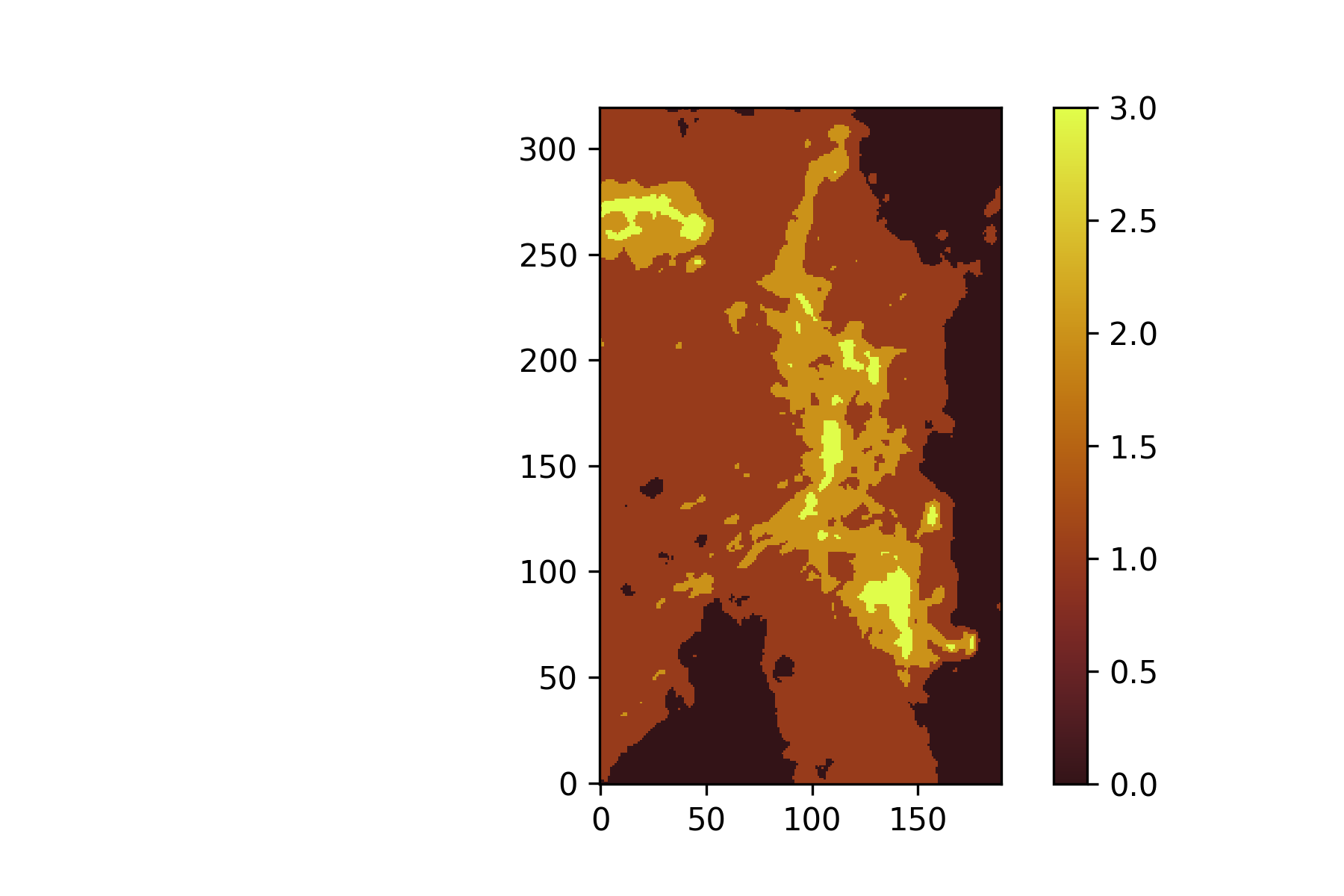}
    \caption{Spatial distribution of the four following masks:
      $1 \le \Av < 2$ in black, $2 \le \Av < 6$ in brown, $6 \le \Av < 15$
      in orange, and $15\le \Av$ in yellow.}
    \label{fig:Av-masks}
  \end{figure}}

\newcommand{\FigLineImportanceAvMasksTrainLatex}{%
  \begin{figure*}
    \centering %
    \includegraphics[width=0.49\linewidth,trim={0 0.6cm 0.25cm 0.3cm},clip]{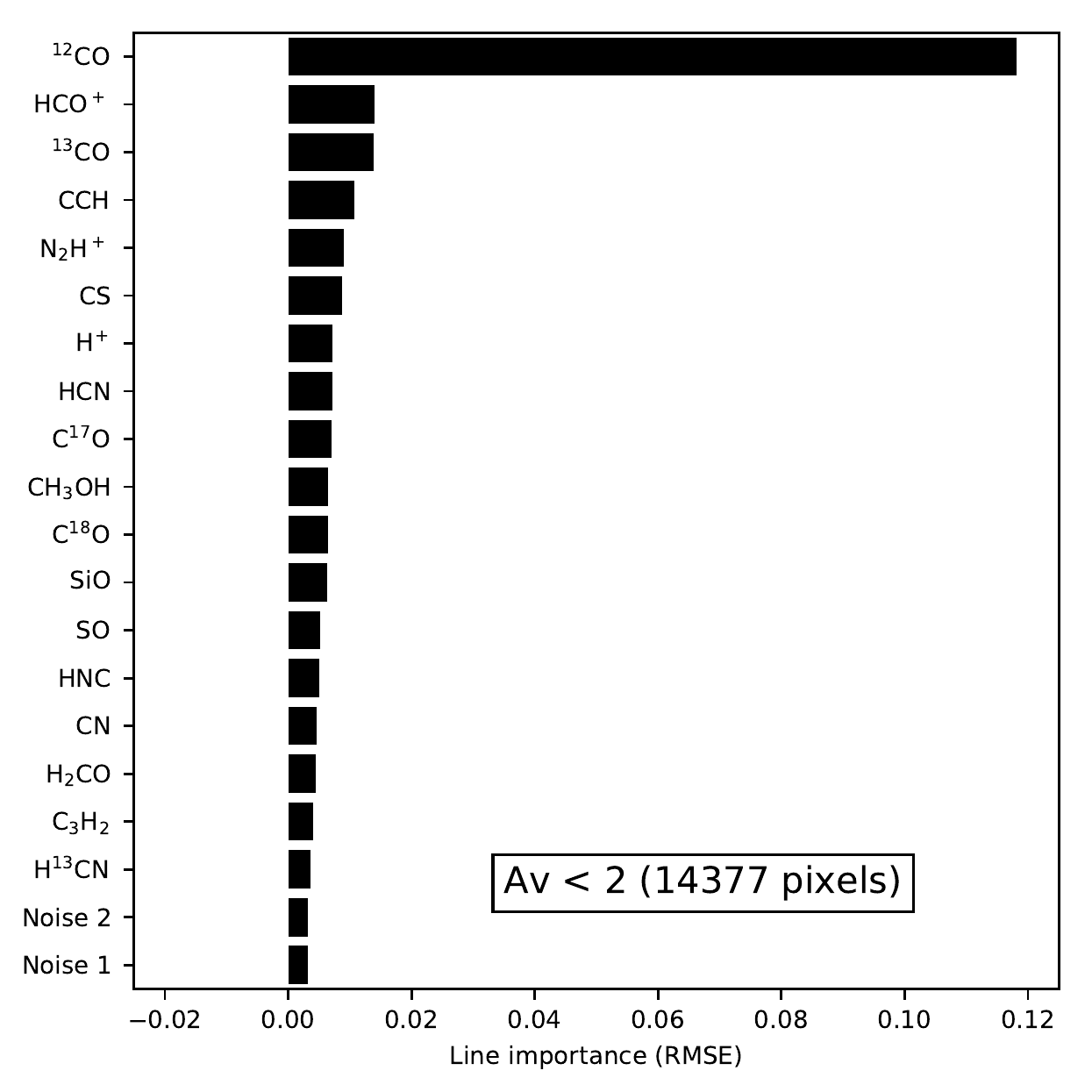}
    \includegraphics[width=0.49\linewidth,trim={0 0.6cm 0.25cm 0.3cm},clip]{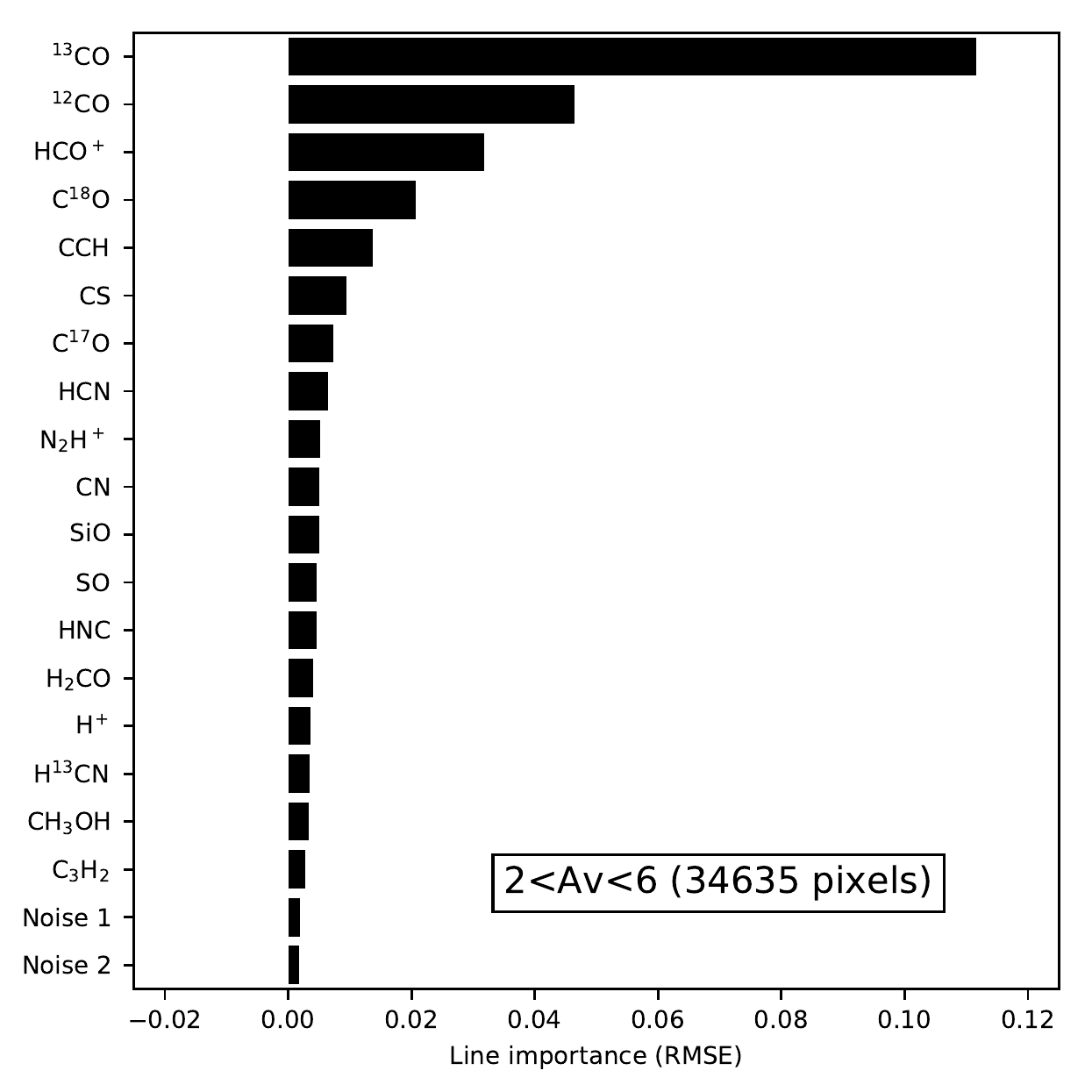}
    \includegraphics[width=0.49\linewidth,trim={0 0.0cm 0.25cm 0.3cm},clip]{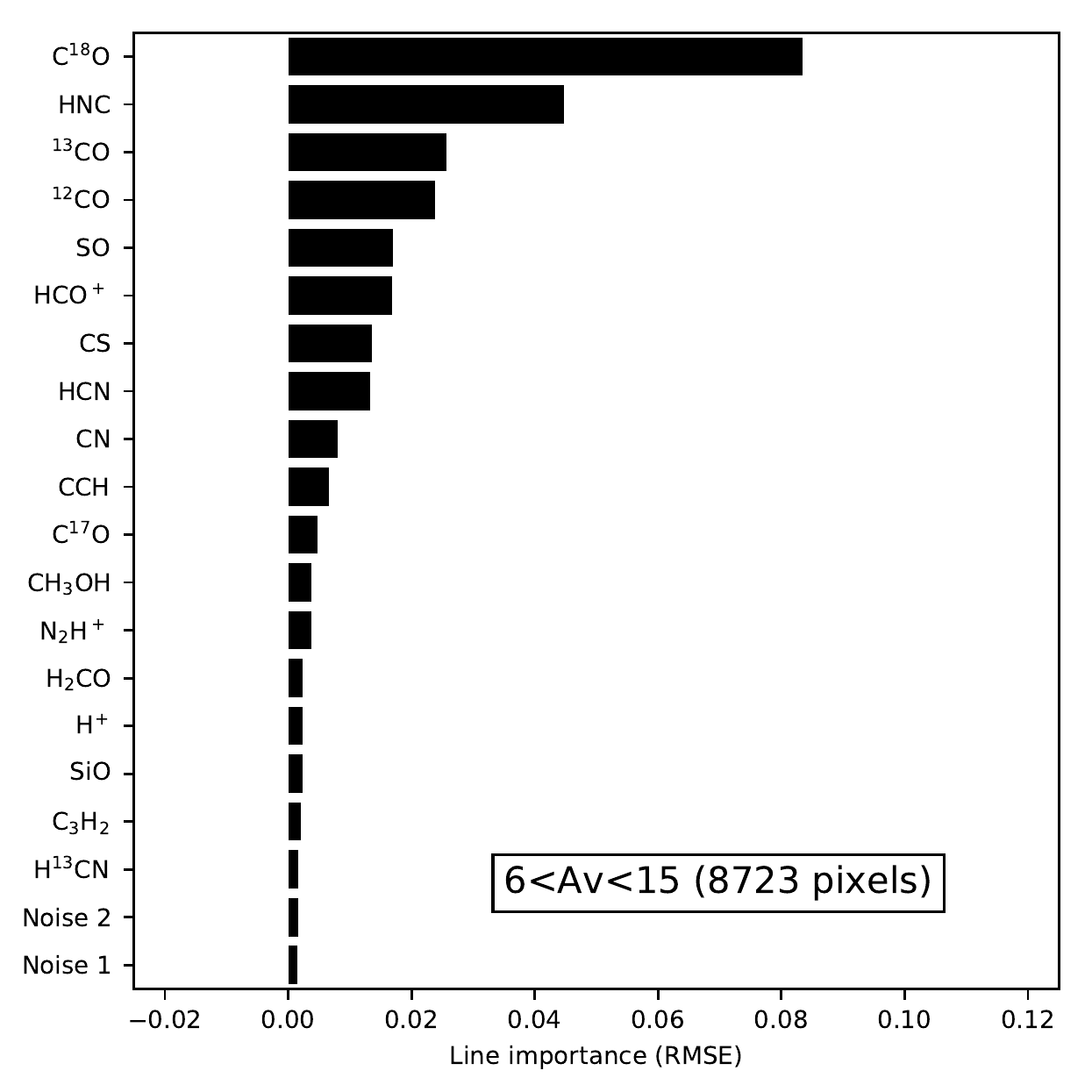}
    \includegraphics[width=0.49\linewidth,trim={0 0.0cm 0.25cm 0.3cm},clip]{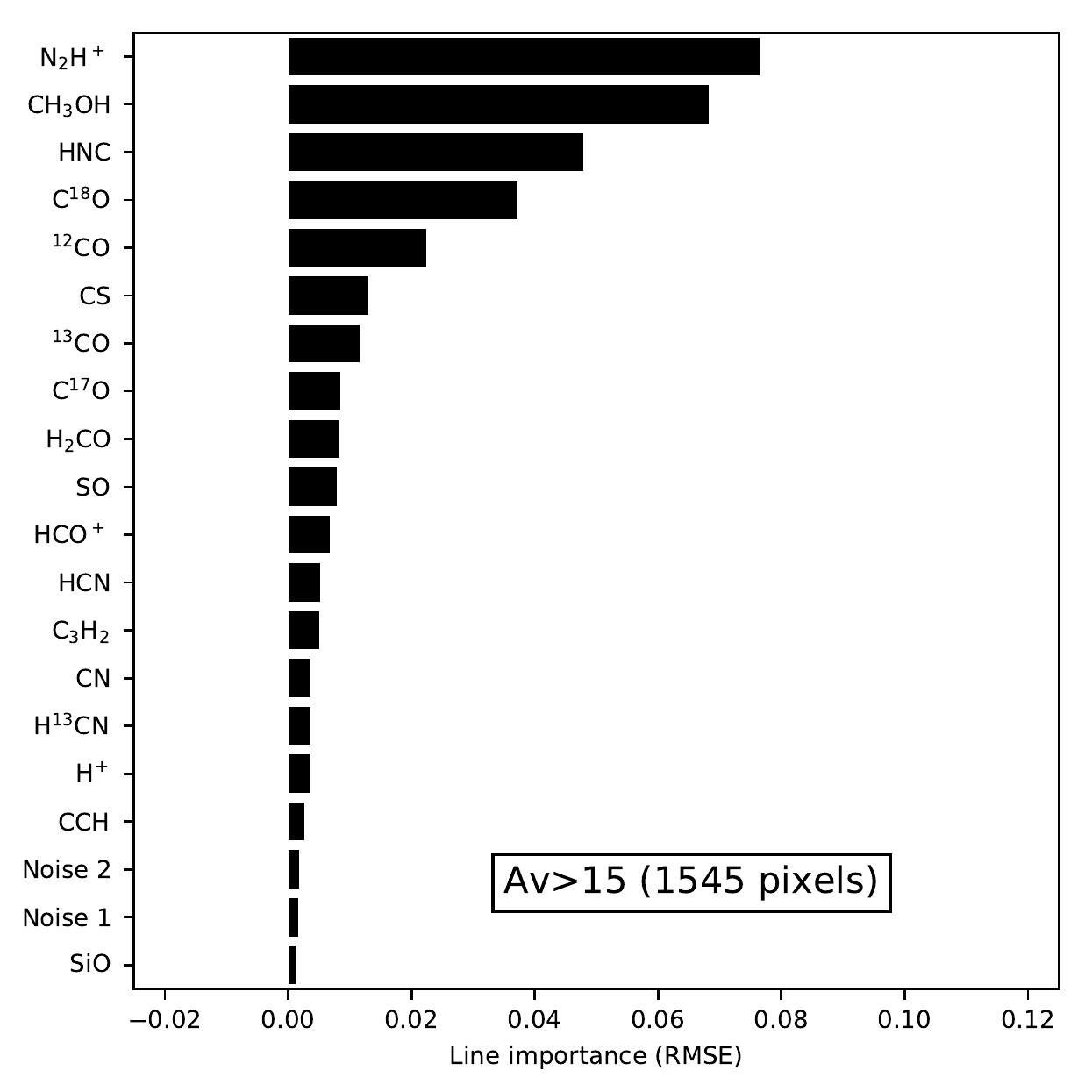}
    \caption{Contributions of the different lines (both integrated
      intensity and temperature peak) to the quality of the $\log \NHt$-fit
      of the training set, depending on the range of visual extinction.
      These results are computed on the training set. Noise \#1 and 2 are
      two additional random sets of input data.}
    \label{fig:line-importance-Av-masks}
  \end{figure*}}

\newcommand{\FigContributionVsAv}{%
  \begin{figure*}
    \centering %
    \includegraphics[width=\linewidth]{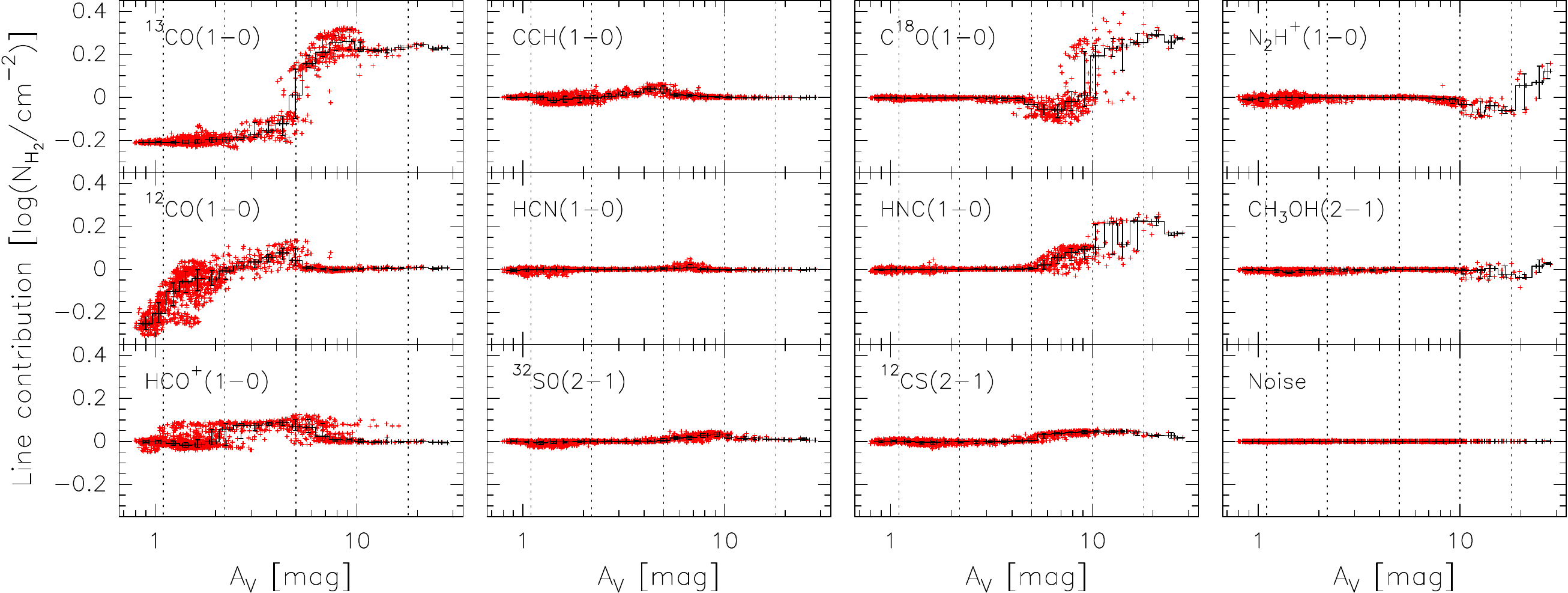}
    \caption{\referee{Contribution of the most important lines to the logarithm of
      the \Ht{} column density around the mean column density of the
      training set as a function of the visual extinction. One red points
      per pixel of the test set is plotted. The black histograms show the
      median values of all data points falling in a regularly sampled
      interval of the logarithm of the visual extinction. The black error
      bars show the range of values where 50\% of the points in the current
      bin are located. The vertical dotted lines show visual extinctions of
      1.1, 2.2., 5.0, 10.0, and 18.0. Except the \thCO{} \Jone{} line that
      contributes at all \Av{} and the noise sample that does not
      contribute at any \Av{}, the other lines contribute inside a given
      \Av{} range. The lines are sorted from top to bottom and then from
      left to right by increasing values of the minimum \Av{} at which they
      start to contribute.}}
    \label{fig:contribution-vs-Av}
  \end{figure*}}

\newcommand{\FigGeneralization}{%
  \begin{figure*}
    \centering %
    \includegraphics[height=6.1cm]{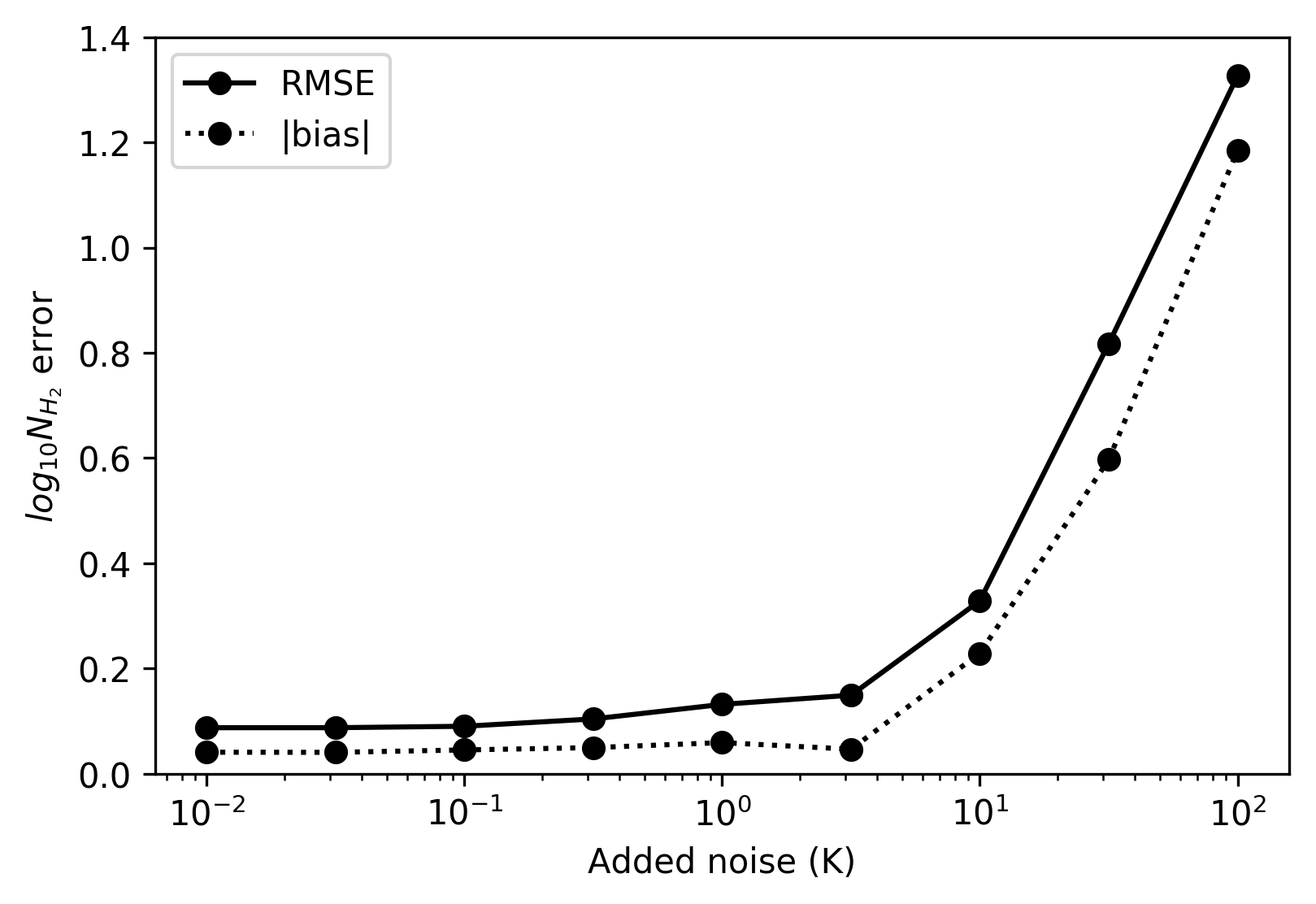}
    \includegraphics[height=6.1cm]{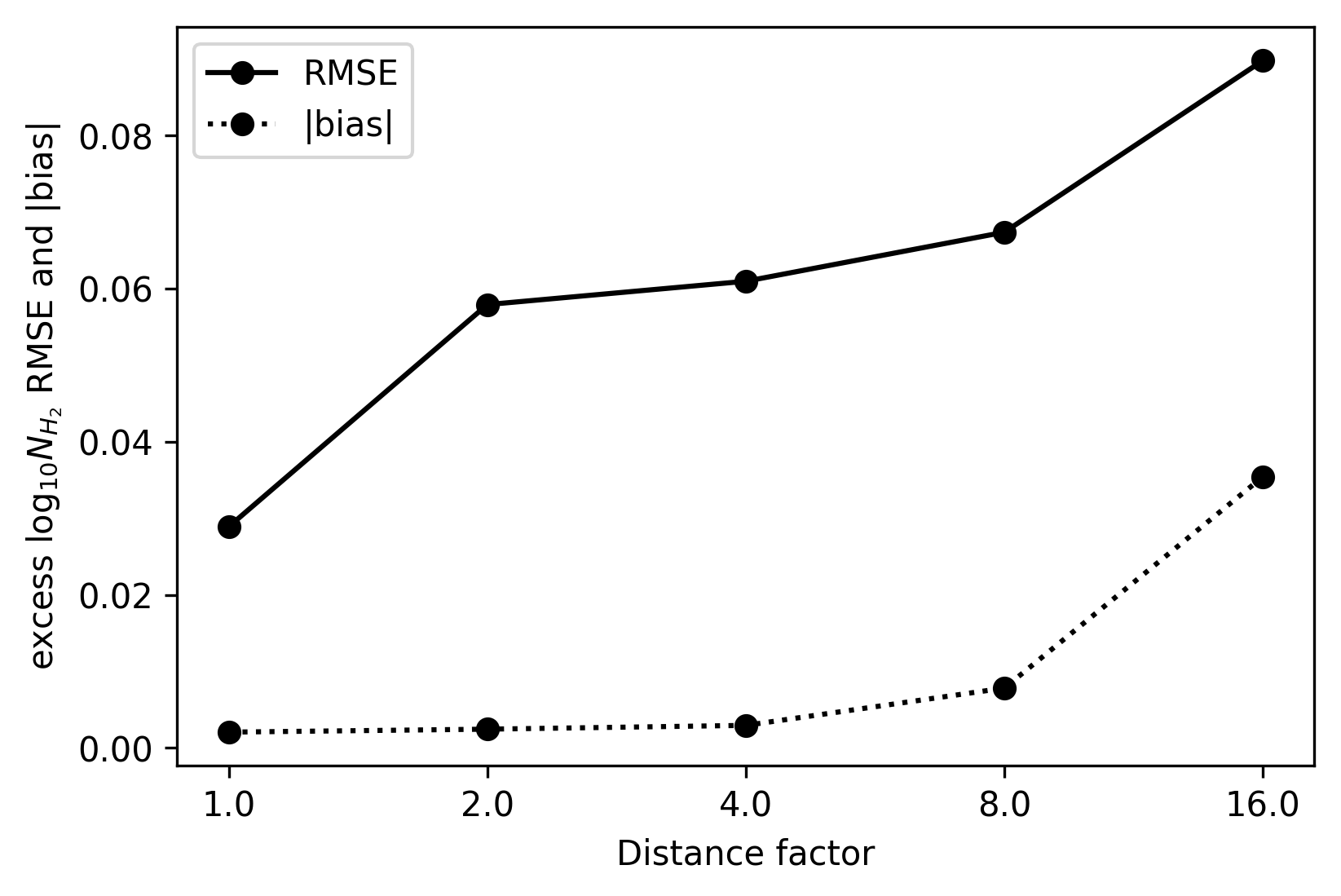}
    \caption{Evolution of the accuracy (RMSE and mean error) of the
      $\log \NHt$ preditcor when changing the condition of
      observations. \textbf{Left:} Gaussian noise of zero mean and a given
      standard deviation is added to the line spectrum. The median noises
      belong to the $[0.03$ and $0.11\K]$ interval depending on the line.
      \textbf{Right:} The line emission maps are smoothed in order to
      simulate an observation at a distance between $0.4$ and $12.8\kpc$.}
    \label{fig:generalization}
  \end{figure*}}

\newcommand{\FigXco}{%
  \begin{figure*}
    \centering %
    \begin{minipage}{0.33\linewidth}
      \centering{} %
      \Large{} %
      Observed
    \end{minipage}
    \begin{minipage}{0.33\linewidth}
      \centering{} %
      \Large{} %
      Predicted
    \end{minipage}
    \begin{minipage}{0.33\linewidth}
      \centering{} %
      \Large{} %
      Predicted/Observed
    \end{minipage}
    \\[\smallskipamount]
    \includegraphics[width=\linewidth,trim={0 0 0 0.75cm},clip]{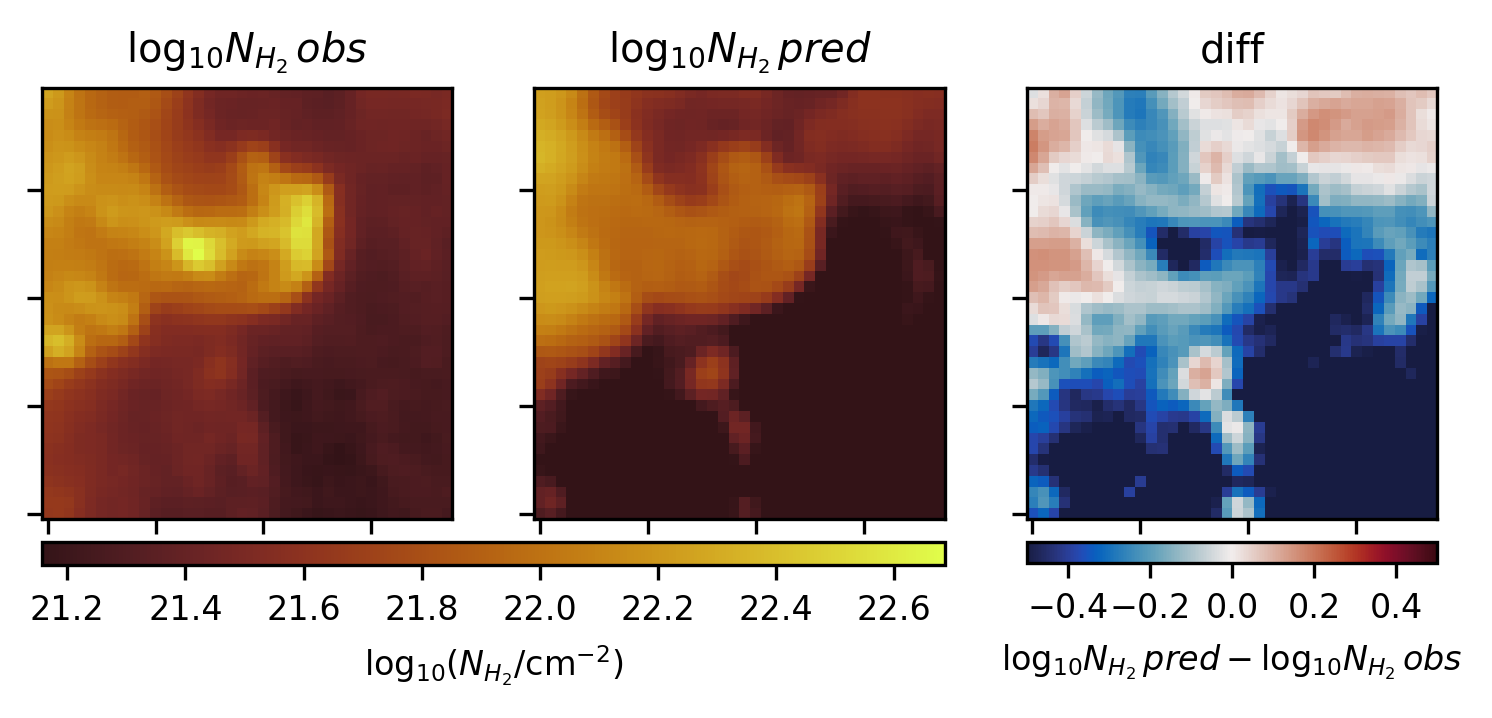}\\[\medskipamount]
    \includegraphics[width=\linewidth,trim={0 0 0
      1.75cm},clip]{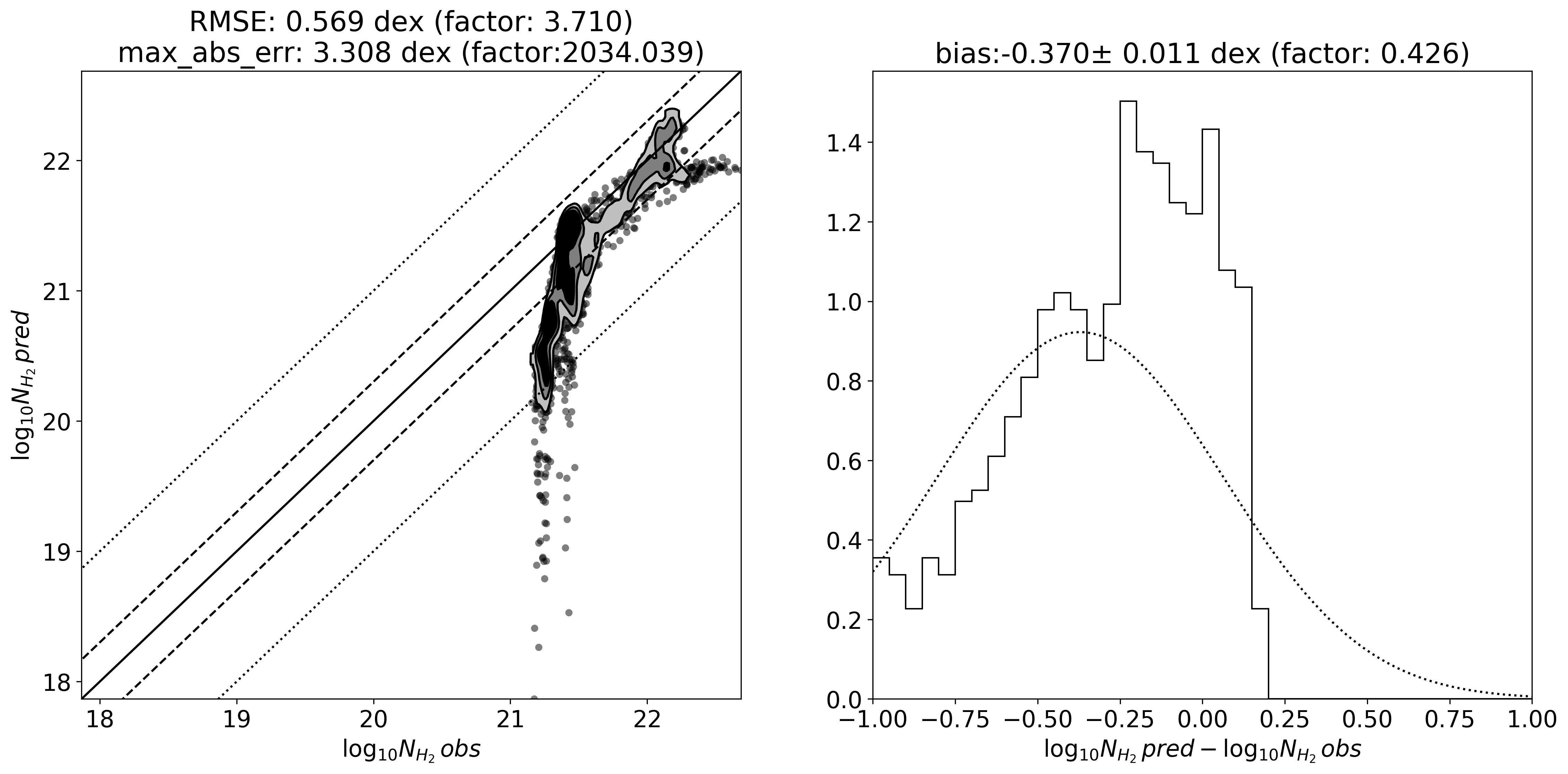}
    \caption{Performance of the standard \Xco{} method to predict the
      column density from the \twCO{} \Jone{} integrated line
      intensity. \textbf{Top:} Spatial distribution of the observed and
      predicted \Ht{} column density (\textbf{left and middle panels}), and
      of the ratio of the predicted column density over the observed one
      (\textbf{right panel}).
      \textbf{Bottom:} Joint histogram of the predicted column density
      as a function of the observed one (\textbf{left panel}), and
      histogram of the ratio of the predicted column density over the
      observed one on a
      logarithmic scale.
      These results are computed on the Horsehead pillar, i.e., the test
      set.}
    \label{fig:Xco}
  \end{figure*}}

\newcommand{\FigCOiso}{%
  \begin{figure*}
    \centering %
    \begin{minipage}{0.33\linewidth}
      \centering{} %
      \Large{} %
      Observed
    \end{minipage}
    \begin{minipage}{0.33\linewidth}
      \centering{} %
      \Large{} %
      Predicted
    \end{minipage}
    \begin{minipage}{0.33\linewidth}
      \centering{} %
      \Large{} %
      Predicted/Observed
    \end{minipage}
    \\[\smallskipamount]
    \includegraphics[width=\linewidth,trim={0 0 0 0.75cm},clip]{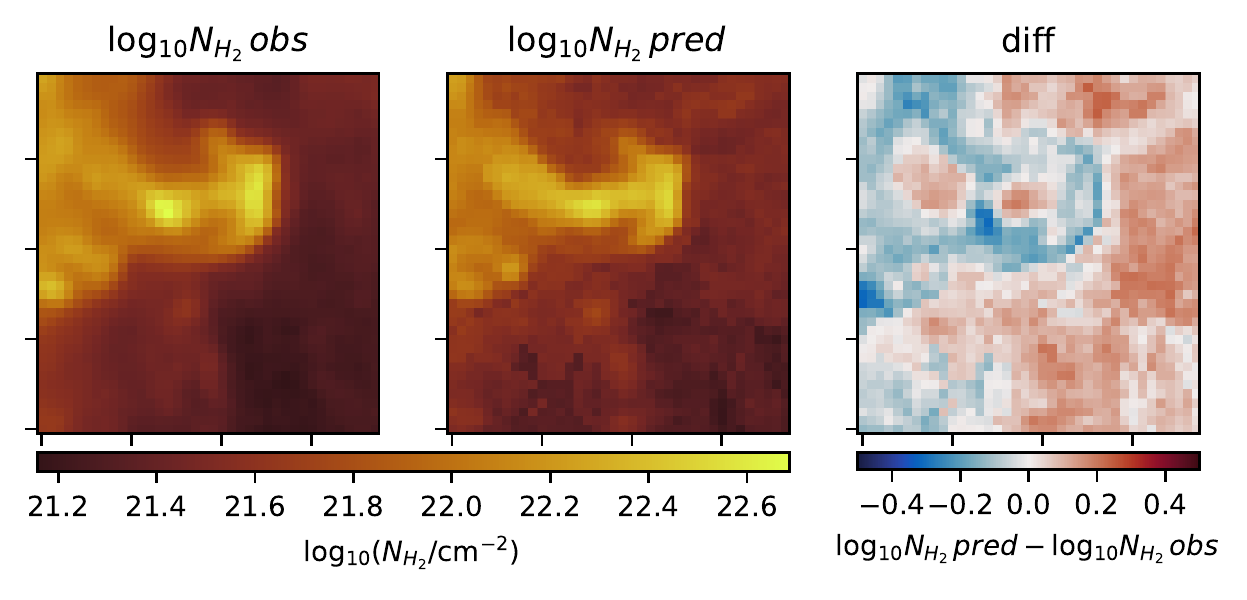}\\[\medskipamount]
    \includegraphics[width=\linewidth,trim={0 0 0
      1.7cm},clip]{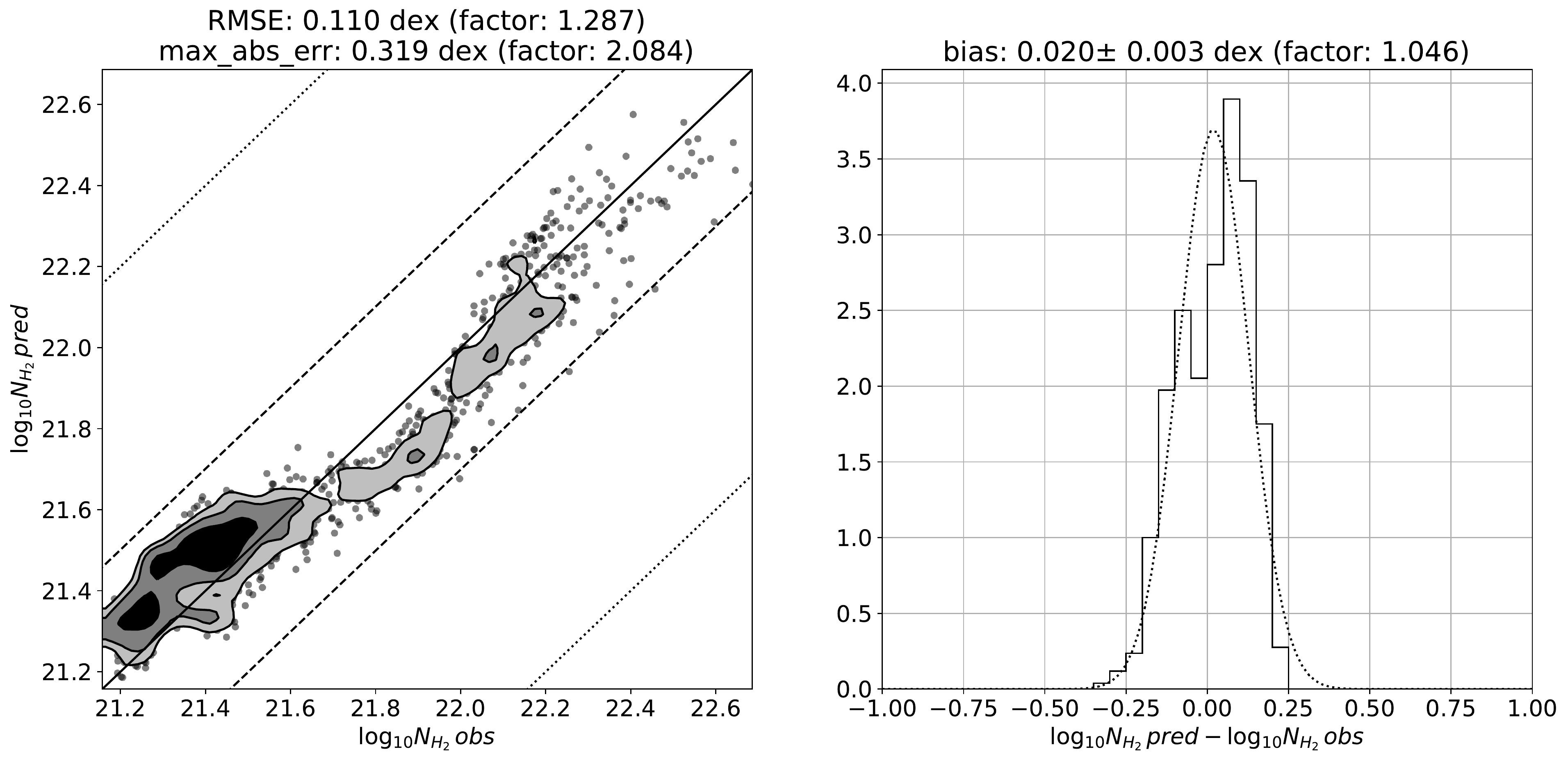}
    \caption{Performance of a random forest regression trained only on the 
      the three main CO isotopologues. The layout of the figure is
      identical to Fig.~\ref{fig:Xco}.}
    \label{fig:COiso}
  \end{figure*}}

\newcommand{\FigGo}{%
  \begin{figure*}
    \centering %
    \begin{minipage}{0.33\linewidth}
      \centering{} %
      \Large{} %
      Observed
    \end{minipage}
    \begin{minipage}{0.33\linewidth}
      \centering{} %
      \Large{} %
      Predicted
    \end{minipage}
    \begin{minipage}{0.33\linewidth}
      \centering{} %
      \Large{} %
      Predicted/Observed
    \end{minipage}
    \\[\smallskipamount]
    \includegraphics[width=\linewidth,trim={0 0 0 0.75cm},clip]{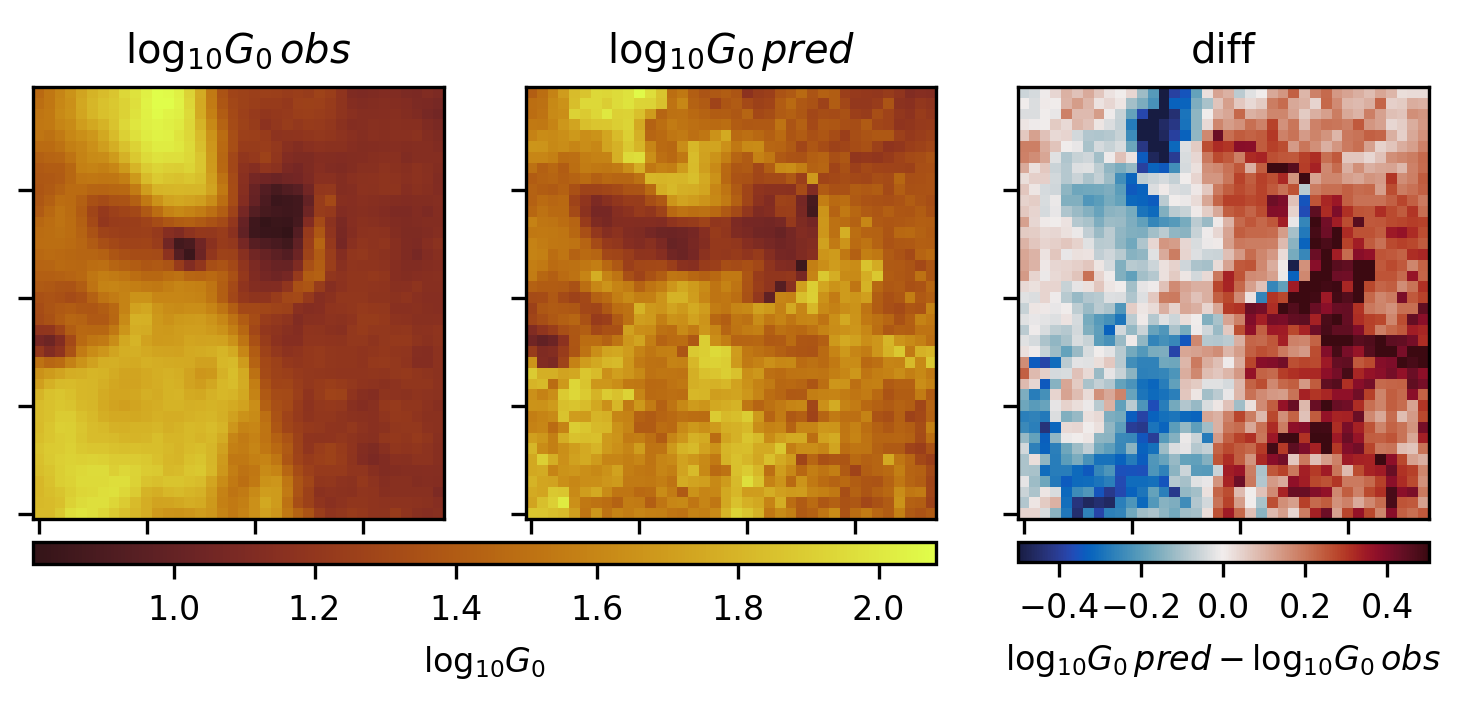}\\[\medskipamount]
    \includegraphics[width=\linewidth,trim={0 0 0 1.7cm},clip]{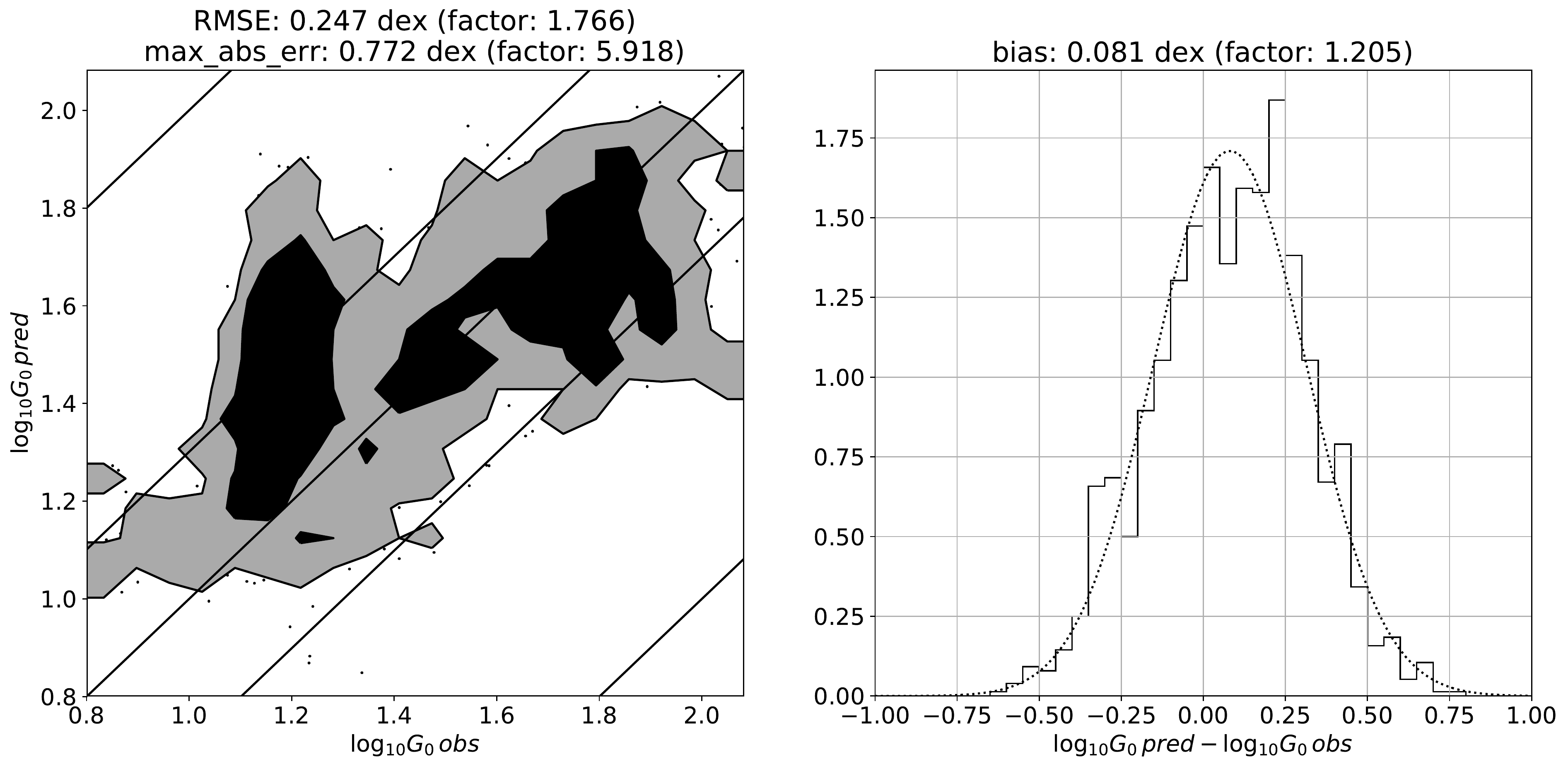}
    \caption{\textbf{Top:} Spatial distribution of the observed and
      predicted far-UV illumination (\textbf{left and middle panels}), and
      of the ratio of the predicted illumination over the observed one
      (\textbf{right panel}).
      \textbf{Bottom:} Joint histogram of the predicted illumination
      as a function of the observed one (\textbf{left panel}), and histogram of the
      ratio of the predicted illumination over the observed one on a
      logarithmic scale.
      These results are computed on the Horsehead pillar, i.e., the test
      set.}
    \label{fig:Go}
  \end{figure*}}

\newcommand{\TabLines}{%
  \begin{table}
    \caption{Spectral properties of the observed lines.}
    \label{tab:lines}
    \centering %
    \resizebox{\linewidth}{!}{%
      \begin{tabular}{lllrrr}
        \hline
        \hline
        Species                            & \multicolumn{2}{l}{Quantum Numbers}                                                                       & Frequency   & s\tablefootmark{a} & Noise\tablefootmark{b} \\
                                           & Simplified & Complete                                                                                     & MHz         & \# & [K] \\  
        \hline				   					                                                                             
        $\ensuremath{\mathrm{^{12}CO}}$    & \Jone{}    & $\ensuremath{\mathrm{J=1 \rightarrow 0}}$                                                    &  115271.202 &  1 & 0.11 \\
        $\ensuremath{\mathrm{^{13}CO}}$    & \Jone{}    & $\ensuremath{\mathrm{J=1 \rightarrow 0}}$                                                    &  110201.354 &  1 & 0.04 \\
        $\ensuremath{\mathrm{C^{18}O}}$    & \Jone{}    & $\ensuremath{\mathrm{J=1 \rightarrow 0}}$                                                    &  109782.173 &  1 & 0.04 \\
        $\ensuremath{\mathrm{C^{17}O}}$    & \Jone{}    & $\ensuremath{\mathrm{J=1 \rightarrow 0}}$                                                    &  112358.982 &  1 & 0.06 \\
        $\ensuremath{\mathrm{H_2CO}}$      & \Jone{}    & $1_{0,1} \rightarrow 0_{0,0}$                                                                &   72837.948 &  3 & 0.11 \\
        $\ensuremath{\mathrm{HCO^+}}$      & \Jone{}    & $\ensuremath{\mathrm{J=1 \rightarrow 0}}$                                                    &   89188.525 &  3 & 0.05 \\
        $\ensuremath{\mathrm{HC^{13}O^+}}$ & \Jone{}    & $\ensuremath{\mathrm{J=1 \rightarrow 0}}$                                                    &   86754.288 &  3 & 0.06 \\
        $\ensuremath{\mathrm{HCN}}$        & \Jone{}    & $\ensuremath{\mathrm{J=1 \rightarrow 0,\, F=2 \rightarrow 1}}$                               &   88631.848 &  3 & 0.06 \\
        $\ensuremath{\mathrm{HNC}}$        & \Jone{}    & $\ensuremath{\mathrm{J=1 \rightarrow 0}}$                                                    &   90663.568 &  3 & 0.06 \\
        $\ensuremath{\mathrm{^{12}CN}}$    & \Jone{}    & $\ensuremath{\mathrm{N=1 \rightarrow 0,\, J=3/2 \rightarrow 1/2,\, F=5/2 \rightarrow 3/2 }}$ &  113490.970 &  1 & 0.07 \\
        $\ensuremath{\mathrm{^{12}CS}}$    & \Jtwo{}    & $\ensuremath{\mathrm{J=2 \rightarrow 1}}$                                                    &   97980.953 &  1 & 0.04 \\
        $\ensuremath{\mathrm{^{32}SO}}$    & (3--2)     & $\ensuremath{\mathrm{J=3 \rightarrow 2,\, N=2 \rightarrow 1}}$                               &   99299.870 &  1 & 0.04 \\
        $\ensuremath{\mathrm{CCH}}$        & \Jone{}    & $\ensuremath{\mathrm{N=1 \rightarrow 0,\, J=3/2 \rightarrow 1/2,\, F=2 \rightarrow 1}}$      &   87316.898 &  2 & 0.05 \\
        c-$\ensuremath{\mathrm{C_3H_2}}$   & \Jtwo{}    & $2_{1,2} \rightarrow 1_{0,1}$                                                                &   85338.890 &  2 & 0.04 \\
        $\ensuremath{\mathrm{N_2H^+}}$     & \Jone{}    & $\ensuremath{\mathrm{J=1 \rightarrow 0,\, F1=2 \rightarrow 1,\, F=3 \rightarrow 2 }}$        &   93173.764 &  3 & 0.05 \\
        $\ensuremath{\mathrm{CH_3OH}}$     & \Jtwo{}    & $\ensuremath{\mathrm{J=2 \rightarrow 1,\, K=0 \rightarrow 0,\, (A+)}}$                       &   96741.375 &  1 & 0.04 \\
        $\ensuremath{\mathrm{SiO}}$        & \Jtwo{}    & $\ensuremath{\mathrm{J=2 \rightarrow 1}}$                                                    &   86846.985 &  1 & 0.06 \\
        $\ensuremath{\mathrm{H^+}}$        & $40\alpha$ & $40\alpha$ recombination line                                                                &   99022.953 &  1 & 0.02 \\
        \hline
      \end{tabular}}
    \tablefoot{%
      \tablefoottext{a}{Number of the IRAM-30m tuning setup the line
        was observed with (see Sect.~\ref{sec:ORION-B} for details).} %
      \tablefoottext{b}{Typical noise level in channels of 0.5\kms{} measured on
        the cubes that were smoothed at an angular resolution of $40''$.} %
    }
  \end{table}}

\newcommand{\TabFeatures}{%
  \begin{table*}
    \caption{Observed spectral lines and dust-traced properties.}
    \label{tab:features}
    \centering %
    \resizebox{\linewidth}{!}{%
      \begin{tabular}{ll|rrrrrrr|rrrr}
        \hline
        \hline
        Species                            & Transitions & \multicolumn{7}{c}{Peak temperature} & \multicolumn{4}{c}{Integrated intensity} \\
        \hline
                                           &             & Max. & Min.  & Max/Min & FoV\tablefootmark{a} & Mean & RMS\tablefootmark{b} & RMS/Mean & FoV\tablefootmark{a} & Mean    & RMS\tablefootmark{b} & RMS/Mean \\
                                           &             & K    & K     & ---     & \%                   & K    & K                    & ---      & \%                   & \Kkms{} & \Kkms{}              & ---      \\  
        \hline				   					                                                                                   
        $\ensuremath{\mathrm{^{12}CO}}$    & \Jone{}     & 61.7 & 0.141 &  437 & 87 & 13.94 & 10.27 & 0.74 & 92 & 51.44 & 45.14 & 0.88 \\
        $\ensuremath{\mathrm{^{13}CO}}$    & \Jone{}     & 34.3 & 0.050 &  686 & 75 &  3.36 &  3.70 & 1.10 & 79 &  7.72 &  9.14 & 1.18 \\
        $\ensuremath{\mathrm{C^{18}O}}$    & \Jone{}     &  6.6 & 0.032 &  206 & 33 &  0.35 &  0.51 & 1.44 & 39 &  0.52 &  0.92 & 1.77 \\
        $\ensuremath{\mathrm{C^{17}O}}$    & \Jone{}     &  1.3 & 0.042 &   31 &  4 &  0.19 &  0.08 & 0.43 & 10 &  0.17 &  0.29 & 1.70 \\
        $\ensuremath{\mathrm{H_2CO}}$      & \Jone{}     &  6.2 & 0.096 &   64 &  5 &  0.33 &  0.18 & 0.55 &  9 &  0.89 &  0.82 & 0.92 \\
        $\ensuremath{\mathrm{HCO^+}}$      & \Jone{}     &  8.8 & 0.051 &  171 & 47 &  0.47 &  0.58 & 1.24 & 68 &  1.67 &  1.80 & 1.07 \\
        $\ensuremath{\mathrm{HC^{13}O^+}}$ & \Jone{}     &  2.1 & 0.053 &   39 &  2 &  0.18 &  0.08 & 0.46 &  3 &  0.38 &  0.30 & 0.80 \\
        $\ensuremath{\mathrm{HCN}}$        & \Jone{}     & 12.4 & 0.045 &  273 & 28 &  0.32 &  0.40 & 1.24 & 43 &  1.74 &  2.33 & 1.34 \\
        $\ensuremath{\mathrm{HNC}}$        & \Jone{}     &  6.3 & 0.051 &  121 & 14 &  0.25 &  0.28 & 1.16 & 22 &  0.88 &  0.98 & 1.11 \\
        $\ensuremath{\mathrm{^{12}CN}}$    & \Jone{}     &  6.9 & 0.011 &  593 & 12 &  0.28 &  0.26 & 0.94 & 22 &  0.59 &  1.37 & 2.32 \\
        $\ensuremath{\mathrm{^{12}CS}}$    & \Jtwo{}     & 15.3 & 0.025 &  605 & 28 &  0.24 &  0.47 & 1.93 & 35 &  0.41 &  1.04 & 2.52 \\
        $\ensuremath{\mathrm{^{32}SO}}$    & (3--2)      &  6.8 & 0.025 &  264 & 18 &  0.18 &  0.25 & 1.38 & 24 &  0.24 &  0.51 & 2.13 \\
        $\ensuremath{\mathrm{CCH}}$        & \Jone{}     &  6.0 & 0.036 &  165 & 14 &  0.18 &  0.17 & 0.92 & 29 &  0.31 &  0.53 & 1.69 \\
        c-$\ensuremath{\mathrm{C_3H_2}}$   & \Jtwo{}     &  1.0 & 0.031 &   34 &  4 &  0.12 &  0.05 & 0.43 & 10 &  0.34 &  0.28 & 0.81 \\
        $\ensuremath{\mathrm{N_2H^+}}$     & \Jone{}     &  5.1 & 0.038 &  131 &  2 &  0.16 &  0.10 & 0.64 &  4 &  0.50 &  0.86 & 1.71 \\
        $\ensuremath{\mathrm{CH_3OH}}$     & \Jtwo{}     &  2.4 & 0.022 &  108 &  3 &  0.11 &  0.06 & 0.51 & 13 &  0.12 &  0.29 & 2.37 \\
        $\ensuremath{\mathrm{SiO}}$        & \Jtwo{}     &  1.0 & 0.053 &   18 &  0 &  0.17 &  0.05 & 0.31 &  1 &  0.32 &  0.24 & 0.76 \\
        $\ensuremath{\mathrm{H^+}}$        & $40\alpha$  &  0.4 & 0.002 &  154 &  1 &  0.06 &  0.02 & 0.33 &  1 &  0.06 &  0.24 & 3.83 \\
        \hline
      \end{tabular}}
    \resizebox{\linewidth}{!}{%
      \begin{tabular}{l|ccrrccc}
        \hline
        Dust-traced properties     & Max. & Min. & Max/Min & FoV\tablefootmark{a} & Mean & RMS & RMS/Mean \\
        \hline				         
        $N(\Ht)$ & \dix{2.5}{23}\pscm{} & \dix{6.2}{20}\pscm{} &     396 & 100\% & \dix{4.0}{21}\pscm{} & \dix{4.9}{21}\pscm{} &  1.22\\
        $G_0$    & \dix{4.5}{ 4}\,ISRF  & \dix{1.5}{ 0}\,ISRF  &   29231 & 100\% & \dix{7.4}{ 1}\,ISRF  & \dix{4.5}{ 2}\,ISRF  &  6.11\\
        \hline
      \end{tabular}}
    \tablefoot{%
      \tablefoottext{a}{Percentage of the Field of View above $3\sigma$.} %
      \tablefoottext{b}{Standard deviation of the data (signal plus noise).} %
    }
  \end{table*}}

\newcommand{\TabComparison}{%
  \begin{table*}
    \caption{Statistical comparison of the performances of three regression
      methods on the test set (the Horsehead pillar).}
    \label{tab:comparison}
    \centering %
    \begin{tabular}{lccccc}
      \hline
      \hline
      Method  & Hyper-parameters & Max. error\tablefootmark{a} & Mean error\tablefootmark{a} & RMSE & MSE \\
              &                  & dex                         & dex                         & dex  & dex \\ 
      \hline
      Linear regression               & 0 & 0.37 & {${0.026 \pm 0.003}$} & 0.14 & {${0.0182 \pm 0.0007}$} \\
      Linear regression on asinh($I$) & 1 & 0.33 & {${0.075 \pm 0.002}$} & 0.11 & {${0.0070 \pm 0.0003}$} \\
      Random forest                   & 3 & 0.26 & {${0.040 \pm 0.002}$} & 0.09 & {${0.0060 \pm 0.0002}$} \\
      \hline
      Method  & Hyper-parameters & Max. error\tablefootmark{a} & Mean error\tablefootmark{a} & RMSE           & MSE \\
              &                  & $10^\emr{dex}$              & $10^\emr{dex}$              & $10^\emr{dex}$ & $10^\emr{dex}$ \\ 
      \hline
      Linear regression               & 0 & 2.34 & 1.062 & 1.38 & 1.043 \\
      Linear regression on asinh($I$) & 1 & 2.14 & 1.190 & 1.29 & 1.016 \\
      Random forest                   & 3 & 1.82 & 1.096 & 1.23 & 1.014 \\
      \hline
    \end{tabular}
    \tablefoot{%
      \tablefoottext{a}{The maximum and mean errors are absolute errors on
        $\log \NHt$.} %
    }
  \end{table*}}

\newcommand{\TabComparisonBis}{%
  \begin{table}
      \referee{
    \caption{Statistical comparison of the performances with two simpler
      regression methods on the test set (the Horsehead pillar).}
    \label{tab:comparison:bis}
    \resizebox{\linewidth}{!}{%
    \begin{tabular}{lccc}
      \hline
      \hline
      Method & Max. error\tablefootmark{a} & Mean error\tablefootmark{a} & RMSE \\
             & dex                         & dex                         & dex  \\ 
      \hline
      \Xco{}-factor                     & 3.31 & $-0.370 \pm 0.011$ & 0.57 \\      
      Random forest on CO isotopologues & 0.32 & $+0.020 \pm 0.003$ & 0.11 \\
      Random forest on all lines        & 0.26 & $+0.040 \pm 0.002$ & 0.09 \\
      \hline
      Method & Max. error\tablefootmark{a} & Mean error\tablefootmark{a} & RMSE           \\
             & $10^\emr{dex}$              & $10^\emr{dex}$              & $10^\emr{dex}$ \\ 
      \hline
      \Xco{}-factor                     & $\sim 2 \times 10^{3}$ & 0.426 & 3.71 \\
      Random forest on CO isotopologues &                   2.08 & 1.046 & 1.29 \\
      Random forest on all lines        &                   1.82 & 1.096 & 1.23 \\
      \hline
    \end{tabular}}
    \tablefoot{%
      \tablefoottext{a}{The maximum and mean errors are absolute errors on
        $\log \NHt$.} %
    }}
  \end{table}}

\begin{document} 

\title{Quantitative inference of the \Ht\ column densities\\
  from 3\,mm molecular emission: A case study towards Orion B}

\titlerunning{Quantitative inference of the \Ht\ column densities from
  3\,mm molecular emission}

\author{Pierre Gratier \inst{\ref{LAB}} %
  \and Jérôme Pety\inst{\ref{IRAM},\ref{LERMA}} %
  \and Emeric Bron\inst{\ref{Meudon}} %
  \and Antoine Roueff\inst{\ref{Marseille}} %
  \and Jan H. Orkisz\inst{\ref{Chalmers}} %
  \and Maryvonne Gerin\inst{\ref{LERMA}} %
  \and Victor de Souza Magalhaes\inst{\ref{IRAM}} %
  \and Mathilde Gaudel\inst{\ref{LERMA}} %
  \and Maxime Vono\inst{\ref{IRIT}} %
  \and S\'ebastien Bardeau\inst{\ref{IRAM}} %
  \and Jocelyn Chanussot\inst{\ref{Grenoble}} %
  \and Pierre Chainais\inst{\ref{Lille}} %
  \and Javier R. Goicoechea\inst{\ref{CSIC}} %
  \and Viviana V. Guzm\'an\inst{\ref{Catholica}} %
  \and Annie Hughes\inst{\ref{IRAP}} %
  \and Jouni Kainulainen\inst{\ref{Chalmers}} %
  \and David Languignon\inst{\ref{Meudon}} %
  \and Jacques Le Bourlot\inst{\ref{Meudon}} %
  \and Franck Le Petit\inst{\ref{Meudon}} %
  \and François Levrier\inst{\ref{LPENS}} %
  \and Harvey Liszt\inst{\ref{NRAO}} %
  \and Nicolas Peretto\inst{\ref{UC}} %
  \and Evelyne Roueff\inst{\ref{Meudon}} %
  \and Albrecht Sievers\inst{\ref{IRAM}}}

\institute{%
  Laboratoire d'Astrophysique de Bordeaux, Univ. Bordeaux, CNRS, B18N,
  Allee Geoffroy Saint-Hilaire,33615 Pessac, France.\\
  \email{pierre.gratier@u-bordeaux.fr}\label{LAB} %
  \and IRAM, 300 rue de la Piscine, 38406 Saint Martin d'H\`eres,
  France. \label{IRAM} %
  \and LERMA, Observatoire de Paris, PSL Research University, CNRS,
  Sorbonne Universit\'es, 75014 Paris, France. \label{LERMA} %
  \and LERMA, Observatoire de Paris, PSL Research University, CNRS,
  Sorbonne Universit\'es, 92190 Meudon, France. \label{Meudon} %
  \and Aix Marseille Univ, CNRS, Centrale Marseille, Institut Fresnel,
  Marseille, France \label{Marseille} %
  \and Chalmers University of Technology, Department of Space, Earth and
  Environment, 412 93 Gothenburg, Sweden
  \label{Chalmers} %
  \and University of Toulouse, IRIT/INP-ENSEEIHT, CNRS, 2 rue Charles
  Camichel, BP 7122, 31071 Toulouse cedex 7, France
  \label{IRIT} %
  \and Univ. Grenoble Alpes, Inria, CNRS, Grenoble INP, GIPSA-Lab,
  Grenoble, 38000, France \label{Grenoble} %
  \and Univ. Lille, CNRS, Centrale Lille, UMR 9189 - CRIStAL, 59651
  Villeneuve d’Ascq, France \label{Lille} %
  \and Instituto de Física Fundamental (CSIC). Calle Serrano 121, 28006,
  Madrid, Spain \label{CSIC} %
  \and Instituto de Astrofísica, Pontificia Universidad Católica de Chile,
  Av. Vicuña Mackenna 4860, 7820436 Macul, Santiago,
  Chile \label{Catholica} %
  \and Institut de Recherche en Astrophysique et Planétologie (IRAP),
  Université Paul Sabatier, Toulouse cedex 4, France
  \label{IRAP} %
  \and Laboratoire de Physique de l’Ecole normale supérieure, ENS,
  Université PSL, CNRS, Sorbonne Université, Université Paris- Diderot,
  Sorbonne Paris Cité, Paris, France \label{LPENS} %
  \and National Radio Astronomy Observatory, 520 Edgemont Road,
  Charlottesville, VA, 22903, USA. \label{NRAO} %
  \and School of Physics and Astronomy, Cardiff University, Queen's
  buildings, Cardiff CF24 3AA, UK. \label{UC} %
} %

\date{}

\abstract
{Molecular hydrogen being unobservable in cold molecular clouds, the column
  density measurements of molecular gas currently rely either on dust
  emission observation in the far-IR, which requires space telescopes, or
  on star counting, which is limited in angular resolution by the stellar
  density. (Sub-)millimeter observations of numerous trace molecules are
  effective from ground based telescopes, but the relationships between the
  emission of one molecular line and the \Ht{} column density is non-linear
  and sensitive to excitation conditions, optical depths, abundance
  variations due to the underlying physico-chemistry.}
{We aim to use multi-molecule line emission to infer the \Ht{} molecular
  column density from radio observations.}
{We propose a data-driven approach to determine the \Ht{} gas column
  densities from radio molecular line observations. We use supervised
  machine learning methods (Random Forests) on wide-field hyperspectral
  IRAM-30m observations of the Orion B molecular cloud to train a predictor
  of the \Ht{} column density, using a limited set of molecular lines
  between 72 and 116\GHz{} as input, and the Herschel-based dust-derived
  column densities as ``ground truth'' output.}
{For conditions similar to the Orion B molecular cloud, we obtain
  predictions of the \Ht{} column density within a typical factor of 1.2
  from the Herschel-based column density estimates. A global analysis of
  the contributions of the different lines to the predictions show that the
  most important lines are \thCO{}\Jone{}, \twCO{}\Jone{}, \CeiO{}\Jone,
  and \HCOp{}\Jone{}. A detailed analysis distinguishing between diffuse,
  translucent, filamentary, and dense core conditions show that the
  importance of these four lines depends on the regime, and that it is
  recommended to add the \NNHp{}\Jone{} and \methanol{}($2_0$--$1_0$) lines
  for the prediction of the \Ht{} column density in dense core conditions.}
{This article opens a promising avenue to directly infer important physical
  parameters from the molecular line emission in the millimeter domain. The
  next step will be to try to infer several parameters simultaneously
  (e.g., the column density and far-UV illumination field) to further test
  the method.}


\maketitle
%

\section{Introduction}

Atoms and molecules have long been thought to be versatile tracers of cold
neutral medium in the universe, from high-redshift galaxies to star forming
regions and proto-planetary disks, because their internal degrees of
freedom bear the signature of the physical conditions in their
environments. Atoms and molecules are affected by many processes:
Photo-ionization and photo-dissociation by far-UV photons, excitation by
collisions with neutrals and electrons, radiative pumping of excited levels
by far-UV or IR photons, gas phase chemical reactions, condensation on
grains, solid state reactions in the formed ice, (non)-thermal desorption,
etc. Moreover, this chemical activity is tightly coupled to the gas
dynamics. Chemistry affects the gas motions because 1) the ionization state
controls the coupling to the magnetic field, and 2) the line radiation from
molecules (mostly rotational lines) and atoms (fine structure lines in the
far-IR) is the main cooling agent of the neutral gas over a broad range of
astrophysical environments, controlling the equation of state and therefore
affecting the dynamics. Conversely, the gas dynamics affects the chemistry
because it produces steep and time-variable density and velocity gradients,
which change the rates of molecule formation and destruction. Numerical
models of interstellar clouds face the difficulty to combine sophisticated
chemical codes (addressing the molecule formation and destruction
processes) with the turbulent gas dynamics. This is a tremendous challenge
given the non-linearity of fluid dynamics, the stiffness of the chemical
reactions and the wide range of time scales involved
\citep{Valdivia.2017,Clark.2019}. It is therefore important to acquire
self-consistent data sets that can be used as templates for this
theoretical work, and at the same time to document the diagnostic
capabilities of molecular lines accurately.

The recent development of spectrometers in the (sub)-millimeter domain
(e.g., IRAM-30m/EMIR, NOEMA, ALMA) opens new avenues to fulfill this
goal. First, wide band spectrometers now allow us to simultaneously observe
tens of lines instead of a single one along each line of sight. The first
studies using these capabilities were sensitive ($\sim3-8\,$mK) unbiased
spectral surveys at 1, 2 and 3\,mm targeting a few specific lines of sight
(e.g., Horsehead WHISPER: \citealt{Pety.2012}, TMC1:
\citealt{Gratier.2016}, ASAI: \citealt{Lefloch.2018}). These show the power
of multi-line studies to constrain the physics and the chemistry of
molecular clouds. Second, the increase in sensitivity now makes it possible
to detect these lines over large areas (several square degrees), paving the
way for an era of quasi systematic hyperspectral imaging in the millimeter
domain. The ORION-B project (Outstanding Radio-Imaging of OrioN-B, co-PIs:
J.~Pety and M.~Gerin) is a Large Project of the IRAM 30m telescope that
aims to improve the understanding of physical and chemical processes of the
interstellar medium by mapping a large fraction of the Orion B molecular
cloud (5 square degrees) with a typical resolution of
$27'' (\sim 50 \unit{mpc}$ at 400\,pc, the typical distance of Orion) and
200\,kHz (or 0.6\kms{}) over the full 3\,mm atmospheric band.

In a first study, \citet{Pety.2017} showed how tracers of different optical
depths like the CO isotopologues allow one to fully trace the molecular
medium, from the diffuse envelope to the dense cores, while various
chemical tracers can be used to reveal different environments. However,
extracting the information contained in these multi-line observations
requires powerful statistical tools. A clustering algorithm applied to the
intensities of selected molecular lines revealed spatially continuous
regions with similar molecular emission properties, corresponding to
different regimes of volume density or far-UV
illumination~\citep{Bron.2018}. In addition, a global Principal Component
Analysis of the line integrated brightnesses revealed that some
combinations of lines are sensitive to the column density, the volume
density, and the UV field~\citep{Gratier.2017}. In this article, we go one
step further by checking whether/how it is possible to build a quantitative
estimate of the \Ht{} column density, based on the molecular emission, and
valid over a large range of conditions. Indeed, this is a prerequisite to
class the interstellar medium into its different phases such as diffuse,
translucent and dense regimes \citep{Pety.2017}, and to identify its
underlying structure, in particular its filamentary nature
\citep{Andre.2010,Orkisz.2019}. Such a method could also be used to
estimate the mass of the different (velocity separated) components of a
giant molecular cloud, for instance, the linear mass of the filaments
relative to their more diffuse environment. The \Ht{} column density is
also required to compute molecular abundances from observed molecular
column densities and compare these with the outputs of astrochemical codes.

To do this, we focus on supervised learning methods. Supervised learning is
a general set of machine learning methods used to learn how to assign a
class or infer the value of a given quantity from a set of measured
observables. These methods need a training set for which we know both the
measured features and the searched class or value. Here, we will use 1) the
emission of selected spectral lines over a fraction of the observed field
of view as input observables, and 2) the dust-traced column density as a
proxy of the gas column density. Indeed, multi-wavelength observations of
dust thermal emission in the submillimeter range and the subsequent fit of
the spectral energy distribution is one of the most successful methods to
derive total column density maps of the interstellar medium. Between 2009
and 2015, the Herschel Observatory instruments PACS (70, 100, 160\mum) and
SPIRE (250, 350, 500\mum) mapped a fraction of the sky with an angular
resolution of $\sim40''$ or better.  In particular, large programs have
been dedicated to this task, for example, the HiGal survey mapping of the
inner Galactic plane ($68^{\circ} > l > -70^{\circ} $ and $ |b|<1^{\circ}$
\citealt{Molinari.2016}), or the Gould Belt survey
\citep{Andre.2010}. However, since the end of the Herschel mission, and
until its potential successor SPICA that could be launched in the 2030s,
only SOFIA/HAWK$+$ is currently able to measure the far-IR dust emission.
Ground-based or ballon-borne submillimeter telescopes will map the
interstellar medium with even higher angular resolution but the lack of
shorter wavelengths data will make the estimation of the dust temperature
highly uncertain. This means that the dust temperature will have to be
guessed when deriving the dust-traced column density. This method will
likely lead to systematic errors. Hence, devising an accurate method for
estimating the \Ht{} column density, which only relies on ground-based
(sub-)millimeter facilities is important.

This article is structured as follows. Section~\ref{sec:data} introduces
the data sets and Sect.~\ref{sec:aims} formalizes the problem.
Sections~\ref{sec:methods:principle} introduces the concepts and
methods. Section~\ref{sec:methods:application} discusses how we applied
them in practice. Section~\ref{sec:results:comparison} compares the
performances of different methods. Section~\ref{sec:results:contributions}
discusses which lines are the most important to infer the \Ht{} column
density. Section~\ref{sec:discussion} discusses first whether the column
density predictor can be used on noisier data or on data from sources more
distant from the Sun than Orion. It also discusses how the method can be
generalized to other physical parameters such as the far-UV illumination.
Section~\ref{sec:conclusions} concludes the article.

\section{Data}
\label{sec:data}

\TabLines{} %

\subsection{Molecular emission from IRAM-30m observations}
\label{sec:ORION-B}

The acquisition and reduction of the molecular dataset used in this study
is presented in detail in \citet{Pety.2017}, but the field of view has been
extended to the North and East by $\sim 60\%$. In short, the data were
acquired at the IRAM-30m telescope by the ORION-B project in only three
frequency tunings: the first one from 92.0 to 99.8\GHz{} (LSB band), and
from 107.7 to 115.5\GHz{} (USB band); the second one from 84.5 to
92.3\GHz{} (LSB band), and from 100.2 to 108.0\GHz{} (USB band); and the
third one from 71.0 to 78.8\GHz{} (LSB band), and from 86.7 to 94.4\GHz{}
(USB band). The data were acquired from August 2013 to February 2015 for
the two first tunings and in August 2016 for the third tuning.

The selection of the studied lines was performed from the spectra averaged
over the observed field of view. Table~\ref{tab:lines} lists the 18
selected lines, and their associated tuning setup. A velocity interval of
80\kms{} was extracted around each line and the spectral axis was resampled
onto a common velocity grid. The systemic velocity of the source is set to
10.5\kms{} and the channel spacing is set to 0.5\kms{}, i.e., the highest
velocity resolution achieved for the \twCO{} $(1-0)$ line.  This implies
that the noise is more and more correlated from one channel to the next as
the rest frequency of the line decreases.

The studied field of view covers $0.9\degr \times 1.6\degr$ towards the
Orion B molecular cloud part that contains the Horsehead nebula, and the
H{\sc ii} regions NGC\,2023, NGC\,2024, IC\,434, and IC\,435. Compared to
\citet{Pety.2017,Gratier.2017,Orkisz.2017,Bron.2018}, it additionally
comprises the northern molecular edge that contains the hummingbird
filament studied in~\citet{Orkisz.2019}. All the cubes were gridded onto
the same spatial grid to ease the analysis.  The projection center is
located on the Horsehead at \radec{05}{40}{54.270}{-02}{28}{00.00}.  The
maps are rotated counter-clockwise by $14\degr$ around this position. The
angular resolution ranges from $22.5$ to $35.6''$. The
position-position-velocity cubes of each line were smoothed to a common
angular resolution of $40''$ to avoid resolution effects during the
analysis. This was done by convolution with a Gaussian kernel of width
$\theta_\emr{kernel} = \sqrt{40^2-\theta_\emr{beam}^2},$ where
$\theta_\emr{beam}$ is the telescope beam for each observed line in arcsec.
A pixel size of $20''$ was used to ensure Nyquist sampling and to avoid too
strong correlations between pixels. At a distance of
400\pc{}~\citep{Menten.2007,Zucker.2019,Zucker.2020}, the sampled linear
scales range from $\sim 80\unit{mpc}$ to $\sim 11\pc$.

From the position-position-velocity cubes, we computed maps of both the
peak temperature and the integrated intensity (moment 0)\footnote{The data
  products associated with this article are available at
  \url{https://www.iram.fr/~pety/ORION-B}.}. The peak temperature is just
the maximum of the spectrum intensity over the 80\kms{} velocity range. The
integrated intensity is computed over a velocity window that is decided as
follows. Starting from the peak intensity velocity, all adjacent channels
whose intensity is larger than zero are added to the velocity
window~\citep{Pety.1999}. This process is iterated 5 times, each time
starting from the next intensity maximum. Up to five velocity components
can hence be present on each line of sight.

While the line integrated intensity can be used as a proxy for the column
density of the species along the line of sight, at least for some column
density interval, we also include the line peak temperature to take into
account the possible effect of the excitation temperature on the
relationship between the column density and the integrated intensity. As
discussed by~\citet{Pety.2017}, we expect variations of the gas temperature
(and thus of the excitation temperature) across the targeted field of view,
because it is exposed to the intense far-UV illumination produced by young
OB stars in the more or less embedded \Hii{} regions.  For optically thick
lines, the line peak temperature can be viewed as a proxy for the line
excitation temperature where the brightness temperature approaches the
excitation temperature. For a (faint) optically thin line, the map of peak
temperature is proportional to the excitation temperature times the column
density per velocity bin. Using both the line area and the peak temperature
partly lifts the degeneracy between excitation and amount of gas along the
line of sight~\citep[see, e.g., the companion article][]{Roueff.2020}.

\subsection{$N(\Ht)$ column density derived from dust thermal emission
  observed with Herschel}
\label{sec:GouldBeltSurvey}

\TabFeatures{} %
\FigPredicted{} %

In order to get an independent measurement of the column density for the
Orion B cloud, we use the dust continuum observations from the
\textit{Herschel} Gould Belt Survey~\citep{Andre.2010,Schneider.2013} and
from the \textit{Planck} satellite~\citep{Tauber.2011}. The fit of the
spectral energy distribution by~\citet{Lombardi.2014} gives us access to
the spatial distributions of the dust opacity at $850\mum$ and of the
effective dust temperature.  As in \citet{Pety.2017}, we converted
$\tau_{850\mum}$ to visual extinctions using
$A_\mathrm{V} = 2.7\times 10^4 \, \tau_{850} \, \mathrm{mag}$, and we use
$N_\mathrm{H}/A_\mathrm{V} = 1.8\times10^{21}\pscm/\mathrm{mag}$ as
conversion factor between visual extinction and hydrogen column density:
$N_\mathrm{H} = N_\mathrm{HI}+2N_\mathrm{H_2}$. Over the observed field of
view, the column density of atomic hydrogen accounts for less than 1 visual
magnitude of extinction~\citep{Pety.2017}.  We thus choose to neglect this
contribution in this study. Figure~\ref{fig:data:dust} shows the spatial
distribution of this dust-traced column density.

We do not claim that the dust-traced column density used here is a perfect
measure of the underlying $N(\Ht)$ column density. We just wish to check
whether the molecular emission alone is able to predict this dust-traced
column density. If this method is efficient, the next step will be to
anchor it on additional sources of information (see
Sect.~\ref{sec:complete}).

\FigDataArea{} %
\FigDataPeak{} %

\subsection{Information content}

Figures~\ref{fig:data:area} and~\ref{fig:data:peak} show the spatial
distribution of the input variables (integrated intensities and peak
temperatures of the lines) and target variable (the dust-traced H$_2$
column density).  Table~\ref{tab:features} lists, among others, the minimum
and maximum values of the peak temperature maps for the 18 selected lines,
as well as the derived dynamic range computed as the ratio of the maximum
over the minimum value. The dynamic range spans values between $\sim 20$
and $\sim 700$. The dynamic range of the column density is $\sim 400$.

The input and targeted variables have different noise properties. Indeed,
the dust-traced \Ht{} column density is derived at high signal-to-noise
ratio over the full field of view. We can safely assume that the targeted
variable is noiseless even though it may be affected by systematic biases
in its derivation. On the contrary, large fractions of the field of view is
measured at low signal-to-noise ratio of the input variables. This is
particularly clear on the maps of the peak temperature, which emphasize the
noise pattern at signal-to-noise ratios lower than 3. The noise pattern
suggests that setups \#1 and \#2 were mainly observed through vertical
scanning while setups \#3 was only observed through horizontal scanning. It
also suggests non-negligible variations of the noise levels either because
of the weather (mostly summer vs winter weather but also degrading weather
during one observing session) or because of the large variation of the
telescope elevation between the beginning and the end of an observing
session.

\section{Astrophysical goal: Is it possible to accurately predict the \Ht{}
  column density based on molecular emission?}
\label{sec:aims}

Figure~10 of \citet{Pety.2017} shows the joint distributions of the dust
visual extinction (proportional to the dust-traced column density of matter
along the line of sight, \Ndust), and of the line integrated intensity
($W_l$) for a selection of the detected lines, $l$. This figure shows a
clear monotonic relationship with low scatter between \Ndust{} and $W_l$
for most of the lines. On one hand, there often exists an interval of
column densities for which the relationship between \Ndust{} and $W_l$ is
linear to a good approximation for one line, but the column density
interval depends on the line. On the other hand, these relationships are in
general non-linear. This is clear when looking at the relationship between
$I_{\twCO(1-0)}$ and \Ndust{}. The integrated intensity stays undetected at
low column density and it saturates at high column density. This property
is generic for any single molecular line, implying that setting up a
predictor of the column density from a single line will always fail in some
regime.

The facts that 1) the relationships between \Ndust{} and $W_l$ are
monotonic and 2) the interval of column densities over which
$\Ndust \propto W_l$ to first order depends on the line, open the
interesting possibility that a joint analysis of the molecular lines will
allow us to devise a predictor of the column density. Building on these
empirical facts, \citet{Gratier.2017} applied the simplest global analysis
of the existing correlations in a multi-dimensional dataset, the Principal
Component Analysis. Figure~10 of \citet{Gratier.2017} shows the joint
histogram of the first principal component and the dust-traced column
density. This histogram shows a tight correlation between these two
quantities with a Spearman’s rank correlation of 0.90 over more than 2
orders of magnitude in column density, from $10^{21}$ to
$10^{23}$\,H$_2\,$cm$^{-2}$. Moreover this correlation does not saturate
anymore at low or high column density. However, it is not exactly linear.

We will now assume that there exists a non-linear continuous function $F$
of the line intensities that predicts the dust-traced column density. Our
goal in this article is to quantitatively estimate an approximation (noted
$f$) of this function $F$. This estimation will be affected by
signal-to-noise issues, since the line intensity measurements are limited
by the sensitivity of the observations, i.e., our measurements are in
general done at intermediate signal-to-noise ratios. We can write our
estimation problem as
\begin{equation}
  \label{eq:problem}
  \Ndust = f(\mathbf{T_l,W_l}) + e,
\end{equation}
where $\mathbf{T_l,W_l}$ is the vector of the peak temperatures and
integrated intensities of lines $l$, and $e$ represents the sum of all
uncertainties. In this estimation problem, the measurements may also suffer
from systematic biases. For instance, the dust-traced column density may
underestimate the amount of matter along the line of sight when the
emission from warm dust hides the emission from colder
dust~\citep{Pagani.2015}. In the absence of a more quantitative knowledge
of such systematic biases, they cannot be separated from the physical
relationship in the estimation $f$. They will therefore be de facto
included in the function $f$ that we try to recover.

\referee{Getting an accurate estimate of the \Ht{} column density from
  molecular line emissions is a long standing quest in the study of stellar
  formation.  Methods followed the development of (sub-)millimeter
  telescopes, receivers, and spectrometers. They can be divided into two
  main categories. The first category relies on an empirical linear
  relationship between the \Ht{} column density and the integrated emission
  of the \Jone{} line of the most abundant tracer of molecular gas, namely
  \twCO. \citet{Bolatto.2013} explain that this method, known
  as the \Xco{}-factor method, relies
  on the fact that Giant Molecular Clouds seems on average close to the
  virial equilibrium. The statistical nature of the basis for this method
  implies that it mostly works at relatively large linear scale
  $(\ga 10-50\pc)$. Several studies \citep{Leroy.2011,Genzel.2012} have shown that the 
  \Xco{} factor depends on the metallicity of the inter-stellar medium to
  take into account varying fraction of CO dark gas, \ie{}, \Ht{} gas
  without enough dust to shield the destruction of CO from the surrounding
  far UV field.}

\referee{The second category of methods invert a radiative transfer model to
obtain the column density of the associated molecular species from the line
intensities or observed spectra. Chemical models estimating the abundances
relative to \Ht{} are then used to infer $N_{\Ht}$ from the molecular column
densities. This method is usually applied on the main CO isotopologues. In this
category, \citet{dickman.1978} and \citet{dickman.1986} derived the mean
\thCO{} abundance relative to \Ht. \citet{frerking.1982},
\citet{bachiller.1986} and \citet{cernicharo.1987} expanded to other CO
isotopologues, showing that the threshold for detecting \CeiO{} is higher than
that for \twCO{} or \thCO{}. \citet{goldsmith.2008} and \citet{pineda.2010}
used a modified version with a variable [CO]/[\Ht] abundance ratio deduced from
the Black and van Dishoeck PDR models \citep[see, e.g.,][]{visser.2009}. By
comparing \twCO{} and \thCO{} \Jone{} emission maps with dust extinction maps,
\citet{ripple.2013} showed that the relationship between the \thCO{} column
density and \Av{} is non-linear, indicating variations of the \thCO{}
abundance. \citet{barnes.2018} analyzed a recent large survey of the main CO
isotopologues to determine a $W_{\twCO{}(1-0)}$-dependent \Xco{} conversion
factor. Their analysis assumes that the excitation temperature is the same for
the \twCO{} and \thCO{} lines, as well as a constant [\thCO]/[\twCO] abundance.
These two assumptions are shown to be incorrect at least in the Orion B
molecular cloud by~\citet{Bron.2018} and~\citet{Roueff.2020}.}

\referee{In summary, the first category search for a direct connection
  between the line intensity and the \Ht{} column density, while the second
  category relies on the estimation of the species abundances. Beam filling
  factor may be an issue in the latter category if it changes the apparent
  abundance. This study belongs to the first category.}

\section{Principle: Regression in machine learning}
\label{sec:methods:principle}

In this section, we briefly define the different machine learning concepts
that we will use in this article. It's mostly an intuitive presentation
aimed at astronomers without machine learning knowledge. The main algorithm
we use in this paper is called Random Forest. Its concept was invented by
Leo Breiman. Detailed theory can be found in
\citet{Breiman.2001}\footnote{See also
  \url{https://www.stat.berkeley.edu/~breiman/papers.html}}. An
introduction can be found in \citet[][chapter~15]{Hastie.2001}.

\subsection{A supervised machine learning method called regression}

Trying to solve Eq.~\ref{eq:problem} for an approximation of $F$ is a
generic machine learning class of problems known as
\textit{regression}. This approximation of $F$, which we note $f$, is
called the \textit{regression model}.  The quantity to be predicted is
often called the \emph{dependent or target variable}. In our case, it will
be the dust-traced column density. The function variables (that is the
observables) are often called \emph{features}. In our case, these will be
the measured molecular integrated intensities and peak temperatures. Each
line of sight (image pixel) constitutes one measurement of
Eq.~\ref{eq:problem}. The regression consists in finding an estimate of $F$
based on the dataset. It is a \textit{supervised} method, meaning that it
needs to be trained on a dataset for which the solution of the problem is
known before applying the trained method to other datasets.

\subsection{Training set, test set, and quality of fit}

We use a standard supervised learning workflow by first dividing the full
dataset into a \textit{training set} comprising the majority of the
observed data on which the best internal parameters of the model are
fitted, and a \textit{test set}, never seen by the fitting/training
algorithm. The \textit{quality of the fit} is checked by computing the mean
square error (MSE) between the value predicted by the model and the
observed quantity. There are two kinds of MSE. On one hand, the training
MSE is used to fit the model on the training set. On the other hand, the
test MSE is computed to assess how the model behaves on data it has not
been trained on.

While the regression fit minimizes the training MSE, the final goal is to
predict correct values for samples outside of the training set, i.e., to
minimize the \textit{test} MSE. Small test and training MSE values are
expected when the model function is well estimated. This indicates that the
model can predict the dust-traced column density based on the values of the
molecular intensities on data that have never been used during the fitting
procedure. However, finding a test MSE much larger than the training MSE is
the symptom of a problem called overfitting. In this case, the estimated
model function learned not only the searched underlying physics, but also
the specifics of the dataset measurements, in particular the noise
properties. This can occur, for instance, when the model complexity (i.e.,
the number of unknown parameters) is too large compared to the information
content of the data.

\subsection{Variance and bias of the model estimation}
\label{sec:variance-bias}

The accuracy of the fitted model, i.e., achieving the minimum test MSE,
always results from a trade-off between the variance and the bias of the
estimated model function (see Sect. 7.3 of \citealt{Hastie.2001}, and
Sect. 3.2 of \citealt{Bishop.2006}).

The \textit{variance} refers to the amount of variation that would affect
our estimate of $F$ if the training dataset was different. There are
different causes of variability in the estimation of $F$. First, the
training set may cover only part of the relevant physical conditions. In
our case, we could have failed to sample well enough one of the different
column density regimes: diffuse or translucent gas, dense cores,
etc. Second, the condition of the observations, e.g., the signal-to-noise
ratio, can be different from one training set to another.

The \textit{bias} refers to the fact that there can be a mismatch between
the actual complexity of $F$ and the complexity implied by a choice of
function family. No matter the amount of data you have in your training
sample, trying to fit a non-linear function by an hyper-plane will result
in a bias.

Minimizing the test MSE requires selecting a supervised learning method
that will find the best trade-off between the variance and the
bias. Achieving a low bias but a high variance is easy: any model that goes
through all the points of the training data set will do this; it is then
completely dependent on the specific noise of the training set and is thus
overfitted. Conversely, achieving a low variance but a high bias is just as
easy: using as model the mean of the training data set will do this. The
actual challenge is to find \emph{the optimal trade-off} between the
variance and the bias.  As a function of model complexity, variance is
minimized at zero complexity, and bias is minimized at infinite complexity
(or large enough to be an interpolation).

\subsection{Decreasing the regressor variance through bagging}

\textit{Bagging} is one of the two standard algorithms to decrease the
variance of a regression method. The other one, called boosting, will not
be discussed in this article. Bagging is an abbreviation for bootstrap
aggregating. It is thus related to the bootstrap method that is well used
in astrophysics to estimate random uncertainties in properties inferred
from a given dataset.

The bootstrap method randomly creates a large number of sub-samples of the
input dataset. In this drawing, replacement is authorized, i.e., the same
data point can be chosen multiple times. A regression is then done on each
sub-sample and the predicted values are computed for each point of each
sub-sample. The aggregation part computes the average of all the
predictions. This average becomes the predicted model. It is intuitive that
the reduction of the variance comes from the averaging process. This
nevertheless assumes that the errors of individual models are uncorrelated.
The bias is conserved in this algorithm. In particular, a low bias method
will give a low bias bagged method.

A practical advantage of bagging is that it is easily parallelizable,
implying short computation times.

\subsection{Regression trees}

\textit{Regression trees} are a regression method that uses a recursive set
of binary splitting on the values of the input variables to estimate the
target variable. The choice of the split point is made to minimize a cost
function. In simpler words, the regression tree asks a series of binary
questions to the data, each question narrowing the possible prediction
until the method gets confident enough that this prediction is the right
one.

In more detail, at one dimension, the training set is made of couples
$(x_i,y_i)$ that are linked by a to-be-estimated function $f$ and the
residuals $e_i$
\begin{equation}
  y_i = f(x_i) + e_i.
\end{equation}
The function $f$ is an approximation of the data in the form of a step
function with steps of variable heights and lengths. To get it, the method
explores all the potential ways to split the values of the $x$ axis into
two categories. For each potential split, it computes the MSE of the two
resulting classes. It then chooses the split value that minimizes the
average of the two MSE weighted by the number of elements in each
class. The outcome of this process is a threshold value of $x$ that splits
the dataset into two classes called tree branches. The decision point is a
node of the tree. The process is then iterated in each branch leading to
new nodes and branches. The process is stopped when the maximum depth of
the tree is reached or a branch contains less than a given number of data
points. The predicted value is then computed as the mean of the $y$ values
for these points. The first decision point is called the root node. By
construction, it's the one that reduces the most the final MSE. It is
called the strongest predictor. Generalization to a multi-dimensional
function is straightforward. All the dimensions are split one after the
other and the first decision is made along the dimension that minimizes the
weighted average of the MSE of the two resulting branches.

Regression trees have several advantages. They make no assumption, either
on the functional form of the learned relationship, or on the shape of the
underlying probability density of the dataset.  They thus belong to the
class of non-parametric methods. Regression trees are non-linear regressors
by nature. They are easy to understand and thus to interpret. In
particular, the relative importance of the features is easy to
extract. Finally, they require neither normalization nor centering of the
data.  This last point is a key advantage for us as \citet{Gratier.2017}
showed that normalization is difficult when the signal-to-noise ratio is
low for a number of features.

Regression trees nevertheless have several drawbacks. A large tree depth
brings high flexibility and thus ensure a low bias, but it also makes these
trees prone to overfitting. They are unstable, meaning that a small
variation in the training set can lead to a completely different regression
tree. This implies that their variance is large and that there is no
guarantee that the outcome is the globally optimal regression
tree. However, most of the drawbacks can be overcome by using
\textit{random forests}.

\subsection{Random forests}

\textit{Random forests} overcome the instability and high variance of a
regression tree by averaging the predictions from many such trees that
include two sources of randomness. The input data set is first
bootstrapped.  However, only introducing this kind of randomness would
produce highly correlated trees in case the data set contains several
strong predictors (i.e., a split decision along a given dimension that
largely decreases the MSE). Indeed, these strong predictors will be
consistently chosen at the top levels of the tree. So random forests
introduce a second source of randomness at each decision point: Instead of
minimizing the MSE along all the dimensions, it minimizes it along a random
subset of the dimensions.  Hence a random forest is a regression method
made of bagged regression trees (first kind of randomness) that split on
random subsets of features on each split (second kind of randomness). This
not only reduces the variance but it will also speed up the computations
because the introduction of randomness is done on both a subset of the data
and a subset of the dimensions.

\FigHorsehead{} %

\subsection{Model complexity vs interpretability}

We later show that linear regressors fall short in capturing all of the
non-linearities of the relationships between molecular emission and
$N(\Ht)$ column density. This calls for a more flexible method. Neural
networks are a well-known example of a method that performs well in complex
machine learning tasks~\citep[see e.g.,][]{Boucaud.2020} but the
interpretation of their output is usually difficult.  Having an
interpretable result is an important criterion for us as we aim at
understanding the properties of the interstellar medium. First, we put more
confidence in the predictions if the interpretation is physically and
chemically sound. Second, the learned relationships between features and
predicted values should provide meaningful insights on the physical and
chemical properties of the molecular interstellar medium. Random forests
represent a good compromise in complexity versus interpretability as they
are able to learn non-linear relationships while keeping properties that
make physical interpretation possible and that allow us to understand
\textit{a posteriori} how the predictions are obtained.

\section{Application}
\label{sec:methods:application}

The random forest implementation that we use is the
\texttt{RandomForestRegressor} class from \texttt{sklearn} python
module~\citep{Pedregosa.2020}.

\subsection{Quality of the regression/generalization: Mean error and RMSE}

We monitor the quality of the regressor and its generalization power by
computing the mean error and the Root Mean Square Error (RMSE) either on
the test or training sets. To do this, we compute the residuals, i.e., the
difference between the observed and predicted values of the column density
for each pixel, and then we compute its mean and RMSE. The RMSE quantifies
the distance between the observations and the predictions, while the mean
error informs us on potential global biases of the regressors/predictors.

While the Mean Square Error (MSE) is used in machine learning because it is
simpler and thus faster to compute as it does not involve the computation
of the square root, the results we will present from this point on use the
root mean square error. This choice enables us to have values that
can be directly compared to the predicted quantity and it gives to first
order an estimation of the uncertainty on the predictions. Moreover, there
is no lost of generality as the square root is a monotonous function.

\subsection{Separation of the data into training and test sets}
\label{sec:train-vs-test}

The molecular emission is spatially coherent over a large number of pixels
for two reasons. First, Nyquist sampling implies that neighbor pixels are
correlated. Second, the underlying physical properties of the molecular
emission are spatially correlated across pixels.  Standard methods to
divide the datasets into training and test sets, which are based on random
draws of the observations, would lead to two correlated sets, weakening the
results obtained on the test set. Furthermore, choosing a test region where
the physical and chemical properties are well known is desirable to ease
the interpretation of the regression results. Figure~\ref{fig:data:dust}
shows our choice of training and test sets. We divide the observed region
into 40 rectangles of $12.7' \times 13.3'$ or $38 \times 40$ pixels.

The test set is chosen as the rectangle containing the Horsehead Nebula
which has been extensively studied
\citep{Pety.2005,Goicoechea.2006,Gerin.2009,Guzman.2011,Guzman.2012,Pety.2012,Gratier.2013,Fuente.2017}
and is shown as a bold white rectangle in
Fig.~\ref{fig:data:dust}. Figure~\ref{fig:Horsehead} zooms in into this
test set in all tracers used in this study.  For the first tuning setup of
the IRAM-30m data set, the noise increases by about a factor two on an
horizontal band towards the southern edge of the test set. This is due to
degrading weather conditions when observing this particular region. The
Horsehead is a pillar that has been sculpted through photo-ionization of
the Orion B molecular cloud by the O star $\sigma$-Ori located about 0.5
degree away or 3.5\pc. The Horsehead is thus surrounded by the IC\,434
H{\sc ii} region that is not completely devoid of diffuse molecular
emission either in the background or in the foreground. As a pillar, it
contains two dense cores: One at the top of its head and the other one in
its throat. These nevertheless exhibit lower column density than some of
the dense cores in the NGC\,2024 region. The Horsehead dense cores are
surrounded by translucent/diffuse gas whose contact with the far-UV
illumination produces several photo-dissociation regions. On the south of
the pillar, there exist a few isolated clumps that contain less column
density, and which are more illuminated in far-UV.

The test set thus contains many different physical/chemical regimes.  This
region is never used during the training part. The 39 other rectangles are
used to train the algorithm.

\subsection{Does the test set belong to the same parameter space as the
  training set?}
\label{sec.gaussianmixture}

\FigGaussianMixture{} %

Supervised learning methods are often biased when used to predict points
outside the span of the training set. It is thus important to be able to
check whether another dataset (e.g., the test set) will belong to the same
parameter space as the training set. We need to compute the likelihood that
a point belongs to the Probability Distribution Function (PDF) of the
training set. To do this, we model this PDF with a sum of simple analytic
functions (instead of, e.g., using kernel density estimators) because this
method is still tractable for high-dimensional data sets.

Gaussian Mixture Models (GMM, see Sect.~6.8 of \citealt{Hastie.2001} or
Sect.~2.3.9 and~9.2 of \citealt{Bishop.2006}) are flexible methods that
give a synthetic probabilistic description of a dataset in terms of a
finite sum of multidimensional Gaussian PDF. We here use the
\texttt{GaussianMixture} class from sklearn to fit a sum of $n$
36-dimensional Gaussians to the training set. The number of Gaussian
components in the mixture is a free parameter that we optimize as
follows. For each $n$ in $\cbrace{1,5,10,20,40,80}$, we train a Gaussian
Mixture Model on two thirds of the training set, selected randomly. We then
compute the mean likelihood, i.e., the average of the values taken by the
GMM for each of the points belonging to the third part not seen during
training. This process is repeated three times to improve the estimation of
the mean likelihood.  The number $n$ of Gaussians that maximizes the mean
likelihood is then selected. We find that $n=10$ is the optimal value. We
also checked different kinds of constraints on the covariance matrix, and
we found that the best results are obtained when the covariance matrix is
left unconstrained.

\FigOptimization{} %

Once the Gaussian mixture model (i.e., the PDF consisting of the sum of the
weighted individual 10 36-D Gaussians) is fitted to the training set, we
compute the value that the Gaussian mixture PDF takes for each point of any
data set. Figure~\ref{fig:gaussian-mixture} shows the histogram of the
negative log$_2$-likelihood values for the training set, the test set, a
random set following the GMM PDF, and a random set uniformly sampling the
hypercube spanning the whole training set.  This latter set is obtained by
sampling independently each parameter from a uniform distribution between
the minimum and maximum values that are defined in the third and fourth
columns of Table~\ref{tab:features}, which list the minimum and maximum
values of the peak temperature for each line. The top panel of this figure
shows that the test and training sets are well related compared to the
random sampling. When zooming on the negative log$_2$-likelihood range that
only contains the test and training sets, the histograms are slightly
different (the test-set histogram has a higher wing at low negative
log$_2$-likelihood), but most of the points of both sets span the same
range of negative log$_2$-likelihood.

Appendix~\ref{sec:appendix} shows that the negative log$_2$-likelihood of a
sample $X$ with respect to a PDF $f$ can be interpreted as the quantity of
information, that is the number of bits necessary to encode $X$ with $f$
(up to the constant offset $-\log_2\Delta$ related to the resolution
$\Delta$ of the quantification). Figure~\ref{fig:gaussian-mixture} can thus
be interpreted as the histogram of the quantity of information necessary to
encode the different sets. The mean cost for a uniformly drawn set of
points (green histogram) is much larger than for the three others samples,
by approximately 3500 bits, computed as the difference between the means of
two histograms. The encoding of this set with the estimated Gaussian
mixture is clearly not adapted. Conversely, the mean number of bits needed
to encode the three other random sample is comparable, i.e., the mean costs
in bits to encode the training set, the test set, or a set of points drawn
from the GMM are similar. The Gaussian mixture is thus adapted for these
three sets.

\subsection{Optimization of the random forest regressor}

Some algorithms have additional parameters, named hyper-parameters, which
can be tuned to improve the quality of the regressor. In the case of random
forests, the generalization performance can be optimized by tuning three
hyper-parameters: 1) the maximum tree depth, 2) the fraction of features
randomly chosen to train each node of each individual regression tree, and
3) the number of trees in the forest.

The values of these hyper-parameters are optimized to obtain the best
generalization behavior of the predictor to previously unseen data. The
goal is to have a predictor that is general enough to learn the complex
non-linear relationship between observed features and predicted quantity
without learning the noise in the dataset. The standard way of tuning the
values of these hyper-parameters is to isolate a part of the training set
as a validation set. We randomly put aside 4 out of 39 training rectangles
as the validation set. The training procedure is repeated for different
(fixed) values of the hyper-parameters. The best hyper-parameter values are
then chosen as the ones that minimize the RMSE on the validation set.  We
also wish to maximize the amount of data used for training in case the
validation set contains rare meaningful events (e.g., dense cores). To
achieve this, random permutations of the validation sets are implemented
and the hyper-parameters are chosen as the ones that are optimal over the
average of the different validation sets. In our case, the best parameters
are obtained as the ones that give the best average performance over 10
such cross-validation draws.

We implemented a grid search to optimize these three hyper-parameters.
Figure~\ref{fig:hyperparams} shows the RMSE averaged over the 10
cross-validation draws as a function of the three hyper-parameters as
described in Sect.~\ref{sec:train-vs-test}.  The red cross shows the values
of the hyper-parameters that minimize the RMSE: 300 for the number of
trees, 32 for the tree maximum depth, and 30 randomly selected features
(i.e., 75\% of the total number of features).  The optimization of the
number of trees and maximum tree depth is particular because we expect that
adding more trees monotonically increases the performance by reducing the
variance. The important information here are that the RMSE does not vary
much anymore when the number of trees is larger than 10, the maximum tree
depth is larger than 16, and the fraction of randomly selected features is
larger than 25\%. This makes the overall random forest estimator robust to
changes of these hyper-parameter values.

\section{Comparison of the random forest prediction with two simpler
  methods}
\label{sec:results:comparison}

\TabComparison{} %
\FigResultsMaps{} %
\FigResultsHistos{} %

In order to show the power of the random forest predictor, we compare its
performance with two other methods. We focus on the generalization
performance of the predictors. This implies that we only check the
prediction power on the test data set that has never been seen during the
training phase.

\subsection{Multi-linear regression with or without a non-linear processing
  of the line intensities}

First, a standard ordinary least square method gives us the minimal
achievable regression accuracy. We use the \texttt{LinearRegression} class
from \texttt{sklearn} with the keyword \texttt{normalize=True} to apply a
standard pre-whitening, i.e., subtracting the mean and dividing by the
standard deviation.

Second, we compare to the result obtained by \citet{Gratier.2017}. In this
study, the linear Principal Component Analysis (PCA) was preceded by the
application of a asinh function to the original dataset
\begin{equation}
  T(x) = a \asinh(x/a),  
\end{equation}
where $a$ is a constant cutoff. This non-linear transformation allowed us
to take care of the specific properties of the histograms of the molecular
emission. They are made of a Gaussian core around zero reflecting the noise
properties, and a power law tail that reflects the signal. The application
of the asinh function had the double advantage of (i) applying a logarithm
transform to the values above the asinh cutoff $a$ to linearize the power
law tail, while (ii) keeping the noise unchanged below the asinh cutoff
$a$. This latter property allowed us to keep all the dataset without noise
clipping in the analysis.  We use a common value of the asinh threshold
$a = 0.08\K$ or $= 0.08\Kkms$ for the peak temperatures and integrated
intensities, respectively, as in \citet{Gratier.2017}. After applying this
transformation, we again use the \texttt{LinearRegression} class from
\texttt{sklearn} with the keyword \texttt{normalize=True} as above.

\subsection{Spatial distributions of the predictions and of the residuals}

The first two columns of Fig.~\ref{fig:results:maps} show the spatial
distribution of the observed and predicted column densities. The last
column shows the ratios of the observed and predicted column densities for
the three methods, on a logarithmic scale. It is thus equal to the
difference between the logarithms of the predicted and observed column
densities.

The residuals (right column of Fig.~\ref{fig:results:maps}) indicate that
all three regressions deliver column density predictions within a typical
factor of two (i.e., $\pm 0.3$\,dex). \textit{This means that it is indeed
  possible to predict the column density of the gas within a factor of two
  based only on the 3\,mm line emission.}  However, the residuals between
the predicted and observed values of $\log \NHt$ never look like random
noise. This implies that the generalization of the column density predictor
is imperfect in the three cases.

The comparison of the spatial distributions of the predicted column
densities and the residuals for the three regression methods clearly shows
that the linear regression is less successful than the other two non-linear
methods. The left/right blue/red pattern indicates that the dense gas
column density is clearly underestimated while the diffuse/translucent gas
column density is overestimated for the linear predictor. The non-linear
predictors perform better overall (lower contrast in the residuals).

The difference between the asinh pre-processing predictor and the random
forest one is more subtle. The contrast of the residual image is slightly
less pronounced for the random forest predictor. It does better than the
asinh pre-processing in the dense cores, under the Horsehead muzzle, and in
the diffuse/translucent gas above the Horsehead pillar. However, the back
of the Horsehead (its mane) appears bluer in the residual maps of the
random forest predictor.

\subsection{Joint distributions of the predicted and observed column
  density and histograms of their ratios}

In order to quantitatively compare the three different methods,
Fig.~\ref{fig:results:histos} shows the joint distributions of the
predicted and observed column densities, as well as the histograms of their
ratios. Table~\ref{tab:comparison} lists the RMSE over the test set, the
maximum root square error, as well as the mean of the ratios. These values
here quantify the performance of the generalization of the method.

A linear regression gives a double-peaked histogram of the column density
log-residuals (i.e., the base-10 logarithm of the ratio of the predicted
over observed column density). This comes from the fact that the prediction
over-estimates the low column densities and under-estimates the high column
densities.  The linear regression on the asinh of the intensities delivers
a much better statistical agreement. Most of the predictions are close to
the measured values even though there still is some scatter in particular
at intermediate column densities. The joint histograms show that the asinh
slightly underestimates the column density around \dix{6.3}{21}\pscm{} and
slightly overestimates it above \dix{1.6}{22}\pscm{}. The random forest
predictor shows the least dispersion around the actual column density. All
these properties translate into the fact that the histogram of the
log-residuals is closest to a Gaussian for the random forest predictor.

These results are quantitatively confirmed by the values of the RMSE and
the maximum absolute error listed in Table~\ref{tab:comparison}. The RMSE
indicates that the predictors infer the column density within 20, 30, and
40\% with a maximum error of a factor 1.8, 2.1, and 2.3, for the random
forest, the asinh pre-processing, and the linear predictor,
respectively. An additional piece of information is that all three methods
over-estimate the column density by 20, 10, and 6\% for the asinh
pre-processing, the random forest and the linear predictors, respectively.
This could be due to the fact that we try to infer a positive quantity from
noisy measurements where the centered noise sometimes hides the signal.

\referee{More quantitatively, we estimated the standard deviation on the
  mean error and MSE as
  \begin{equation}
    \sigma_\emr{Mean\,\,error} = \sqrt{\emr{var}/N}
    \quad \mbox{and} \quad
    \sigma_\emr{MSE} = \sqrt{2/N}\,\emr{var},
  \end{equation}
  where $N$ is the number of pixels in the test set, and var is the
  variance of the error, i.e., the difference between the logarithm of the
  predicted and observed values. The values listed in
  Tab.~\ref{tab:comparison} shows that the difference between the mean
  errors associated with the random forest, and the asinh pre-processing
  methods is much larger than the sum of the associated standard deviations. A
  similar result is obtained for the mean square errors of the two
  methods. These two results mean that the random forest method yields a
  significantly better prediction than the asinh pre-processing.}

\subsection{Uncertainty}

\FigUncertainty{} %

The fact that a random forest is an ensemble method can be leveraged to
associate an uncertainty value to each predicted quantity. At the
prediction step, the regression trees yield a set of 300 values and the
mean of these values is the prediction of the random forest algorithm.  It
is thus possible to compute the value at a given percentile of the
cumulative distribution of these 300 values. Specifically, we compute the
values for the \{2.3\%, 15.9\%, 84.1\%, 97.7\%\} percentiles, which would
corresponds to 1 and $2\sigma$ uncertainty intervals for a Gaussian
distribution.

Figure~\ref{fig:rf:uncertainty} compares the median prediction surrounded
by the uncertainty intervals that comprise 68 and 95\% of the trees with
the column density of each pixel with the observed value. This confirms
that column densities are well estimated between \dix{2.5}{21}\pscm{} and
\dix{1.6}{22}\pscm{}, and slightly over-estimated outside this
interval. Nevertheless, the observed values are within the
95\%--uncertainty interval for more than $90\%$ of the test set.

The largest discrepancies between the observed and predicted values happen
when the observed column density is lower than \dix{2.5}{21}\pscm{}. In
this range, the column densities are most often over-predicted for the
three methods tested in this section. This could be related to the fact
that \twCO{} \Jone{} is over-luminous in the diffuse gas as already
observed by~\citet{Liszt.2012b}. Indeed, diffuse gas is actually present
around the Horsehead pillar in the second velocity component between 2 and
8\kms{} as mentioned in~\citet{Pety.2017}. This behavior may have not been
well learned due to the lack of clean example in the training set.

\section{Contribution of the different lines to the performance of the
  predictor}
\label{sec:results:contributions}

\FigLineImportanceLatex{} %
\FigResultsMapsOneByOne{} %

In the previous section, we showed that the random forest predictor is able
to predict the column density with a precision of 20\% on data points that
belong to the same parameter space as the training set. We now try to
quantify the contribution of the different molecular lines to this
result. This question can be answered on both the training and test sets,
as they belong to similar distributions of the input observables and
targeted physical quantity. The advantage of the training set is that it
contains 40 times as many points. The advantage of the test set is that the
spatial variations of the input and targeted variables show shapes that are
easy to describe. We thus use the training set to get global trends and the
test set to discuss finer trends.

\subsection{Which lines contribute the most?}
\label{sec:importance}

A first interpretative tool is the quantification of the line importance,
i.e., how much each line contributes to the prediction of \Ndust. To do
this, we first keep the value of the root mean square error computed on the
training set as a reference. We then randomly permute the values of the
intensities for a given line, all other intensities being kept constant.
We finally compute the root mean square error on the prediction using this
shuffled dataset. The increase in RMSE is the importance associated with
the line. This importance has the same unit as the predicted quantity and
it measures how much the performance would be degraded when the species is
replaced by a noise that keeps the shape of the probability distribution
function. For each line of sight, we simultaneously shuffle the values of
the peak intensities and of the integrated intensity values of a given
molecular line to estimate the overall importance of this line. Moreover,
we try to check whether all lines have a significant contribution. We thus
added two random data sets as additional input features to check which
lines bring more information than plain noise. We used two different random
data sets to check that the result is not biased by any given random
drawing.

The top panel of Fig.~\ref{fig:line-importance} shows the line importance
by decreasing value for our dataset. We first see that all lines bring more
information than plain noise. Second, the $(J=1-0)$ line of \thCO, \twCO,
\CeiO, \HCOp, HNC, \NNHp{}, CCH, and the $(J=2-1)$ line of \twCS{} have the
largest line importance. Shuffling the \thCO{} samples increases the RMSE
by $\sim 0.22\,$dex (i.e., a factor 1.65 multiplying the reference factor
of 1.2). The least important of these 8 lines still increases the RMSE by
0.01\,dex (or a factor 1.02), when shuffling its samples. Shuffling the
data of any other line increases the RMSE by less than 0.01\,dex.

Another way of visualizing this effect is to build predictors with an
increasing number of lines ordered by decreasing feature importance.  The
bottom panel of Fig.~\ref{fig:line-importance} clearly shows that adding
more than the 4 most important lines only marginally increases the overall
performance of the prediction. In other words, it seems that only 4 lines
(the $J=1-0$ line of the three main CO isotopologues, and \HCOp) can predict
\Ndust{} almost as well as when all the lines are included, but this
statement will be refined in Sect.~\ref{sec:regime}. Adding more than the
first 8 lines even seems to bring no added value.  This also shows that the
random forest method is rather insensitive to the presence of ``noisy''
data in the input features.

\FigContributionMaps{} %

\subsection{Where does each line contribute to the prediction?}

To confirm these quantitative measures,
Fig.~\ref{fig:results:maps:onebyone} shows the evolution of the spatial
distribution of the predicted column density and of the associated
residuals around the Horsehead pillar when building random forest
predictors that are trained on an increasing number of lines ordered by
their decreasing importance.

We see that the predictor trained only on the \thCO{} \Jone{} line is able
to recover the shape of the Horsehead pillar.  This implies that this line
contributes to differentiate the column density between translucent and
denser gas. Adding the contribution of the \twCO{} \Jone{} line changes the
residual maps mostly in the regions made of diffuse gas (red part on the
right side). However, a predictor trained on these two lines alone provide
a rather poor estimation of the column density in either the halo that
surrounds the Horsehead pillar or the dense cores (e.g., the dense core at
the top of the Horsehead or in its throat).  Adding the \CeiO{} \Jone{}
line is important to improve the estimation within the denser parts
(Horsehead spine and dense cores), while including the \HCOp{} \Jone{} line
improves the estimation of the column density in the far-UV-illuminated
transition region between the translucent and denser gas. Completing the
line sample with the HNC and \NNHp{} \Jone{} lines slightly changes the
contrasts of the predicted images but this seems a second order effect as
the shape of the residuals does not change much when adding these
lines. The most important change comes from the contribution of the \NNHp{}
\Jone{} line to the dense cores inside the Horsehead pillar.

\FigAvMasks{} %

Looking at the structure of regression trees, and random forests, it is
possible to compute the contribution of each line $l$ to the \Ht{} column
density. Indeed, it is possible to show\footnote{A demonstration can be
  found at \url{http://blog.datadive.net/interpreting-random-forests/}.}
that the predicted value of the column density at pixel $(i,j)$ can be
written as
\begin{equation}
  \log N(i,j) = \log N_0 + \sum_{l=1,L}\, \log N_l(i,j),
\end{equation}
where $\log N_0$ is the mean of the column density over the training
dataset, and $\log N_l$ is the quantitative contribution of each line
(either integrated line profile or peak temperature) to the predicted value
of the column density at pixel $(i,j)$. Figure~\ref{fig:contribution-maps}
shows the spatial distribution of the contribution of the eight lines most
important to predict the column density.  The first striking result is that
all contribution maps show well-behaved structures (with extremely few
exceptions), even though all lines are not detected over the full field of
view. This suggests again that the column density estimate is rather
insensitive to noise.

The contribution maps also allow us to quantitatively refine the above
picture. They confirm that the \thCO{}, \twCO{}, \CeiO{} and \HCOp{}
\Jone{} lines are the first order corrections to the mean value of the
column density. The integrated intensity of \thCO{} contributes the most to
the predictor. Its contribution map shows that it is important over the
whole area. It contributes positively where the column density is high and
negatively in the most far-UV illuminated regions. This coincides with the
visual impression that the contribution map recovers well the shape of the
Horsehead pillar. This is expected because the \thCO{} line traces most of
the gas without being too optically thick.  The second biggest contributor
is the \twCO{} \Jone{} line. It also contributes positively where the gas
is translucent (see for instance the clumps south of the Horsehead
pillar). It is almost neutral (white or light blue around the Horsehead)
where the gas is diffuse and it contributes negatively (dark blue) where
the H{\sc ii} region dominates. The overall visual impression is that the
\twCO{} line is important to predict the diffuse to translucent part of the
column density along the line of sight. The next two main contributors are
the \HCOp{} and \CeiO{} \Jone{} lines. They contribute in two complementary
physical regimes. The \CeiO{} line contributes mostly where the gas is
dense with a positive contribution at the highest densities (Horsehead
spine) and a negative contribution at lower densities (nose, mane, feet).
Conversely, the \HCOp{} line contributes mostly on the photo-dissociation
regions that surround more or less dense gas. The comparison of the
contribution maps of the integrated intensity and peak temperature for
these four lines shows that the integrated intensities contribute more to
the estimation of the column density.  The peak intensities provide second
order corrections, sometimes of opposite signs (clearest on the \CeiO{}
line contributions) to the prediction of $\log \NHt$. This suggests that
these two parameters indeed play different roles in the estimation of the
column density.

The next four most important lines are the $J=1-0$ lines of HNC, \NNHp{},
CCH, and the $J=2-1$ line of \twCS{}. HNC, and \NNHp{} contribute mostly on
the densest parts of the Horsehead pillar, i.e., the dense cores and their
surroundings. A striking feature is that the HNC peak temperature
contributes one of the most important corrections in the places where the
gas is dense, while its integrated intensity does not play a role in the
prediction. This property is probably related to the detection pattern of
this line in Fig.~\ref{fig:Horsehead}, which shows that the spatial
distribution of the peak temperature is more contrasted/structured than the
integrated line emission for the HNC line. Indeed, its integrated intensity
varies between about 1 and 3\Kkms{} (it appears mostly red) everywhere it
is detected in the horsehead pillar, while its peak temperature varies from
about 0.3 to 2\K{} (its color varies from green to white) on the same
region.  Another striking feature is that the \NNHp{} (1-0) peak
temperature contribution map is structured even in regions where this line
is not obviously detected. This probably means that the random forest has
learned that a correction is needed when the \NNHp{} line stays undetected.
Finally, the CCH \Jone{} and \twCS{} \Jtwo{} lines contribute smaller
corrections in photo-dissociation regions and UV-shielded dense gas,
respectively. The peak temperature and integrated intensity of both lines
contribute corrections of similar magnitude.

\FigLineImportanceAvMasksTrainLatex{} %
\FigContributionVsAv{} %
\FigGeneralization{} %

\subsection{Which physical regime does each line contribute to?}
\label{sec:regime}

The line importance discussed in Sect.~\ref{sec:importance} is computed on
the full training set that indifferently mixes all physical regimes. But
the contribution maps discussed in the previous section clearly showed that
some lines are more important in some particular physical regimes. We thus
now ask whether some other lines are important for a given physical regime
of column densities. Using the same categories as discussed in
\citet{Pety.2017}, we compute the line importance on four subsets of the
training set (see Fig.~\ref{fig:Av-masks}): $1 \le \Av < 2$ (diffuse gas:
14\,377 pixels), $2 \le \Av < 6$ (translucent gas: 34\,635 pixels),
$6 \le \Av < 15$ (filaments: 8\,723 pixels), and $15\le \Av$ (dense cores:
1\,545 pixels).  Figure~\ref{fig:line-importance-Av-masks} shows the line
importance diagrams for the four different intervals of visual extinction.

The \thCO{} line is important for the estimation of the column density in
all kinds of environments except in the dense cores. However, it is the
most important only in the translucent gas. In diffuse gas, it contributes
much less to the accuracy than the \twCO{} \Jone{} line. In the filamentary
gas, it contributes less than the \CeiO{} and HNC \Jone{} lines. The
\twCO{} \Jone{} line is important in all regimes. However, while it
completely dominates the estimation in diffuse gas, its importance
regularly decreases when the visual extinction increases.  The CCH \Jone{}
line plays a role in diffuse and translucent gas and almost no role in the
two larger visual extinction regimes.  The \HCOp{} \Jone{} line plays an
important role in diffuse and translucent gas (third line after the \twCO{}
and \thCO{} \Jone{} lines), and some role in relatively dense gas (6th line
for filaments) when it is exposed to far-UV illumination. Its role is minor
in the prediction for dense cores.

The \CeiO{} and HNC \Jone{} lines play major roles in relatively dense gas,
i.e., filaments and dense cores. Finally, the \NNHp{} and \methanol{} lines
are the most important ones to accurately predict the column density in
dense cores where the CO isotopologues are depleted. This finding clearly
indicates that line importance (as defined here) must be interpreted with
caution when the populations of the different physical regimes are
unbalanced in the training data set. A change in the RMSE value appears
larger when a physical regime that is impacted is over-represented and
vice-versa.

\referee{Figure~\ref{fig:contribution-vs-Av} shows the contribution of the
  most important lines to the logarithm of the \NHt{} column density as a
  function of the visual extinction. One red points per pixel of the test
  set is plotted. The black histograms show the median values of all data
  points falling in a regularly sampled interval of the logarithm of the
  visual extinction. As the test set does not contain many pixel at high
  visual extinction, the typical contribution of each line at $\Av \ga 10$
  is not as well constrained as at lower visual extinction.
  The scatter around each typical value of the histogram comes from two
  sources. Tunings \#2 and \#3 are noisier than tuning \#1 (see
  Tab.~\ref{tab:lines} for the list of concerned lines). However, noise
  does not explain all the observed scatter. A large fraction of the
  scatter comes from the fact that the combined measurement of several
  lines is indeed required to yield an accurate value of the column density
  at any given \Av{}.}

\referee{For each line, the contribution is computed as the sum of the
  contribution of the line peak temperature and integrated intensity. These
  contributions have to be added to the mean logarithm of the column
  density computed over the training set to get the column density value of
  the considered pixel of the test set. A contribution value of zero
  implies that the line has no impact at the associated visual
  extinction. This is clearly the case of the noise feature.  A value of
  -0.2 or 0.2 implies that the line intensities require to multiply the
  column density by 0.63 or 1.58, respectively. This is the case of the
  \thCO{} \Jone{} line that requires to multiply the average column density
  by a factor $\sim 0.6$ below $\Av \sim 5$ and by a factor $\sim 1.6$
  above $\Av \sim 10$.
  Except the \thCO{} \Jone{} line that contributes at all \Av{} and the
  noise sample that does not contribute at any \Av, the other lines
  contribute inside a given \Av{} range. The lines are sorted from top to
  bottom and then from left to right by increasing values of the minimum
  \Av{} at which they start to contribute. The \twCO{} \Jone{} line
  contributes at $\Av \la 5$, \HCOp{} \Jone{} line in the range
  $1 \la \Av \la 10$. The \CeiO{} and HNC \Jone{} lines contribute at
  $\Av \ga 5$, and the \NNHp{} \Jone{} and \methanol{} \Jtwo{} lines start
  to contribute at $\Av \ga 10$ and contribute even more at $\Av \ga 18$.}
  
\subsection{Comparison with previous works on the Orion\,B molecular cloud}

The results of this analysis shed a new light on the role of the molecular
lines in tracing different gas density regimes in previous studies of the
Orion\,B molecular cloud.

The main lines contributing to an accurate estimation of the \Ht{} column
density -- the $J=1-0$ lines of \twCO{}, \thCO{}, \CeiO{}, and \HCOp --
had already been identified as effective tracers of the density regime by
\citet{Bron.2018}. In particular, the three main CO isotopologues traces
well the transition from diffuse ($\sim 10^2\pccm$) to relatively dense
($\sim 10^3\pccm$) gas, with an increasing importance of the rarer
isotopologues at higher densities. \citet{Bron.2018} also showed that
adding the \HCOp{} \Jone{} line brings sensitivity 1) to higher density
regions (up to $\sim 10^5\pccm$), and 2) to far-UV illuminated regimes in
the regions of lower densities ($\le 10^3\pccm)$. While the latter result
is confirmed by this study, the former result, i.e., the role of \HCOp{} in
detecting regions of higher densities is in contrast with the Random Forest
results, where this tracer only has a minor role in the dense medium. This
probably comes from methodological differences between the two
studies. First, \citet{Bron.2018} used a discrete clustering approach,
while we here use a continuous method. Second, \citet{Bron.2018} worked on
a subset of the lines studied here.

The qualitative study of the correlation between molecular line intensities
and column densities performed by~\citet{Pety.2017} also contains results
that are quantitatively confirmed by the current analysis. The \thCO{}
\Jone{} and \CeiO{} \Jone{} lines were already identified as overall good
tracers of the \Ht{} column density, i.e., they have a monotonous
relationship with low scatter over a broad range of column densities.  The
contribution of the different molecular lines to the Random Forest
estimator of the column density in various extinction regimes (see
Fig.~\ref{fig:line-importance-Av-masks}) is consistent with the earlier
results from \citet{Pety.2017}. In particular, the \HCOp{} \Jone{} line is
confirmed to contribute at low extinctions, despite its high critical
density, while the actual best tracers of dense gas are the \NNHp{} \Jone{}
and \methanol{} \Jtwo{} lines. This latter point was already observed by
\citet{Gratier.2017}, who noted the strong anti-correlation of these lines
with the emission of CO isotopologues in dense gas where CO depletion
occurs. The role of the CCH \Jone{} as a tracer of diffuse, far-UV
illuminated regions was noted by \citet{Pety.2017,Gratier.2017,Bron.2018},
and by earlier studies of the Horsehead photo-dissociation
region~\citep[e.g.,][]{Pety.2005,Pety.2012,Pilleri.2013,Guzman.2015}.

Previous studies of Orion B have also used a single molecular tracer as a
simple proxy for the \Ht{} column density, either for the bulk of the
cloud~\citep[\thCO{} $J=1-0$ in][]{Orkisz.2017} or for the dense filaments
and pillar regimes~\citep[\CeiO{} $J=1-0$
in][]{Hily-Blant.2005,Orkisz.2019}.
Figure~\ref{fig:line-importance-Av-masks}
\referee{and~\ref{fig:contribution-vs-Av}} shows that these qualitative
choices of tracers are \referee{relatively} well-suited for the targeted
density regimes.

\section{Comparison, generalization, limitations, and perspectives}
\label{sec:discussion}

\FigXco{} %
\FigCOiso{}%
\TabComparisonBis{}%
\FigGo{} %

In this section, \referee{we first discuss how the found Random Forest
  predictor of the column density compares with simpler approaches as the
  standard \Xco{}-factor method.} We then discuss how the found Random
Forest predictor of the column density can be used on noisier or smoothed
data sets. We also try to generalize the method to predict the far-UV
illumination. And we discuss the fact that other physical variables partly
control the 3\,mm line strengths and we finally propose additional sources
of information that could help to better constrain the physical conditions
inside a giant molecular cloud.

\subsection{Comparison with simpler approaches to infer the \Ht{} column
  density from molecular lines}

\referee{As explained in Sect.~\ref{sec:aims}, the three main CO
  isotopologues are often used to derive the \Ht{} column density because
  they are among the easiest detectable molecular lines in molecular
  clouds. Figures~\ref{fig:Xco} and~\ref{fig:COiso} show the performance to
  predict $N_{\Ht}$ of two simpler approaches. We first use the standard
  \Xco-factor method,
  \begin{eqnarray}
    \mbox{i.e.,} \quad & N_{\Ht} = \Xco \, W(\twCO{}\,J=1-0) \\
    \mbox{with}  \quad & \Xco = 2\times 10^{20} \pscm\,(\K\kms)^{-1}.
  \end{eqnarray}
  Second, we trained a random forest predictor using only \Jone{} lines
  of the three main CO isotopologues, \ie{}, \twCO{}, \thCO{}, and
  \CeiO{}. Table~\ref{tab:comparison:bis} lists the maximum error, mean
  error, and RMSE for these two methods and the random forest trained on
  all lines used in this article.}

\referee{The \Xco{}-factor method overall yields a poor inference of the
  column density in the Horsehead pillar. This is a well known behavior:
  The \Xco{}-factor method only brings reasonable results when considering
  large fractions of a Giant Molecular Cloud. By fitting the value of the
  \Xco{} factor, we could in principal improve the mean error of the method
  on the test set but the large dispersion of the results would stay
  identical. A random forest trained on the three main CO isotopologues
  yield a much better inference of the column density. Its mean error is
  even lower than the random forest trained on all our lines. This implies
  that the simpler method is less biased but the histogram of the
  difference between the logarithms of the predicted and observed column
  densities is not centered on zero. This means that the predicted values
  are more often either under or over-estimated. This also implies a
  significantly larger RMSE.  Lines such as the \Jone{} lines of \HCOp{},
  CCH, HNC, and \NNHp{} are important to yield more consistent results over
  the full range of visual extinction.}

\subsection{Noise and distance effects}

Our training and test sets have been observed with a decent signal to noise
ratio. We thus wonder whether it is possible to use the random forest
prediction of the column density on observations having low signal-to-noise
ratios. To explore the predictive power of the method when confronted with
noisy observations, we have computed the mean error and RMSE on the
predicted column density when adding increasing Gaussian noise to each
channel of each spectra of the test set. The same noise rms value is used
for all lines simultaneously.

The left panel of Fig.~\ref{fig:generalization} shows the results of this
procedure. The effect is negligible when the added noise rms is lower than
0.1\K{} in channels of $0.5\kms$. Both the mean error and the RMSE increase
slowly for values of added noise between $0.1$ and $\sim 3\K$. They then
increase quickly for larger noise values. These noise values have similar
orders of magnitudes than the mean peak temperatures listed in
Table~\ref{tab:features}. The mean peak temperatures belong to the
$[0.1,0.5\K]$ interval for all lines except H$_\emr{40\alpha}$ that anyway
plays a negligible role here, and the $J=1-0$ line of \thCO{} and \twCO{}
whose mean temperature are 3.4 and 14.0\K, respectively. This means that
the detection of a given line is enough to make a first estimate of the
\Ht{} column density. This also implies that the column density predictor
could be used on most observations taken today in millimeter astronomy as
the ORION-B data set was obtained at the highest telescope velocity
possible at the IRAM-30m telescope, i.e., between 16 and
18$''\unit{s^{-1}}$ (this limit coming from the data rate).

Our training and test sets also belong to one of the closest giant
molecular clouds. We also wonder up to which distance it is possible to use
random forest prediction of the column density. We assume that we observe
other molecular clouds at larger distances from the Sun with the same
telescope. This means that the linear resolution decreases with the
distance, i.e., the images of the input and predicted variables are
consistently smoothed with increasing Gaussian kernels and then
downsampled. The right panel of Fig.~\ref{fig:generalization} shows the
results. We conclude that it is possible to use the random forest predictor
trained at the highest angular resolution up to a distance 8 times larger
with a similar precision. The predicted column density is then of course
the beam-diluted column density.

\subsection{Generalization to the prediction of the far-UV illumination}

The far-UV illumination is another key parameter in the physics of
molecular clouds.  We thus check whether it is possible to quantitatively
infer the far-UV illumination from the 3\,mm molecular emission, i.e., to
check whether the procedure described above is generalizable to other
physical quantities.

In far-UV illuminated regions, the dust emission is closely linked to the
far-UV photon flux. \citet{Pety.2017} converted the dust temperature map
into an approximate map of the far-UV radiation field $G_0$ in units of the
Habing Interstellar Standard Radiation Field (ISRF, \citealt{Habing.1968}),
using the simple approximation of~\citet{Hollenbach.1991} for face-on PDRs
\begin{equation}
  G_0 = \left ( \frac{T_\emr{dust}}{12.2\K} \right )^5.
\end{equation}
\citet{Shimajiri.2017} compared this estimation with another estimation
directly using the far-IR intensities at 70 and $100\mum$. Both estimates
agree within 30\%.  Given the complex geometry of molecular clouds with
respect to the illuminating stars and the presence of far-UV shielded dust
emission, the deduced values of $G_0$ should only be trusted at
order-of-magnitude levels. As for the observed column density used to train
the Random Forest (see Sect.~\ref{sec:GouldBeltSurvey}), we do not claim
that our dust-traced estimation of the far-UV illumination is a perfect
measure of $G_0$. Another way of looking at this is to state that we only
here try to predict the dust temperature from the 3\,mm molecular emission
alone, as we try to predict $\log G_0$ that is linearly related to
$\log T_\emr{dust}$. However, our long-term goal is to infer the values of
the parameters that control the underlying physics. That is why we here
state that we wish to check whether the 3\,mm molecular emission alone is
able to predict this dust-traced $G_0$ with all its caveats (see also
Sect.~\ref{sec:complete}).

We used the same input features, the same training and test sets, and the
same methods as the ones used to learn how to infer the \Ht{} column
density. Figure~\ref{fig:Go} shows the results. Quantitatively, the mean
error on $\log G_0$ is 0.081\,dex (i.e., a factor of 1.21), the RMSE is
0.25\,dex (a factor of 1.8), and the maximum absolute error is 0.78\,dex (a
factor of 5.9). The prediction of the observed far-UV illumination is
typically off by a multiplicative factor of 1.2 with a multiplicative
scatter of 1.8 and errors of up to a factor 6. These values must be
interpreted keeping in mind that the observed far-UV illumination around
the Horsehead pillar spans slightly more than one order of magnitude. The
joint distribution of the predicted and observed values of $G_0$ shows a
large scatter in the ranges $10-15$ and $50-80$. The histogram of the
predicted over observed values shows a maximum around a factor 1.6, a
secondary maximum around a factor 1.0, plus two other over-densities
compared to a Gaussian of same mean and width at factors of 0.5 and
2.5. The comparison of the spatial distributions shows that the predicted
$G_0$ has much less smooth variations than the observed $G_0$ with large
errors in diffuse gas and dense photo-dissociation regions.

These results are not as good as for the column density. We interpret the
lower quality prediction of $G_0$ by the fact that the far-UV illumination
is related to the third principal component in the work of
\citet{Gratier.2017}. This one only explains $\sim 5\%$ of the correlations
present in the input dataset, compared to 60\% for the first principal
component that is correlated to the column density. This means that $G_0$
is more difficult to extract from the current set of line intensities
detected in the 3\,mm band. This set mainly contains one rotational
transition per molecular species, most often the ground level one. It is
known \citep[see, e.g., the companion paper][]{Roueff.2020} that these
transitions probe the molecular column densities quite well, but that they
are less efficient for constraining the excitation conditions of the
species. It is expected that the sensitivity to $G_0$ would be increased by
adding information on the molecular excitation in warmer gas, since the
populations in higher energy levels are altered in regions illuminated by a
high radiation field.  This means that the line set should be complemented
by higher level transitions of a subset of the species probed in the 3\,mm
band. This ``excitation signature'' will complement the ``chemical
signature'' already present in the 3\,mm line sample.

\subsection{Confounding variables}

The intensity of a given line depends not only on the total column density
of matter, but also on the molecule abundance, and the excitation
conditions (kinetic temperature, density of neutrals, and electrons). The
latter quantities can thus be confounding variables in the relationship $F$
between the total column density $\Ndust$ and the input intensities
$I_l$. The lack of knowledge of these confounding variables implies that
somewhat different values of the dust-traced column density are possible
for a given set of line intensities, and thus acts as a source of
uncertainty (of unknown distribution). In the absence of ancillary
knowledge besides the line intensities, the best estimation of the column
density predictor nevertheless exists: It is the conditional expectation of
the dust-traced column density, given the knowledge of the line
brightnesses, mathematically $<\Ndust | I_1,...,I_L>$. In this paper, we
ignore the effect of the confounding variables on the ground that the
correlation between the line intensities in the PCA analysis are largely
dominated by the first component that is, in turn, highly correlated with
the column density.  The presence of confounding variables induces a small
additional uncertainty on the relationship between the dust-traced column
density and the line intensities, that would need additional information to
be lifted.

\subsection{Other potential sources of information to improve the
  predictions of the gas column density and far-UV illumination}
\label{sec:complete}

The dust-traced column density (\Ndust) used in this article to train the
Random Forest represents an approximate measure of the actual gas column
density. Given the range of extinction, it is assumed that all gas is
molecular and \ion{H}{I} represents only a small fraction of the total gas
column. Indeed, the column density of atomic-hydrogen-dominated gas (i.e,
very low molecular fraction) accounts for less than 1 visual magnitude of
extinction in the studied field of view~\citep{Pety.2017}.  A second
assumption is the constancy of the dust-to-gas ratio over the whole
region. Our study shows that the knowledge of the emission of a small
number (six to eight) of 3\,mm molecular lines is sufficient to predict the
dust-traced column density in regions where the visual extinction is mostly
associated with molecular gas.

In other regions, and especially when dealing with larger spatial scales,
the dust thermal emission is associated with both atomic and molecular gas
and is therefore used to determine the total gas column. At large scales
$(> 10\pc),$ and because the molecular gas is more concentrated than the
atomic gas, the total gas column is often dominated by the contribution of
atomic hydrogen. This applies to the large scale halos around giant
molecular clouds in our Galaxy as well as in external
galaxies~\citep[e.g.][]{Leroy.2009}. \citet{Remy.2017,Remy.2018a,Remy.2018b}
shows that a fraction of the total gas column, called the CO-dark, remains
unaccounted for when tracing the atomic gas with the \ion{H}{I} 21\,cm line
and the molecular gas with the \twCO{} \Jone{} emission using a simple
linear method. The composition of the CO-dark gas (or more generally dark
neutral medium) is actually a mixture of atomic gas when the 21\,cm
\ion{H}{I} line becomes optically thick, and molecular gas with low CO
abundance~\citep[e.g.,][]{Liszt.2018,Liszt.2019}. The regions dominated by
\ion{H}{I} or by the CO-dark gas have low to moderate extinctions. The
emission in \twCO{} \Jone{} is too weak in this gas to be easily detected
at the sensitivity of typical observations.

While the random forest method is efficient for molecular cloud conditions like
Orion B ones, \referee{it will have to be tested in other regions to check how
it behaves on other datasets. Datasets covering a parameter space too far from
the ORION-B one as defined in Sect.~\ref{sec.gaussianmixture} will have to be
joined to the ORION-B data to generalize the training. When the physical space
will be correctly sampled, users should be able to apply the predictor. More
generally, this method still needs to be anchored on more diverse datasets.}
These could be grids of models of photo-dissociation regions, which predict the
line brightness of many species as a function of the column density for
different physical regimes~\citep{Le-Petit.2006}. Dust extinction maps derived
from star counts~\citep{Capitano.2017}, or from maps of the gamma ray
fluxes~\citep{Remy.2017,Remy.2018a,Remy.2018b} would also provide independent
estimates of the total column density, although at a lower spatial resolution
than the molecular data. Finally, it would be interested to complement the
3\,mm molecular observations with velocity-resolved observations of the key
\ion{H}{I} 21\,cm and 158\unit{\mu m} $[\ion{C}{II}]$ lines. Both lines would
provide complementary information about the neutral gas not emitting in \twCO{}
\Jone. As one of the strongest cooling line, $[\ion{C}{II}]$ is also expected
be a good probe of the far-UV illumination in combination with the 3\,mm
lines~\citep{Pabst.2017,Pabst.2019}.

\section{Conclusions}
\label{sec:conclusions}

In this article we have shown that it is possible to derive an estimator of
the \Ht{} column density from a set of molecular line observations. We have
used observations from the ORION-B data set, training the random forest on
both the line integrated intensities and their peak temperatures.  We
obtained the following results.
\begin{itemize}
\item When compared to linear regression on raw intensities or after a
  non-linear (asinh) processing of the intensities, the random forest
  regression delivers the best statistical agreement and thus provides the
  best generalization power. Indeed, the mean biases have a similar
  magnitude but the error variances and the maximum errors are the smallest
  for the random forest predictor.
\item On average, eight lines play the strongest role in the prediction of
  the \Ht{} column density. The $J=1-0$ lines of the three main CO
  isotopologues and \HCOp{} dominate the performance of the prediction. The
  $J=1-0$ lines of HNC, \NNHp{}, CCH, and the $J=2-1$ line of \twCS{}
  contribute as second order corrections.
\item A deeper analysis shows that the \twCO{} \Jone{} line is the most
  important line in diffuse gas $(\Av \la 2)$, the \thCO{} \Jone{} line in
  translucent gas $(2 \la \Av \la 5)$, the \CeiO{} \Jone{} line in the
  filament gas $(5 \la \Av \la 15)$, and the \NNHp{} \Jone{} and
  \methanol{} $(2_0-1_0)$ lines in dense cores $(15 < \Av)$.
\item \referee{The accuracy of the method over a large range of visual
    extinction depends on the number of lines measured. In particular, the
    intensity from a single line can not alone bring an accurate \Ht{}
    column density under all the physical regimes, \eg{}, when the gas is
    more or less far UV-illuminated.}
\item The prediction of the far-UV illumination field using the same method
  is less successful, probably because the set of lines we use is not
  sensitive enough to the excitation conditions of the gas.
\end{itemize}
This work gives further support to the use of the CO isotopologues for
deriving the total column density in molecular gas. It indicates that
acquiring the three main isotopologues should be preferred over targeting a
single CO line because they are sensitive to different ranges of visual
extinction (e.g., diffuse, translucent, dark lines of sight). To further
progress on the understanding of the physical conditions of molecular
clouds, a detailed understanding of the variations of the CO isotopologue
excitation conditions and relative abundances is needed. This will be the
subject of further articles in this series, \referee{starting with a
  companion article by~\citet{Roueff.2020}, which derives accurate
  excitation temperatures and column densities of the three main CO
  isotopologues.}

\begin{acknowledgements}
  This work is based on observations carried out under project numbers
  019-13, 022-14, 145-14, 122-15, 018-16, and finally the large program
  number 124-16 with the IRAM 30m telescope. IRAM is supported by INSU/CNRS
  (France), MPG (Germany) and IGN (Spain). This research also used data
  from the Herschel Gould Belt survey (HGBS) project
  (http://gouldbelt-herschel.cea.fr). The HGBS is a Herschel Key Programme
  jointly carried out by SPIRE Specialist Astronomy Group 3 (SAG 3),
  scientists of several institutes in the PACS Consortium (CEA Saclay,
  INAF-IFSI Rome and INAF-Arcetri, KU Leuven, MPIA Heidelberg), and
  scientists of the Herschel Science Center (HSC).

  We thank CIAS for their hospitality during the many workshops devoted to
  the ORION-B project. This work was supported in part by the Programme
  National ``Physique et Chimie du Milieu Interstellaire'' (PCMI) of
  CNRS/INSU with INC/INP, co-funded by CEA and CNES.  This project has
  received financial support from the CNRS through the MITI
  interdisciplinary programs. JRG thanks Spanish MICI for funding support
  under grant AYA2017-85111-P.
\end{acknowledgements}

\bibliographystyle{aa} %
\bibliography{ms.bib}

\appendix %

\section{Interpreting the negative log$_2$-likelihood of a set of points}
\label{sec:appendix}

If $X$ is a discrete random variable whose probability distribution is
$p_i=P(X=x_i)$ with $\sum_i p_i=1$, $Q(X=x_i)=-\log_2 p_i$ can be
interpreted as the quantity of information associated to the event
$X=x_i$~\citep[see, e.g.,][]{Cover.1991}. For instance, having knowledge of
an unlikely event corresponds to a great amount of information.

Moreover, the quantity of information of a couple of independent events is
simply the sum of both quantities. Indeed,
$Q(X=x_i, Y=y_j) = Q(X=X_i) + Q(Y=y_j)$, because
$P(X=x_i, Y=y_j)=P(X=x_i) P(Y=y_j)$. For a source of information $X$, the
statistical mean of $Q(X)$ is
\begin{equation}
  \langle Q(X) \rangle_p
  = -\langle \log_2 p_X \rangle_p
  = - \sum_i p_i \log_2 p_i.
\end{equation}
According to the Shannon coding theory, the quantity
$H(X) = - \sum_i p_i \log_2 p_i$, called the (Shannon) entropy, is equal to
the number of bits required to encode the information. It thus allows one
to quantify the quantity of information of a source $X$.

In practice the distribution $p_i$ of $X$ may be unknown, but we can
postulate that the source $X$ is described by another probability
distribution $q_i$. The expectation
\begin{equation}
  \langle Q_q(X) \rangle_p
  = -\langle \log_2 q_X \rangle_p
  = - \sum_i p_i \log_2 q_i 
\end{equation}
is the number of bits necessary to encode the same source (i.e., $X$), but
with $q$ (which is known) instead of $p$ (which is unknown).  It is
straightforward to show that
\begin{equation}
  \langle Q_q(X) \rangle_p
  =  -\langle \log_2 p_X \rangle_p + \langle  \log_2 \frac{p_X}{q_X} \rangle_p.
\end{equation}
We thus yield
\begin{equation}
  \langle Q_q(X) \rangle_p
  = H(X) + {\cal K}\left(p|q\right),
\end{equation}
where $H(X)$ is the unknown entropy of the source (i.e., the number of bits
necessary to encode the source $X$) and ${\cal K}\left(p|q\right)$ is the
Kullback-Leibler divergence. One can show that
${\cal K}\left(p|q\right)\geq 0$. Thus the entropy is the minimum amount of
information required to encode the source $X$ and the Kullback-Leibler
divergence represents the number of bits that needs to be added when one
encodes the source $X$ with $q$ instead of $p$, i.e., the cost (in bits) of
choosing the wrong probability density $q$.

When we quantify a continuous random variable $X$ with resolution $\Delta$
to get a discrete random variable $X^{\Delta}$, we can show
\begin{equation}
  H(X^\Delta)+\log\Delta \mapsto h(X)
  \quad \mbox{when} \quad
  \Delta \mapsto 0,
\end{equation}
where
\begin{equation}
  h(X)=-\int f_X(x) \log f_X(x) dx 
\end{equation}
is called the differential entropy. An important difference between entropy
and differential entropy is that $h(X)$ can be negative because of the
$\log\Delta$ offset. Nevertheless, the previous interpretation of the
Kulback-Leibler divergence (i.e., the cost of choosing a different
probability density instead of the actual one) remains valid up to a
constant offset related to the quantification resolution. In others words,
if we assume that there exists an unknown density $f_X$ of $X$, but we use
the density $g$ instead of $f$, we get
\begin{equation}
  -\langle \log_2 g_X \rangle_f = h(X)+{\cal K}(f|g),
\end{equation}
where the Kullback-Leibler divergence
${\cal K}(f|g)=\int f \log \frac{f}{g}$ is always positive.

Figure~\ref{fig:gaussian-mixture} shows the histograms of the negative
log$_2$-likelihood of the estimated Gaussian mixture for different data
sets.  It thus represents (within an unknown offset) the quantity of
information that is necessary to encode the data sets with the Gaussian
Mixture Model. The green histogram shows that the information cost is very
high for a uniformly distributed random set, while it is similar for the
training set and the test set.

\end{document}